\documentclass[fleqn,usenatbib]{mnras}

\usepackage{newtxtext,newtxmath}

\usepackage[T1]{fontenc}

\DeclareRobustCommand{\VAN}[3]{#2}
\let\VANthebibliography\thebibliography
\def\thebibliography{\DeclareRobustCommand{\VAN}[3]{##3}\VANthebibliography}

\usepackage{graphicx}	
\usepackage{amsmath}	
\usepackage{amssymb}	
\usepackage{cite}
\usepackage{hyperref}
\usepackage{color}
\usepackage{bm}

\usepackage{float}
\usepackage{CJKutf8}            
\definecolor{darkgreen}{rgb}{0,0.35,0}
\definecolor{blue}{rgb}{0,0,1}

\newcommand{\be}{\begin{eqnarray}}
\newcommand{\ee}{\end{eqnarray}}

\renewcommand{\vec}[1]{{\bm #1}}

\newcommand\ys{\bgroup\markoverwith{\textcolor[rgb]{1.0, 0, 1.0}{\rule[0.5ex]{8pt}{1.5pt}}}\ULon}

\defcitealias{Li2021}{L21}

\title[Accretion of Giant Planets in Protoplanetary Discs]{3D Global Simulations of Accretion onto Gap-opening Planets: Implications for Circumplanetary Disc Structures and Accretion Rates}

\author[Li, Chen \& Lin]{
Ya-Ping Li\begin{CJK*}{UTF8}{gbsn} (李亚平)\end{CJK*}$^{1,2}$\thanks{E-mail: liyp@shao.ac.cn},
Yi-Xian Chen\begin{CJK*}{UTF8}{gbsn} (陈逸贤)\end{CJK*} $^{3}$,
Douglas N. C. Lin\begin{CJK*}{UTF8}{gbsn} (林潮)\end{CJK*} $^{4,5}$
\\
$^1${Shanghai Astronomical Observatory, Chinese Academy of Sciences, Shanghai 200030, People’s Republic of China}\\
$^{2}${University of Chinese Academy of Sciences, 19A Yuquan Road, Beijing 100049, People’s Republic of China}\\
$^3${Department of Astrophysics, Princeton University, Princeton, NJ 08544, USA} \\
$^4${Department of Astronomy \& Astrophysics, University of California, Santa Cruz, CA 95064, USA}\\
$^5${Institute for Advanced Studies, Tsinghua University, Beijing 100086, People’s Republic of China}
}
\date{Accepted XXX. Received YYY; in original form ZZZ}

\pubyear{2023}

\begin{document}

\label{firstpage}
\pagerange{\pageref{firstpage}--\pageref{lastpage}}
\maketitle

\begin{abstract}
We perform a series of 3D simulations to study the accretion of giant planet embedded in protoplanetary discs (PPDs) over gap-opening timescales. We find that the accretion mass flux mainly comes from the intermediate latitude above the disc midplane. The circumplanetary disc (CPD) for a super-thermal planet is rotation-supported  up to $\sim$20-30\% of the planet Hill radius. While both mass inflow and outflow exists in the CPD midplane, 
the overall trend is an outflow that forms a meridional circulation with high-latitude inflows. 
We confirm the absence of accretion outburst from disc eccentricity excited by massive planets in our 3D simulations, contrary to the consensus of previous 2D simulations. 
This suggests the necessity of 3D simulations of accretion even for super-Jupiters. The accretion rates of planets measured in steady-state can be decomposed into the ``geometric" and ``density depletion" factors. 
Through extensive parameter survey, we identify a power-law scaling for the geometric factor $\propto q_{\rm th}^{2/3}$ for super-thermal planets ($q_{\rm th}$ being the thermal mass ratio), 
which transforms to $\propto q_{\rm th}^{2}$ for less massive cases. 
The density depletion factor is limited by the disc accretion rate for mildly super-thermal planets, and by gap-opening for highly super-thermal ones.
Moderate planetary eccentricities can enhance the accretion rates by a factor of $2-3$ through making the gap shallower, 
but does not impact the flow geometry. 
We have applied our simulations results to accreting protoplanet system PDS 70 and can satisfactorily explain the accretion rate and CPD size in observations.
\end{abstract}


\section{Introduction}\label{sec:intro}

In the widely accepted core accretion model for planet formation \citep{Bodenheimer1986,Pollack1996,Ida2004}, the quasi-steady
atmosphere accretion driven by slow Kelvin–Helmholtz contraction is followed by unstable runaway accretion, 
when an atmosphere’s mass grows beyond a critical core mass and thermal equilibrium is disrupted \citep[e.g.,][]{Pollack1996,Lee2014,Ormel2015,Chen2020,Zhong2022}. 
As the planet to star mass fraction grows past the disc thermal ratio $h^3$ where $h$ is the protoplanetary disc (PPD) aspect ratio, its tidal interaction with the disc produces a gap in the disc surface density profile \citep{LinPapaloizou1986}. 
In early paradigms of gap formation, gas depletion in the gap is severe and is able to quench the planet's subsequent growth, limiting a gas giants' final masses to be no more than $\sim M_{\rm th} =h^3 M_* $ \citep{LinPapaloizou1993,Bryden1999}, where $M_{\star}$ is the host star mass.

On the other hand, in the recently developed new gap opening paradigm (\citealt{Kanagawa2018}, see also recent review by \citealt{Paardekooper2022}), numerical studies verified that as long as disc viscosity is non-negligible,  there is always a residue minimum density within the planet proximity maintained by diffusion of materials across the gap \citep[e.g.,][]{Lubow1999,Kley2001,Bate2003,Duffell2013,Fung2014,Szulagyi2014,Durmann2015}.This may allow planets to acquire large masses through accretion, enough to excite the disc eccentricity \citep{Goldreich2003}. Indeed in many 2D simulations, 
it has been found that the gap carved by a very massive planet becomes eccentric, 
which can induce accretion burst onto the planet and rapidly produce super Jupiters or brown dwarfs \citep{Papaloizou2001,Kley2006,Li2021,Tanaka2022}, unless the PPD dissipates very quickly. 

However, the accretion geometry of low mass embedded planets are quite different in 3D compared to 2D.
A robust flow pattern of polar inflow accompanied with midplane outflow were found in early 3D simulations for embedded planets \citep{Machida2008,Wang2014,Ayliffe2012,Tanigawa2012}. This meridional circulation comes from high-latitude horseshoe flows which penetrate deep into the atmosphere before outflowing in the midplane \citep{Machida2008,Fung2015}. 
Such kind of circulation could happen for different planet masses and different equations of states \citep[e.g.,][]{Ormel2015,Lambrechts2017,Schulik2019,Zhu2021}.
In contrast to 2D, 
in these 3D global simulations we do not observe excitation of mass accretion rate even up to $\gtrsim 10\ M_J$ \citep{DAngelo2003,Bodenheimer2013,Choksi2023}. This dichotomy suggests that for the relevant parameter space it could be much harder to excite disc eccentricity in 3D than compared to 2D.
Interestingly, such eccentric cavities has been seen in 3D magnetohydrodynamics simulations for circumbinary simulations \citep[e.g.,][]{Shi2012},
which suggests that the growth of eccentricity is a robust
property of discs around accreting binaries with large enough mass ratio.

In this study, we run high resolution, long-term 3D global simulations of super-thermal-mass gas giant accretion to revisit the structure and dynamics of isothermal circumplanetary discs (CPDs). We confirm that in 3D mass flux comes from higher characteristic altitudes as the planet mass reduces from the super thermal to sub-thermal limits. 
The flow pattern thus gradually transforms from Hill accretion towards Bondi accretion.
By comparing the gap profiles and streamlines' eccentricities with those of 2D simulations, we identify the 
main cause for the absence of eccentricity excitation in 3D simulations may be associated with the 
gap being narrower near the planet's azimuth, and therefore damping of eccentricity is more effective. 
As a by-product of the parameter study that explores the planet's accretion-rate dependencies on the disc 
scale height and viscosity, we also construct a simple accretion-rate scaling function in terms of 
the planet's mass and ambient density. 
We are aware that very recently, \citet{Choksi2023} have
concluded an accretion rate with similar numerical procedures. 
Their simulations usually do not reach 100 orbits before the gap becomes steady, while our study, 
which are 
computed to quasi steady-states on much longer timescales, 
can serve as cross-validation and supplements 
to their results.

Another effect is the orbital eccentricity of the planet itself. 
Most existing simulations assume a circular orbit
for the planet, with the exception of those carried out by \citet{Bailey2021,Li2022,ChenYX2022}.
However, \citet{Bailey2021,ChenYX2022} focused on the moderate planetary thermal mass regime with $q_{\rm th} 
\lesssim 6$ and did not implement any sinkhole prescription for accretion onto the planet. 
While \citet{Li2022} had considered the accretion for the embedded object in disc, 
they mainly adopted model parameters in the sub-thermal regime. Those 2D simulations 
are mostly relevant for embedded 
stellar-mass black holes (sBHs) in active galactic nucleus (AGN) discs. Such eccentricity is hard to 
maintain in PPDs.  In this study, we relax the circular-orbit assumption in the low-orbital-
eccentricity limit.  Using 3D simulations, we investigate the robustness of these accretion-rate 
measurements for planets with both sub-thermal and super-thermal masses.

This paper is organized as follows. We briefly overview the accretion theory for the embedded objects in discs in 
\S~\ref{sec:theo}, and present the numerical method for our simulations in \S~\ref{sec:method}. The results are shown in \S~\ref{sec:results}, which is followed by observational implication for accreting protoplanets and embedded 
stellar-mass black holes in AGN discs in \S~\ref{sec:imp}. 
The conclusion and discussion are presented in \S~\ref{sec:con}.

\section{Modified Bondi Accretion}\label{sec:theo}

In the runaway phase of giant planet formation, the dynamical interaction between the embedded planet and the surrounding protostellar becomes vital for the gas accretion and the growth of protoplanet. 
The dynamical accretion process could be also relevant for the accretion of the embedded sBHs in AGN discs. 
For simplicity, we use planet to refer the embedded object in disc to avoid confusion. 
There exist several important length scales which are relevant to the accretion processes of the embedded planet.
The first length scale is planet's Bondi radius $R_{\rm B}$,

\begin{equation}\label{eq:rb}
R_{\rm B}\equiv \frac{GM_{\rm p}}{c_{\rm s}^{2}},
\end{equation}
where $G$, $M_{\rm p}$, and $c_{\rm s}$ are the gravitational constant, planet mass, and the sound speed. 
This describes the regime within which the planet gravity dominates over the thermal pressure forces of the gas. 
The second one is the planet's
Hill's radius $r_{\rm h}$,
\begin{equation}\label{eq:rh}
r_{\rm h}\equiv r_{\rm 0}\left(\frac{q}{3}\right)^{1/3},
\end{equation}
where $r_{\rm 0}$ is the companion's semi-major axis and characteristic distance (for low eccentricity) to the central object, $q$ is the mass ratio between the planet mass and the central object mass $M_{\star}$.  $r_{\rm h}$ is determined by the balance between the planet gravity and that of star gravity.  Another natural scale is the disc scale height $H$ based on the vertical hydrostatic equilibrium, 
\begin{equation}\label{eq:h}
H\equiv \frac{c_{\rm s}}{\Omega_0},
\end{equation}
where $\Omega_0$ is the local Keplerian frequency at $r_0$. 

Whether a planet can form a circumplanetary disc or an envelope depends on
the balance between the gravitational tidal forces and the pressure gradient \citep{Szulagyi2016,Szulagyi2017}. This define a planet thermal mass parameter,
\begin{equation}\label{eq:qth}
q_{\rm th}\equiv \frac{R_{\rm B}}{H_{\rm p}}=\frac{q}{h_{0}^{3}},
\end{equation}
where $h_{0}=H_{\rm p}/r_{\rm 0}$, and $H_{\rm p}=H_{0}$ is the disc scale height at $r_{\rm 0}$.

Now we discuss accretion onto the embedded object like planets in PPDs. For the less massive planets with $R_{\rm B}/r_{\rm h}=(3q_{\rm th}^{2})^{1/3}\lesssim1$, i.e., $q_{\rm th}\lesssim \sqrt{3}/3$, this sub-thermal limit 
also leads to $r_{\rm h}<H_{0}$. Therefore, the tidal force becomes less important, and the accretion could take the form of Bondi formulae from a uniform medium without angular momentum.  In this case, the planetary accretion could be approximated by

\begin{eqnarray}\label{eq:mdot_mb}
\dot{m}_{\rm B} &\equiv& \pi R_{\rm B}^{2} \times \rho_{\rm p}c_{\rm s}, \nonumber \\
&=& \pi q_{\rm th}^{2} h_{0}^{3} \rho_{\rm p}r_{0}^{3}\Omega_{0},
\end{eqnarray}
where $\rho_{\rm p}$ is the characteristic gas density at the planet location.

Another circumstance, e.g., $r_{\rm h}/H_{\rm 0}=(q_{\rm th}/3)^{1/3}>1$,  which happens for very massive planets in the super-thermal limit 
with $q_{\rm th}>3$, and $R_{\rm B}>r_{\rm h}$, is the more relevant regime explored in this work. The Hill's sphere of the planet pops out from the PPD, then the accretion proceeds as \citep{Rosenthal2020,Choksi2023}

\begin{eqnarray}\label{eq:mdot_mh}
\dot{m}_{\rm H} &\equiv& \pi r_{\rm h}H_{0} \times \Omega_{0}r_{\rm h}\rho_{\rm p} , \nonumber \\
&=& \pi\left(\frac{q_{\rm th}}{3}\right)^{2/3}h_{0}^{3}\rho_{\rm p}r_{\rm 0}^{3}\Omega_{0},
\end{eqnarray}
which suggests a sublinear scaling of planetary accretion rate $\dot{m}_{\rm p} \propto q_{\rm th}^{2/3}$ for super-thermal planets' accretion if we neglect the gap density dependence here. This is termed as Hill accretion hereafter. Note that there are some simulations which suggest a scaling relation of $\dot{m}_{\rm p} \propto q_{\rm th}^{4/3}$ for the marginally super-thermal regime $q_{\rm th}>1$ \citep[e.g.,][]{Maeda2022}. We will come back to this issue in Section~\ref{sec:scaling_relation}.

The accretion rates discussed above only takes into account the ``geometric factors". In the super-thermal context, 
we also expect gap formation from tidal effects to significantly reduce $\rho_0$ from the local density of an unperturbed PPD towards $\rho_{\rm p} \ll \rho_0$ \citep{LinPapaloizou1993}, 
and therefore $\dot{m}_{\rm p}$ measured in the natural units of $\rho_0 r_0^3 \Omega_0$ will not be linear anymore.
\citet{Duffell2013,Kanagawa2018} proposed a scaling relation for this ``density depletion factor" $\rho_{\rm p}/\rho_0$ (or the column-integrated analog $\Sigma_{\rm p}/\Sigma_0$) dependent on the disc viscosity, aspect ratio and planet mass ratio from 2D simulations, and one can include such a factor in the empirical formula for Hill accretion \citep{Tanigawa2016,Rosenthal2020}.
However, we note that self-consistency of this scaling relies on the assumption that the gap is relatively flat and shallow, which might break down at high $q_{\rm th}$ \citep{Chen2020b}. 
To separate this density factor in the study of mass accretion rate, 
we also provide measurements of $\dot{m}_{\rm p}$ in the units of $\rho_{\rm p} r_0^3 \Omega_0$ with $\rho_{\rm p}$ being an average reduced density in steady state, see \S \ref{sec:scaling_relation} for details. {However, 
it should be noted that this depletion factor assumes that planetary gas removal does not significantly alter the accretion structure of the global disc. 
When appropriate inner boundary condition is considered, 
the inner disc is expected to experience significant depletion when the rate of planetary accretion approaches that of the PPD. 
This in turn regulates gap formation to ensure accretion rate onto the planet is capped by the outer disc replenishment, 
and the steady-state value of $\rho_{\rm p}/\rho_0$ becomes more complicated \citep{Rosenthal2020}.
This factor, however, does not affect the scaling of accretion rate when measured in the units of $\rho_{\rm p} r_0^3 \Omega_0$.
We will discuss this in more details in Appendix~\ref{app:inner_bc}.}

\section{Method}\label{sec:method}

We use Athena++ code \citep{Stone2020} to simulate the gravitational interaction of an embedded planet with a disc.

The disc around a pre-main sequence (PMS) star
with a mass of $M_{\star} = 1.0\ M_{\sun}$  is initialized with a power-law gas surface density profile. 
The initial density profile at the midplane is 

\begin{equation}
    \rho(r,z=0)=\rho_{0}(r_0,z=0)\left(\frac{r}{r_{0}}\right)^{p}
\end{equation}
with $p=-1.5$ unless otherwise stated, where $\rho_{0}=\Sigma_{0}/\sqrt{2\pi} H_{0}$ is the midplane density at $r_{0}$, 
The locally isothermal temperature profile is initialized as 
\begin{equation}
T(r)=T(r_{\rm 0})\left(\frac{r}{r_{0}}\right)^{\zeta},
\end{equation}
where $\zeta=-1.0$ unless otherwise noted. This is to mimick an irradiated disc with a cooling time much shorter than its the dynamical time.
The vertical distributions of the initial density and velocity profiles 
are set according to the hydrostatic equilibrium \citep{Nelson2013}:

\begin{equation}
\rho = \rho_{0}(r,z=0)\exp\left[\frac{GM_{\star}}{c_{s}^2} \left(\frac{1}{\sqrt{R^2+z^2}}-\frac{1}{R}\right) \right],
\end{equation}
and 
\begin{equation}
v_{\phi}(R,z) = v_{\rm K}\left[ (p+\zeta)\left(\frac{c_{s}}{v_{\rm K}}\right)^{2}+1+\zeta-\frac{\zeta R}{\sqrt{R^2+z^2}}\right]^{1/2}, 
\end{equation}
where $v_{\rm K}=\sqrt{GM_{\star}/R}$ is the local Keplerian velocity, $R, z$ are defined in the cylindrical coordinate system.
The disc density is low enough to ensure that the disc self-gravity can thus be neglected. 
Note that the simulations are scale-free so $r_{0}$ and $\rho_{0}$ can be appropriately chosen to be applicable to the observed systems.

We choose a locally isothermal equation of state (EoS) with the sound speed $c_{\rm s}$ given by 
$\frac{c_{\rm s}}{v_{\rm K}}=\frac{H}{r}=h_{0}$,
for our temperature profile, where $h_{0}=0.05$ is adopted as a typical value for the disc aspect ratio, $H$ is the disc scale height. 
The disc viscosity is adopted from the Shakura-Sunyaev prescription $\nu_{\rm g}=\alpha c_{\rm s}H$ with
a constant $\alpha$ across the whole disc \citep{Shakura1973}. { In the absence of embedded accreting planets, 
the steady-state accretion rate 
\begin{equation}
    {\dot M}_{\rm d} = 3 \pi \Sigma \nu_{\rm g} = \sqrt {18 \pi^3} \alpha h^3 \rho \Omega r^3
    \label{eq:steadymdot}
\end{equation}
is a constant at all radii throughout the global disc. The unperturbed steady-state value of ${\dot M}_{\rm d}$ at $r_0$, 
${\dot M}_{\rm 0} = \sqrt {18 \pi^3} \alpha h_0^3 \rho_0 \Omega_0 r_0^3$, 
can be used to calibrate 
${\dot M}_{\rm d}$ and ${\dot m}_{\rm p}$ for models with embedded accreting planets.}
We choose $\alpha=0.001$ as our fiducial model parameter, and also explore the dependence of $\alpha$ for some specific models.

The planet is usually fixed at a circular orbit with distance $r_{\rm p}=r_{0}$. For a moderate orbital eccentricity of the planet ($e\lesssim0.1$), we approximate the eccentric
orbital motion to first order in eccentricity $e$, $r_{\rm p}=r_{0}(1-e\sin\Omega_{0}t)$, $\phi_{\rm p}=-2e\cos\Omega_{0}t$.
This treatment can avoid computational complexity from integrating Kepler’s equations. 

In calculating the gravitational potential of the planet at $\vec r$, we use a smoothed potential 
of the form \citep[e.g.,][]{GT1980}
\begin{equation}
\phi_{\rm p}=-\frac{G M_{\mathrm{p}}}{(\left|\vec{r}_{\rm p}-\vec{r}\right|^2+\epsilon^2)^{1/2}}+q \Omega_{\rm p}^{2} \vec r_{\rm p} \cdot \vec r
\label{eq:potential}
\end{equation}
where $\vec{r}_{\rm p}$ indicates the location of the planet, $\epsilon=0.1\ r_{\rm h}$ is the softening length for most cases.
The second term on the 
right hand side of the above equation corresponds to the indirect term due to our choice of heliocentric coordinate system. Hereafter we refer to $\left|\vec{r}_{\rm p}-\vec{r}\right|$ as $\delta r$.

To model the active accretion of the embedded planet, we follow previous works and implement a sink hole around the planet (\citealt{Li2021}, see also \citealt{Kley2001,DAngelo2003}). Accretion is determined by the sink hole radius $r_{\rm a}$, and the removal rate $f$ in unit of local Keplerian frequency $\Omega_{0}$. We remove a uniform fraction of mass in every cell within $\delta r < r_{\rm a}$ each numerical timestep, such that when the density profile within the sink hole settles to a steady state, 
the removal rate converges with the integrated mass flux into the sink hole and roughly $100\%$ of the mass within the sink hole is accreted within a timescale of $f^{-1}$. Because the simulation domain is only half a global disc, we further multiply this rate by a factor of 2.
In our fiducial setup, we set $r_{\rm a}=0.1\ r_{\rm h}$ and $f=5\Omega_{0}$. Numerical convergence of planet accretion rate $\dot{m}_{\rm p}$ for different accretion parameters has been verified in the Appendix~\ref{app:acc}.

A static (adaptive for eccentric planets) mesh refinement is adopted to resolve the region around the planet. We use a base grid with 128 radial grids spaced  uniform between $r_{\rm min}=0.5\ r_{0}$, $r_{\rm max}=2.5\ r_{0}$, 16 uniform grids within 4 disc scale heights from the midplane, and 512 uniform grids in azimuth. 
Only half disc above the midplane is simulated to save the computation expense after considering the symmetry. Four levels of mesh refinement is adopted within the region $\delta r<r_{\rm h}$, details see \citet{ChenYX2022}. This treatment significantly reduces computation cost and make large parameter surveys in 3D possible.
For our fiducial case of planet mass $q=0.001$, the Hill sphere can be resolved by about 50 cells in each dimension. 
For certain models we have extended the radial and vertical domains to test the effect of the boundary, as shown in Appendix~\ref{app:bd}.
In most cases, 
we adopt a fixed boundary condition in the inner and outer radial edge,
and a reflecting boundary for $\theta$ in the direction.
To avoid the wave reflections, we apply wave-killing regions in inner and outer radial boundaries \citep{deValborro2006}.

The fixed boundary conditions are necessary to ensure that the gap profile and planetary accretion rates are able to evolve into steady states within a computationally feasible timescale. 
However, such boundary conditions cannot capture the reduction of disc accretion rate $\dot{M}_{\rm d}$ across the gap due to the removal by the planet - mass is somewhat artificially generated in proximity to the inner boundary to maintain the fixed disc accretion rate. 
In Appendix \ref{app:inner_bc} we present a few test cases with a modified inner boundary condition (where flow into the computation domain is prohibited), 
just to demonstrate the point that realistically the absolute value of $\dot{m}_{\rm p}$ will always be capped by the outer disc accretion rate (a pre-specified ${\dot M}_{0}$), while the inner disc will be gradually depleted over a few viscous timescales. 
This difference in the inner boundary numerical treatment, however, does not affect the flow structure in the circumplanetary region.

\subsection{2D models}

For 2D models presented in \S \ref{sec:higq}, we use the FARGO3D code \citep{Benitez-Llambay2016} to simulate the gravitational interaction of an embedded circular/eccentric planet with a disc. The temperature scales with distance $R^{-1}$, and
the surface density profile is 
\begin{equation}\label{eq:gas}
  \Sigma_{\rm g}(R)=\Sigma_{0}\left(\frac{R}{r_{\rm 0}}\right)^{-\gamma},
\end{equation}
where $\gamma=0.5$, consistent with the density profile in our 3D simulations. The choice of $\Sigma_{0}$ is arbitrary and has no consequence on the normalized accretion rates we measure.

\begin{figure}
\centering
\includegraphics[width=0.45\textwidth,clip=true]{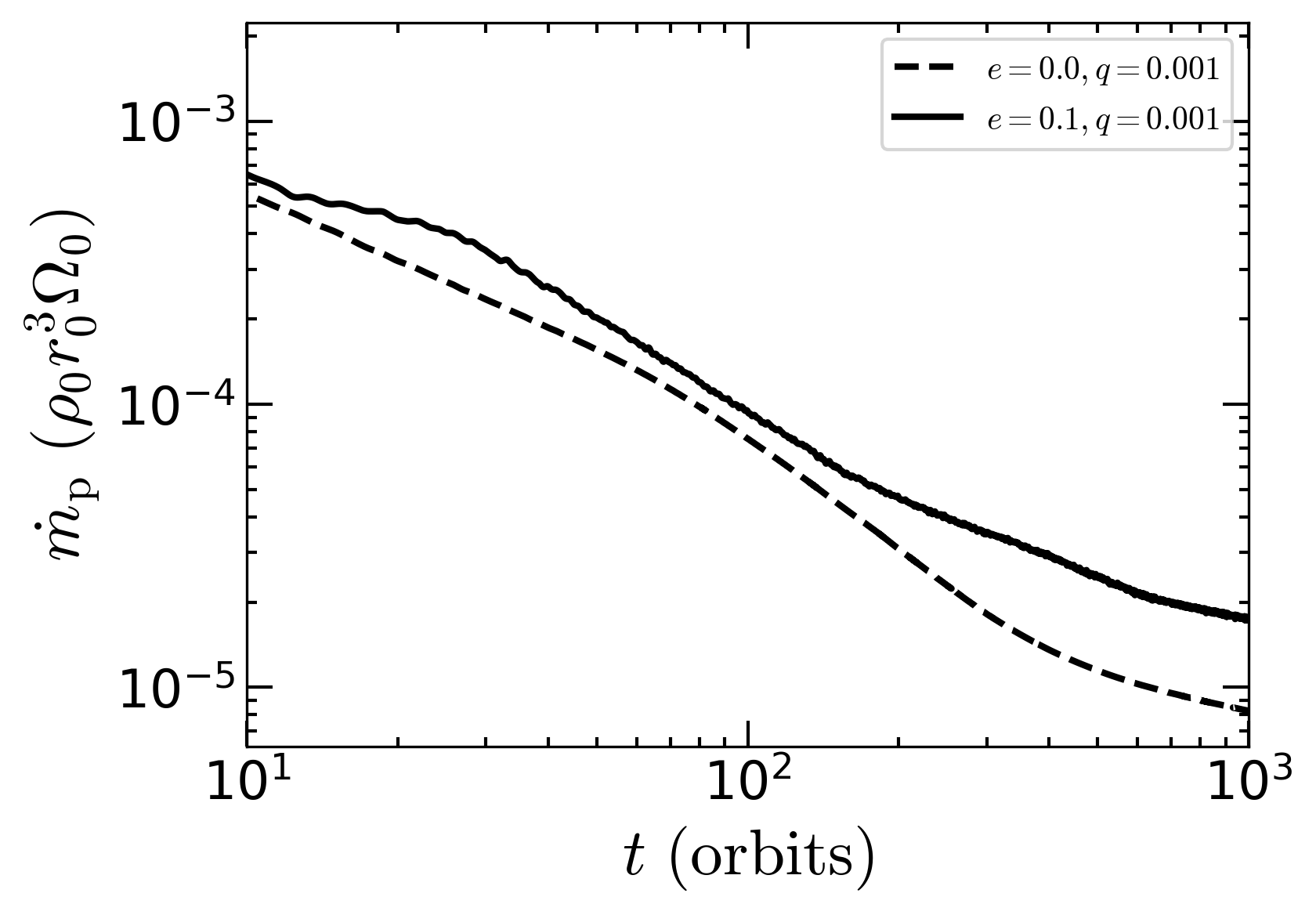}
\includegraphics[width=0.45\textwidth,clip=true]{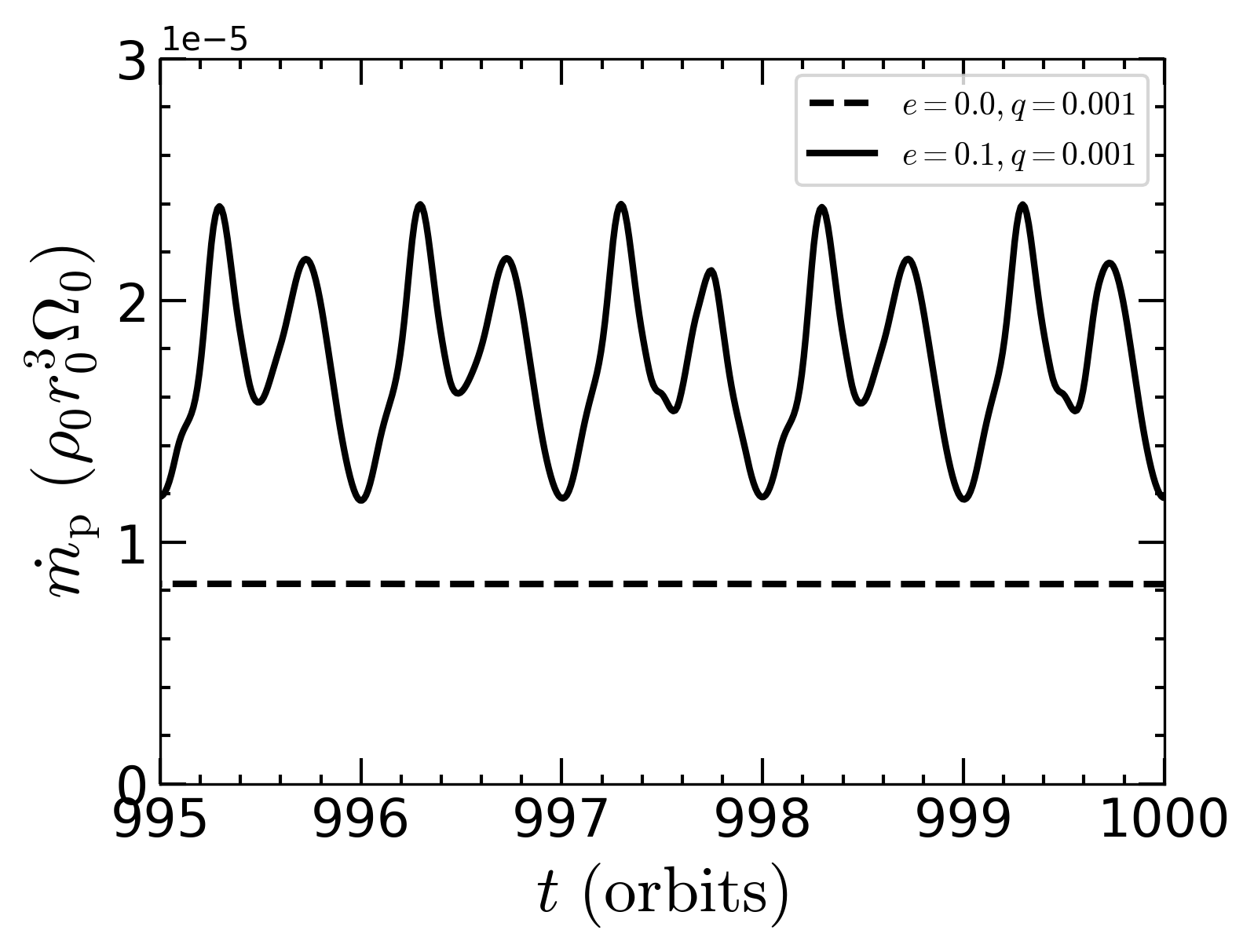}
\caption{The evolution of planetary accretion rate, measured in scale-free unit $\rho_{\rm 0}r_{\rm 0}^3\Omega_{\rm 0}$. The accretion rates in the upper panel have been smoothed with a time averaging, while the accretion rate for $e=0.1$ case in the lower panel without the time-averaging shows strong variability.}
 \label{fig:mdot_ecc}
\end{figure}

\begin{figure}
\centering
\includegraphics[width=0.45\textwidth,clip=true]{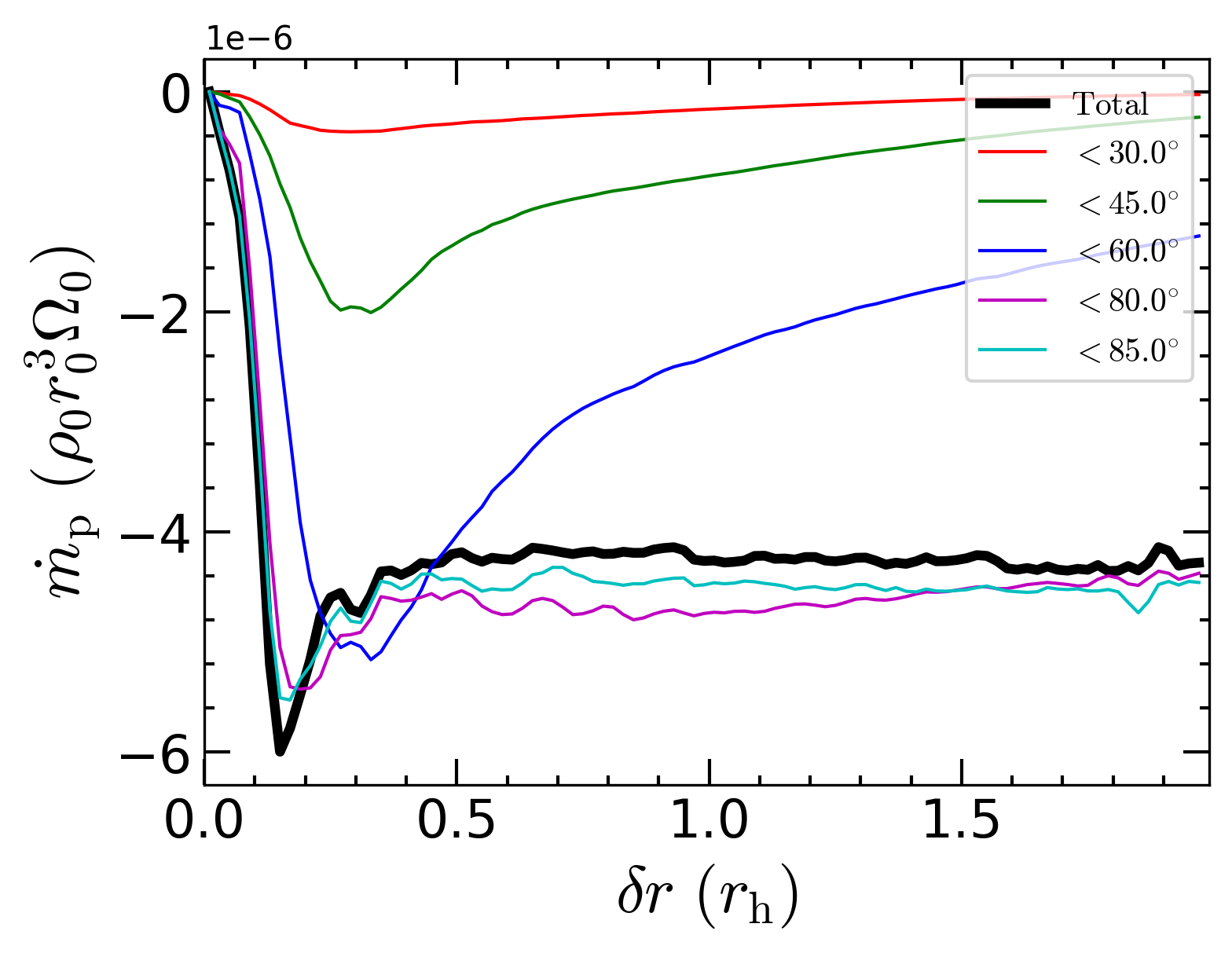}
\caption{The radial profile (in spherical polar coordinates relative to the planet) of mass flux onto the planet for our fiducial circular planet with $q=0001$, $h_{0}=0.05$, $\alpha=0.001$. The black solid line integrates over the half disc ($90^{\rm o}$ polar angle), which can be compared with the mass accretion rates shown in Figure~\ref{fig:mdot_ecc} (dashed lines) by multiplying 2. Lines with different colors show the contribution within different polar-angles latitude in the spherical coordinate relative to the planet. 
}
 \label{fig:fluxmass1d_fid}
\end{figure}

We solve the hydrodynamics equations with a high-resolution uniform 2D polar grid
of $(n_{R},n_{\phi})=600\times3072$ in the radial and azimuthal direction. The radial grid is refined in the region of $[0.6,1.4] r_{0}$ with a resolution of $0.002\ r_{0}$. The whole radial domain is in between $[0.4-4.0]\ r_{0}$. We have tested that a smaller inner boundary of $0.3r_{0}$ does not change the results. The same softening scale as 3D simulations is adopted to make quantitative comparison.
The outer boundary condition is similar to 3D in both radial and azimuthal directions.

\section{Results}\label{sec:results}

\subsection{Circular Planet with Intermediate $q_{\rm th}=8.0$}

We first present the accretion history for a circular planet with $q_{\rm th}=8.0$. The corresponding model parameter is $q=0.001$, $h_{0}=0.05$, and $\alpha=0.001$.

\subsubsection{Time Evolution of Accretion Rates}

The long-term and short term (in steady-state) accretion rates onto the planets are shown in Figure~\ref{fig:mdot_ecc}. 
We evolve the system about 1000 planetary orbits to ensure the accretion onto the planet reach a quasi-steady state.
Following an initial accretion burst associated with the dense environment, the accretion rate for our fiducial circular 
($e=0$) planet gradually decreases and then settles down to its stable state both in short- and long-time scale.  After the evolution of the gap carved by the planet becomes insignificant, 
the accretion rate evolution also ceases to evolve significantly.

To show the steady state within the circumplanetary disc, 
we plot the azimuthal-averaged radial profile (in spherical polar coordinates $\delta r$) 
of the mass inflow rates, within various polar-angle latitudes, as a function of distance to the planet in 
Figure~\ref{fig:fluxmass1d_fid} for the circular fiducial case (black solid line, see next Section for analysis of different vertical components). 
Note that the total inflow rate (obtained from integrating mass flux over all $0-90^{\rm o}$ polar angles centered on the 
planet) evaluated at $\delta r \simeq 3r_{\rm a} = 0.3 r_{\rm h}$ , is consistent with half of the converged planetary accretion rate measured from sink hole removal, both being half of the value shown in Figure~\ref{fig:mdot_ecc}. This mass conservation is a good indication that our simulations have reached a steady state.

\begin{figure*}
\centering
\includegraphics[width=0.45\textwidth,clip=true]{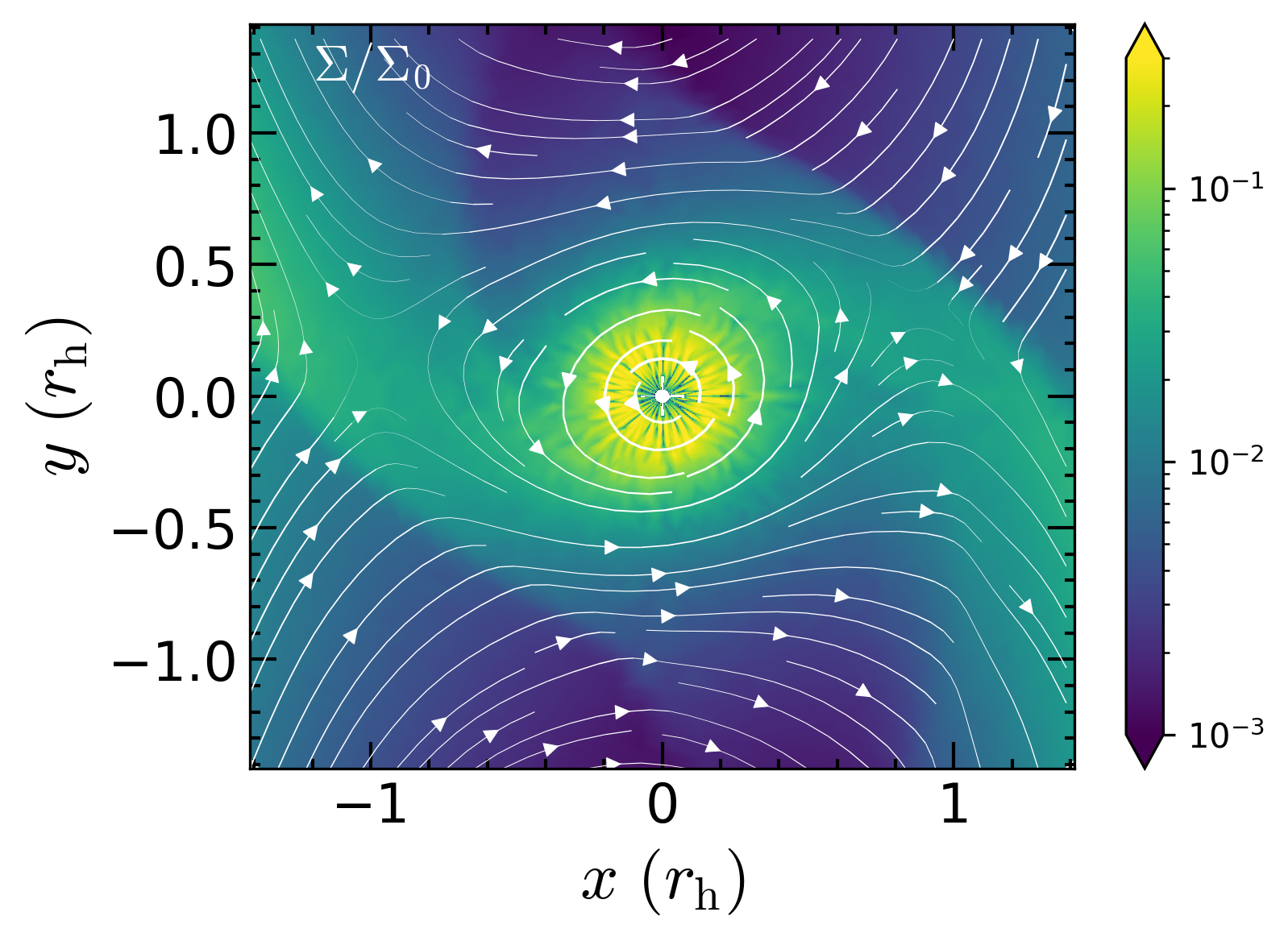}
\includegraphics[width=0.45\textwidth,clip=true]{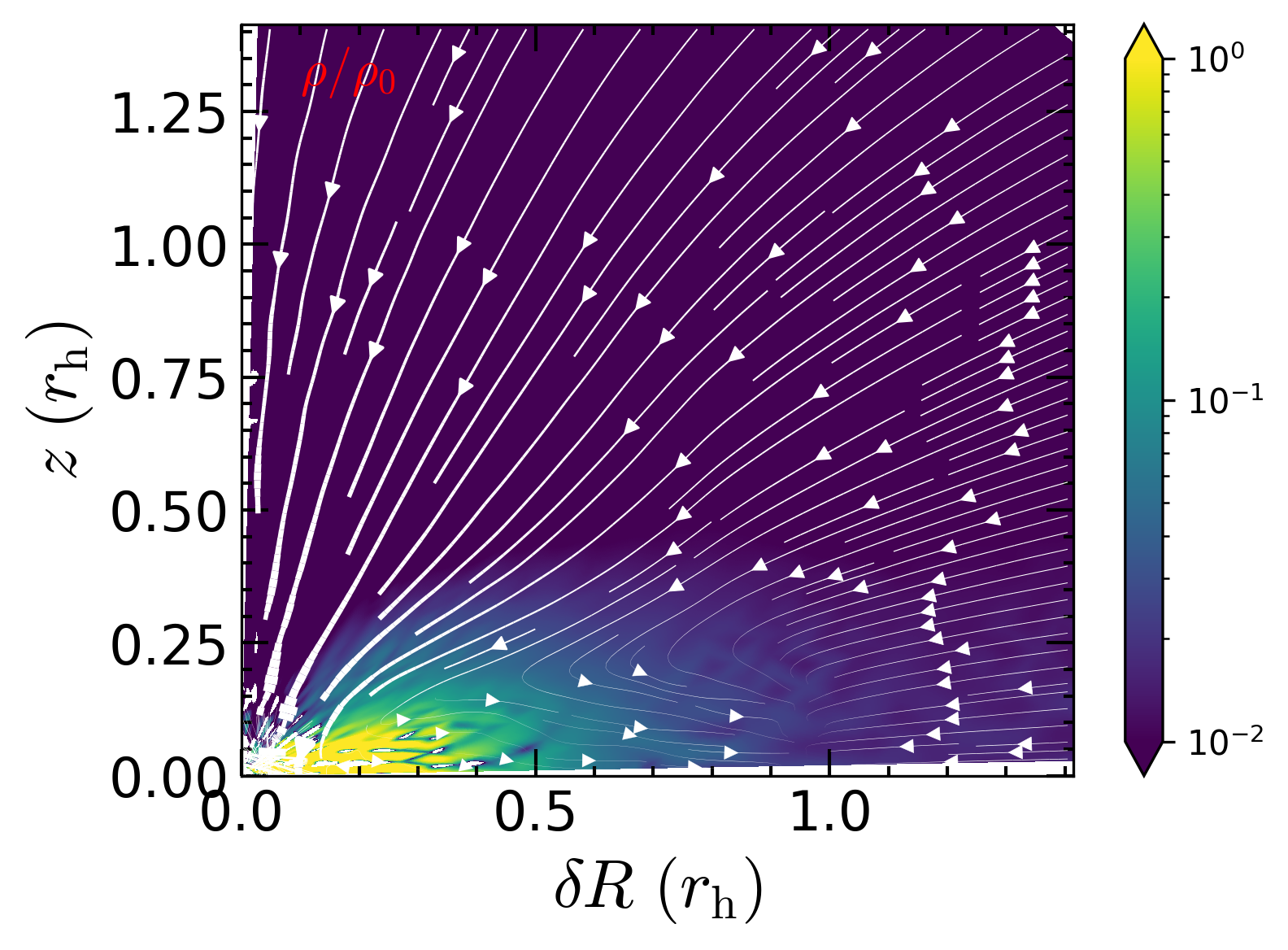}
\includegraphics[width=0.45\textwidth,clip=true]{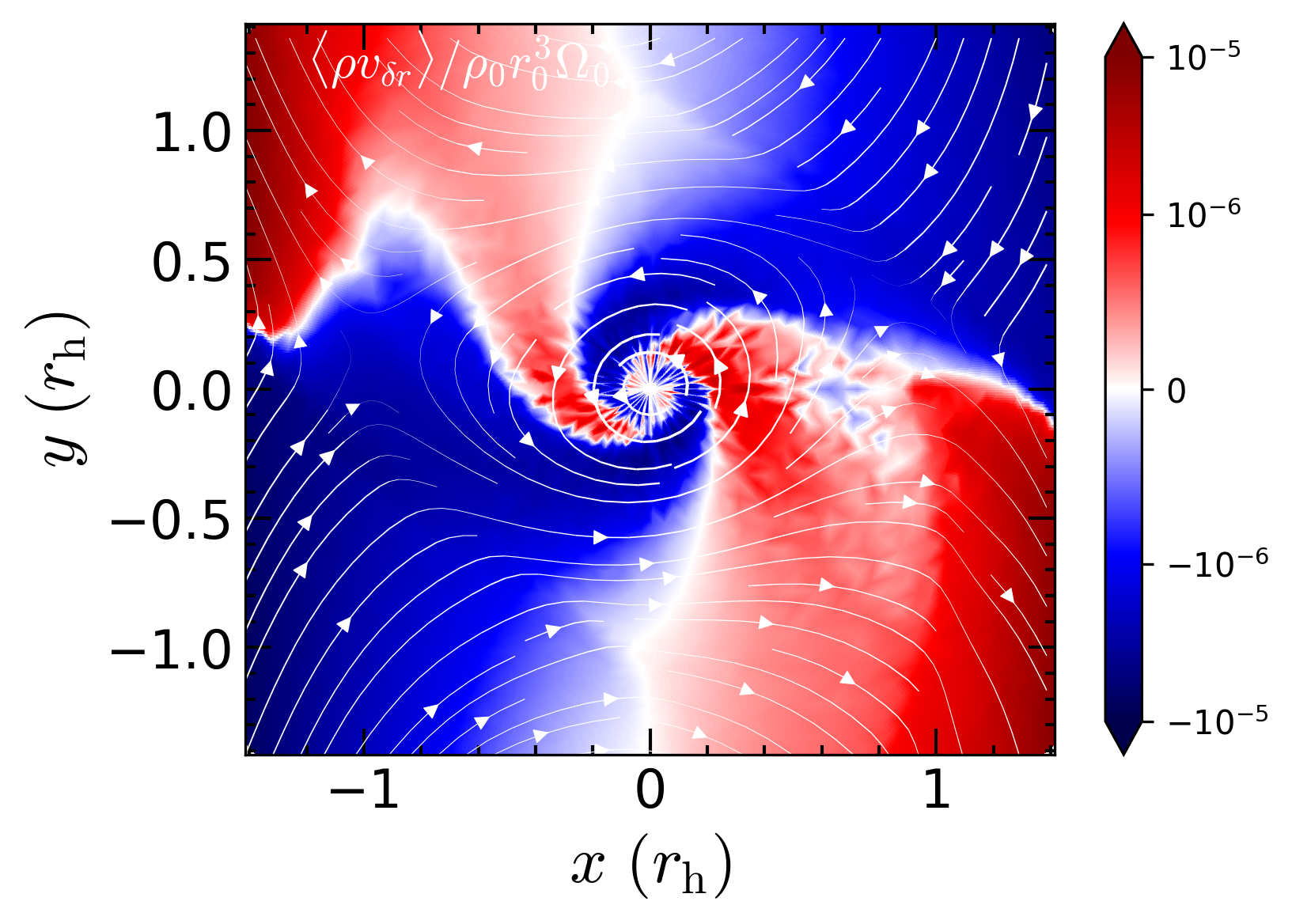}
\includegraphics[width=0.45\textwidth,clip=true]{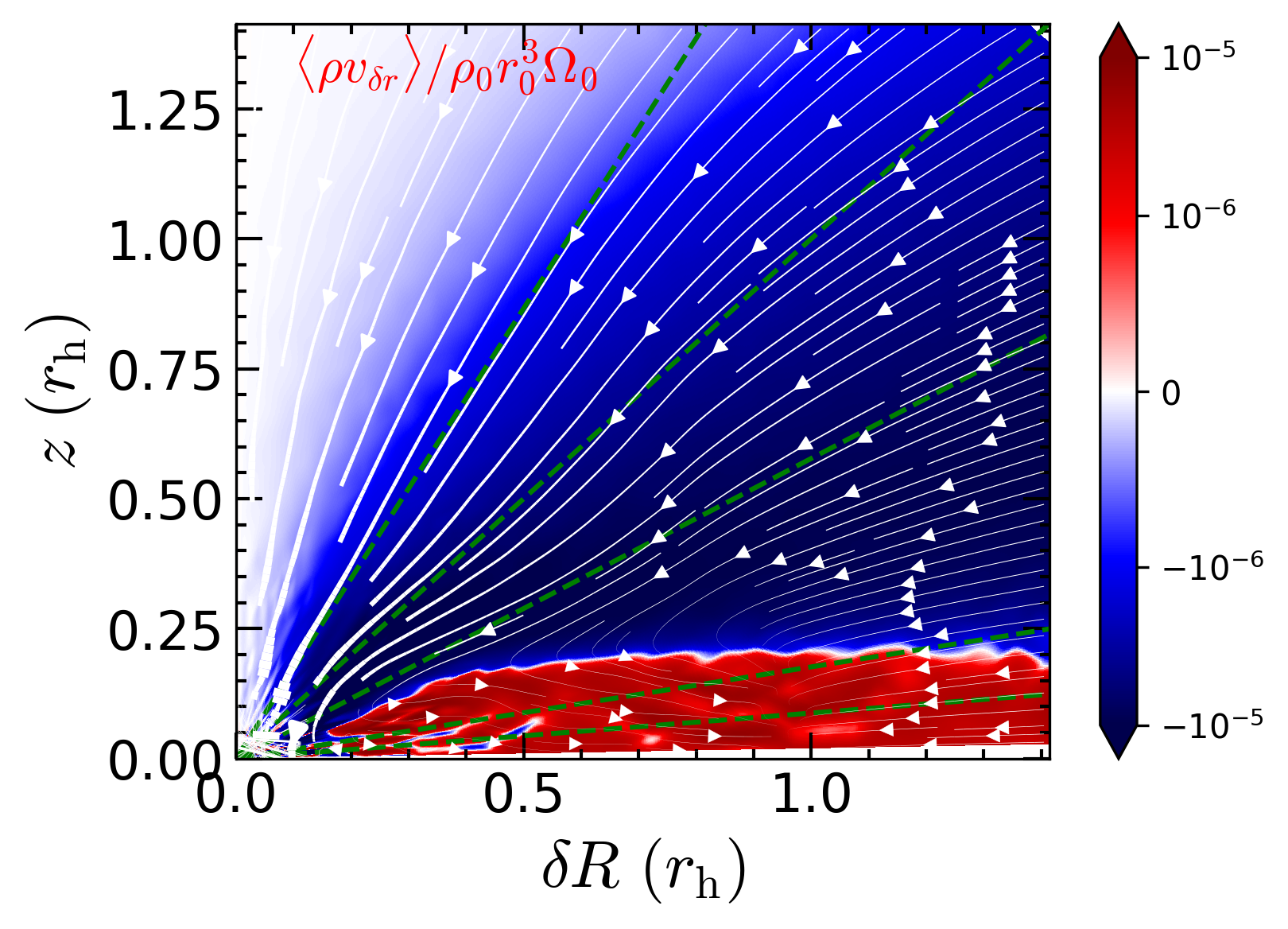}
\caption{Upper left panel: time-averaged  surface density for our fiducial model. Upper right: time-, and azimuthally- averaged density distribution in the $\delta R-z$ plane.  Lower left: time-averaged vertically-integrated mass flux onto the embedded planet, negative (represented by the blue or black colors) indicates accretion and positive (represented by the red color) 
indicates decretion. Lower right: time-, and azimuthally-averaged mass flux onto the planet. 
All plots are shown at $\approx$1000 orbits and time-average is over one orbit. 
The arrows in the left panels show the $x-y$ velocity vectors in midplane plane relative to the planet, 
while the arrows in the right panels show the the azimuthally-averaged $\delta R-z$ velocity vectors with streamlines. 
The thickness of these streamlines indicates the velocity magnitude. 
The direction and magnitude of the mass flux are indicated by the color. Horseshoe flow pattern is
clearly visible at the top and bottom of the left panels.  Near the midplane, 
the radial velocity can converge while the mass flux is still going outwards relative to the planet 
at $\delta R \sim r_{\rm H}$.
This mismatch is associated with the non-axisymmetric density and velocity around this region. A net mass influx into 
the planetary proximity is mostly channeled through the mid-latitude. The dashed lines show the integration boundaries of mass flux over different lattitude ranges as shown in Figure \ref{fig:fluxmass1d_fid}. }
 \label{fig:fluxmass_fid}
\end{figure*}

\subsubsection{Density and Mass Flux Distribution}

To diagnose the accretion structure of CPDs, we show the density and mass flux distribution close to the planet.
In the left panels of Figure~\ref{fig:fluxmass_fid}, we show the time-averaged surface density distribution (measured in the unit of $\Sigma_0$) in the upper panel, and the time-averaged mass flux $\langle\rho v_{\delta r}\rangle$ in the lower panel, 
where $v_{\delta r}$ is the fluid velocity in the $\delta r$ direction. 
The mass flux is also integrated over the vertical direction and measured in code unit $\rho_0 r_0^3 \Omega_0$.
The time-averaged velocity streamline is overlaid in each plot, 
with the thickness of the streamline proportional to the velocity magnitude. Note that the jagged patterns in the very inner region around the planet are numerical artifacts.

\begin{figure}
\centering
\includegraphics[width=0.45\textwidth,clip=true]{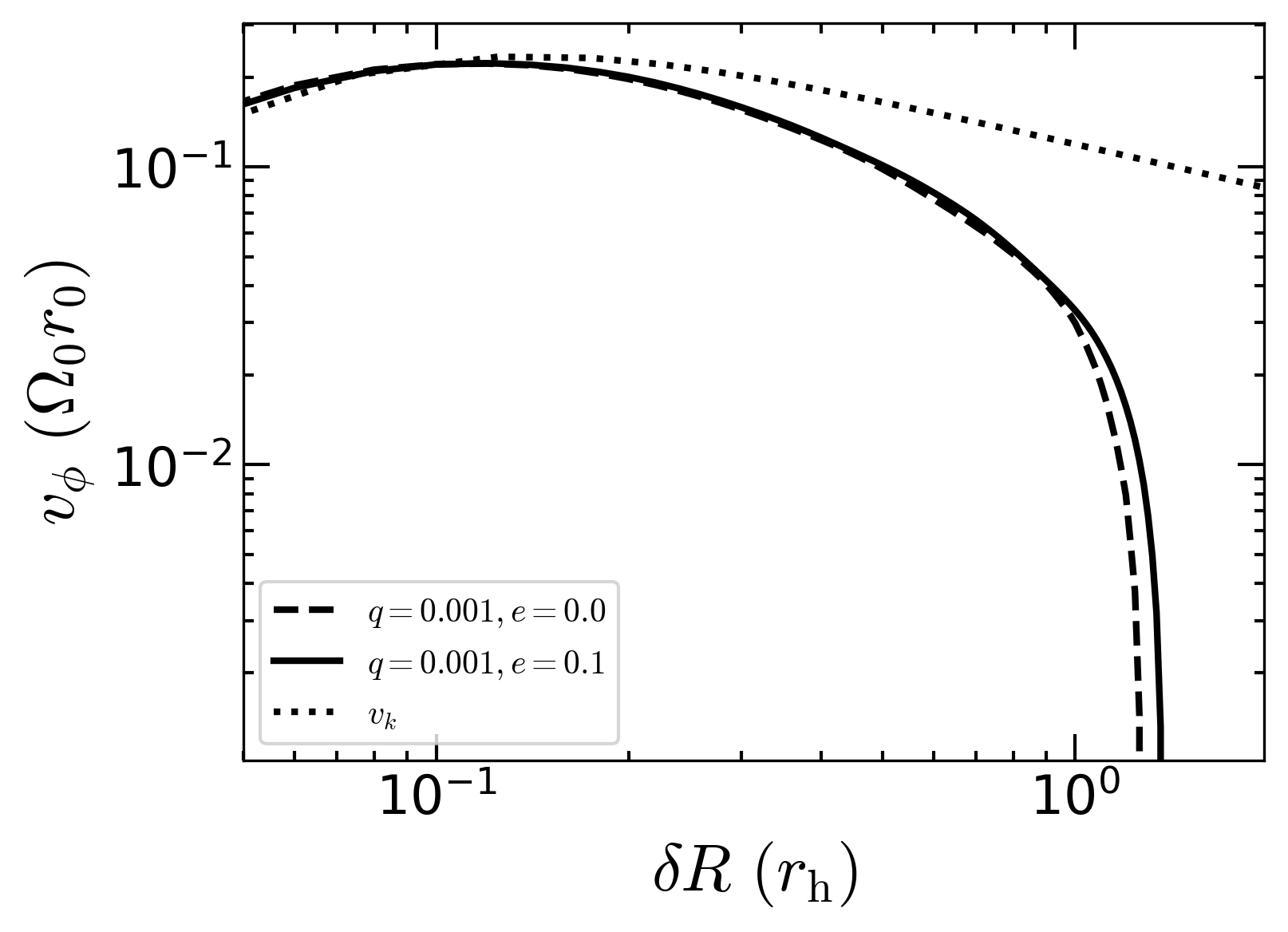}
\caption{Azimuthally averaged midplane rotation velocity in the CPD in the co-moving frame of the planet, as a function of 
the cylindrical radius $\delta R$. 
The comparison Keplerian velocity $v_{\rm k}$ relative to the
planet includes contribution from the softening parameter but neglects the 
stellar gravity.
This plot is analogous to Figure 3 of the 2D simulation 
by \citet{Li2022}. The CPD is dominated by nearly Keplerian rotation up to $0.2\sim0.3\ r_{\rm h}$. 
It is no longer predominantly supported by a centrifugal balance near $r_{\rm h}$ and the disc  
flow is interrupted beyond $r_{\rm h}$ by the horseshoe streamlines in the global disc.}
 \label{fig:vphi_cpd}
\end{figure}

\begin{figure}
\centering
\includegraphics[width=0.45\textwidth,clip=true]{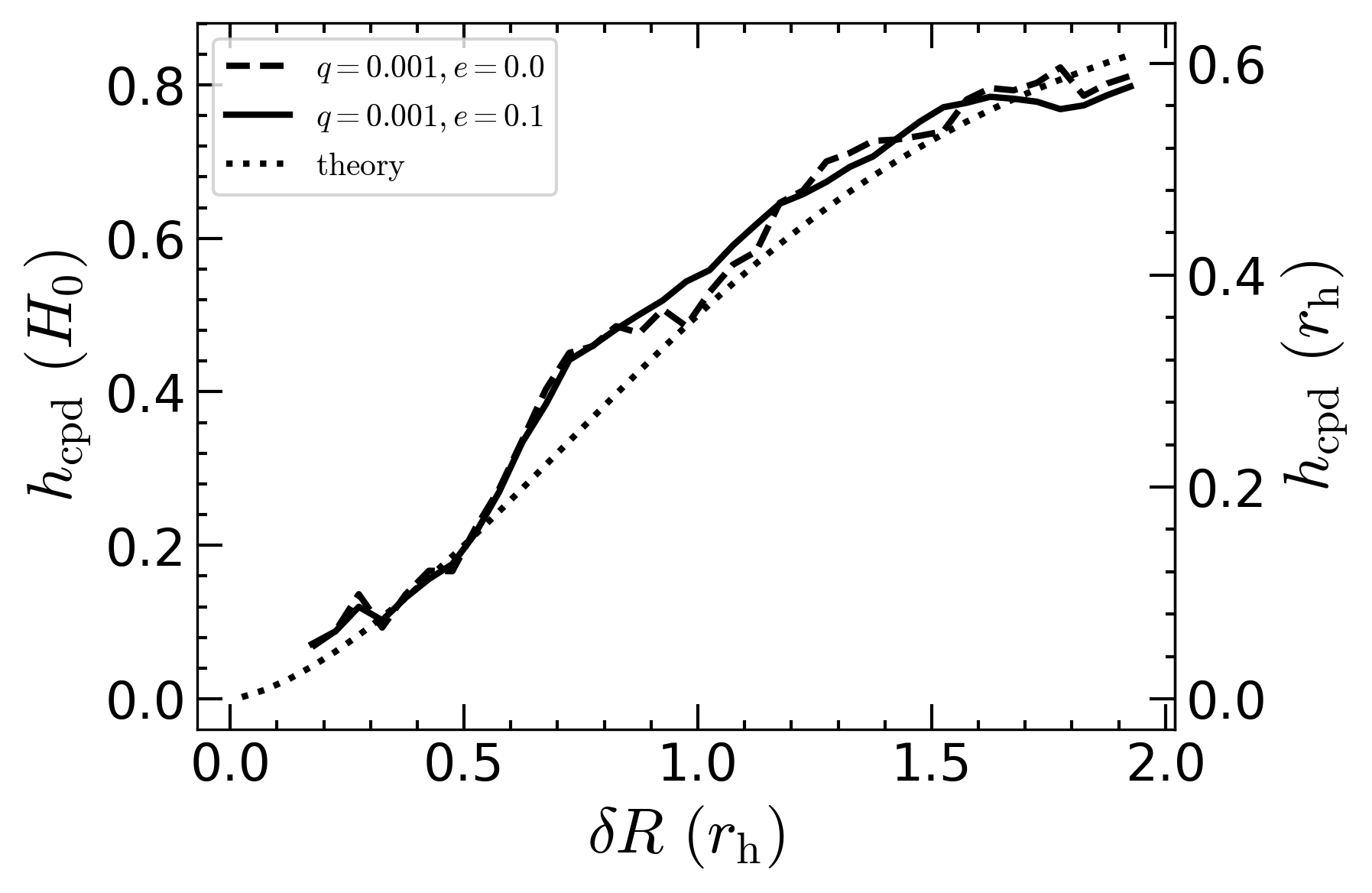}
\caption{Scale height of the CPD. Dotted line is theoretical scale height based on Equation~\ref{eq:h_cpd}. The left y-axis is in unit of global disc scale height, while the right y-axis is in unit of $r_{\rm h}$. The CPD's aspect ratio $h_{\rm cpd}/\delta R \sim 0.3$
and $h_{\rm cpd}$ transitions to $H_0$ at $\delta R \gtrsim 2 r_{\rm h}$.
} \label{fig:hcpd}
\end{figure}

\begin{figure*}
\centering
\includegraphics[width=0.45\textwidth,clip=true]{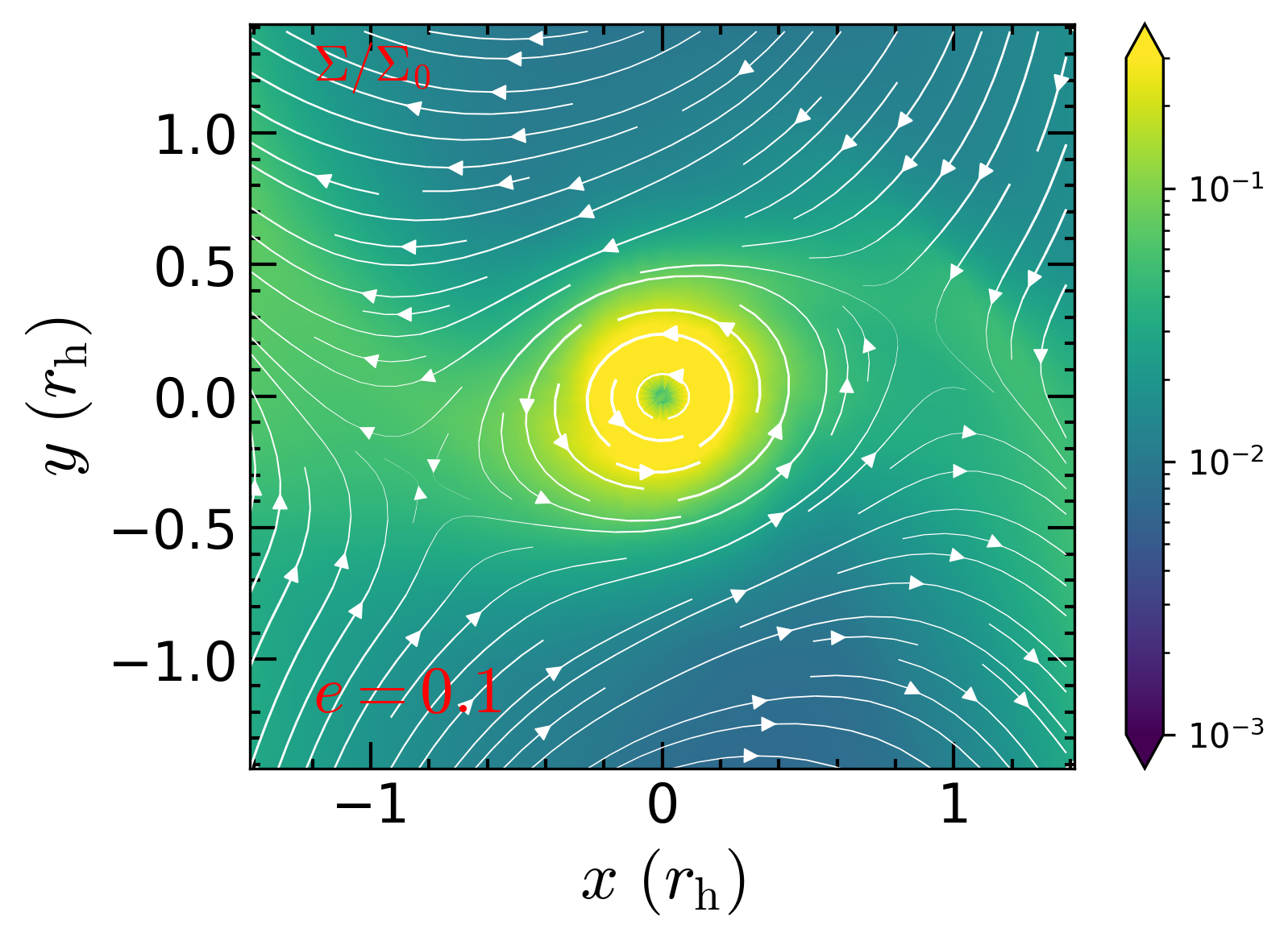}
\includegraphics[width=0.45\textwidth,clip=true]{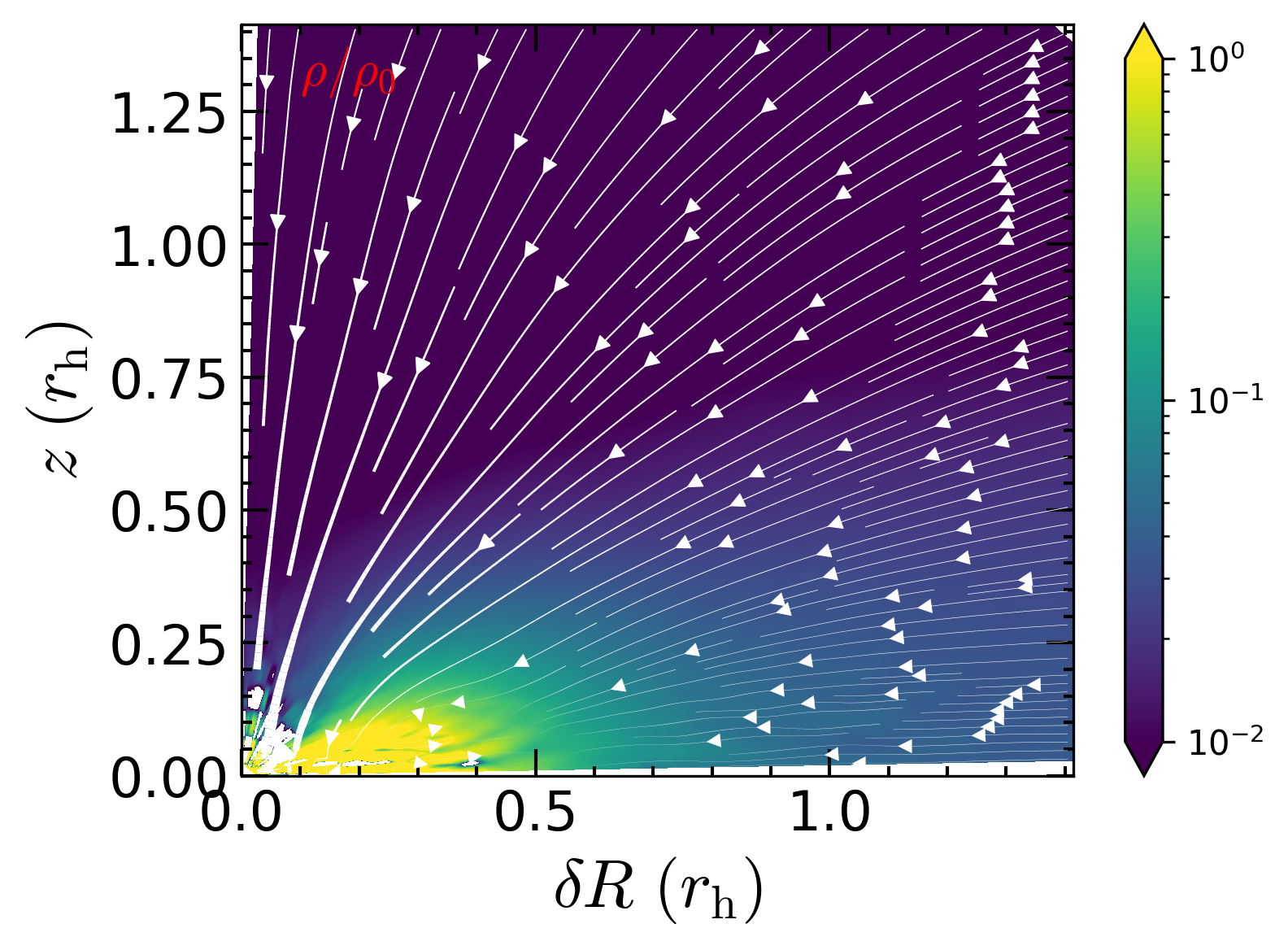}
\includegraphics[width=0.45\textwidth,clip=true]{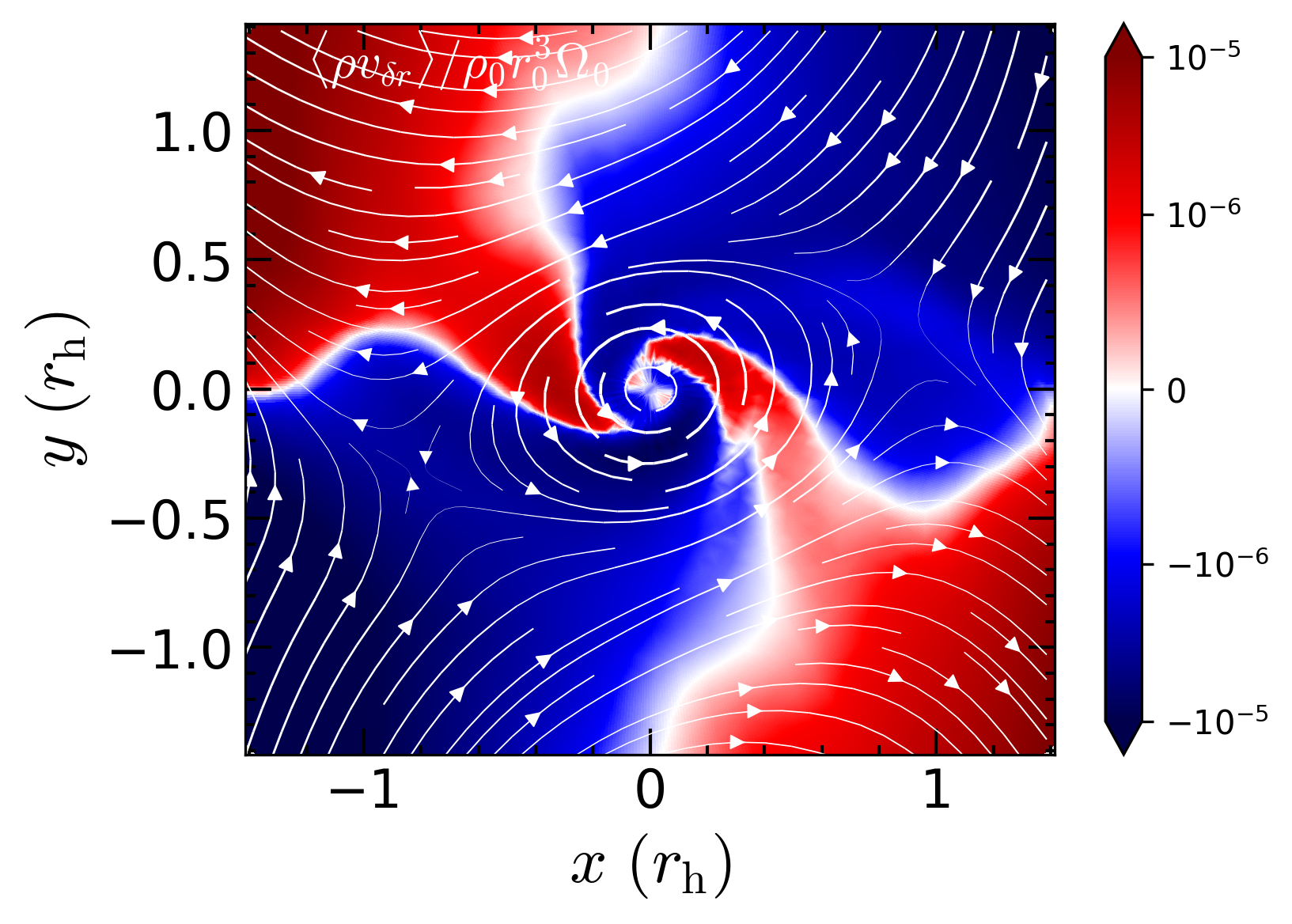}
\includegraphics[width=0.45\textwidth,clip=true]{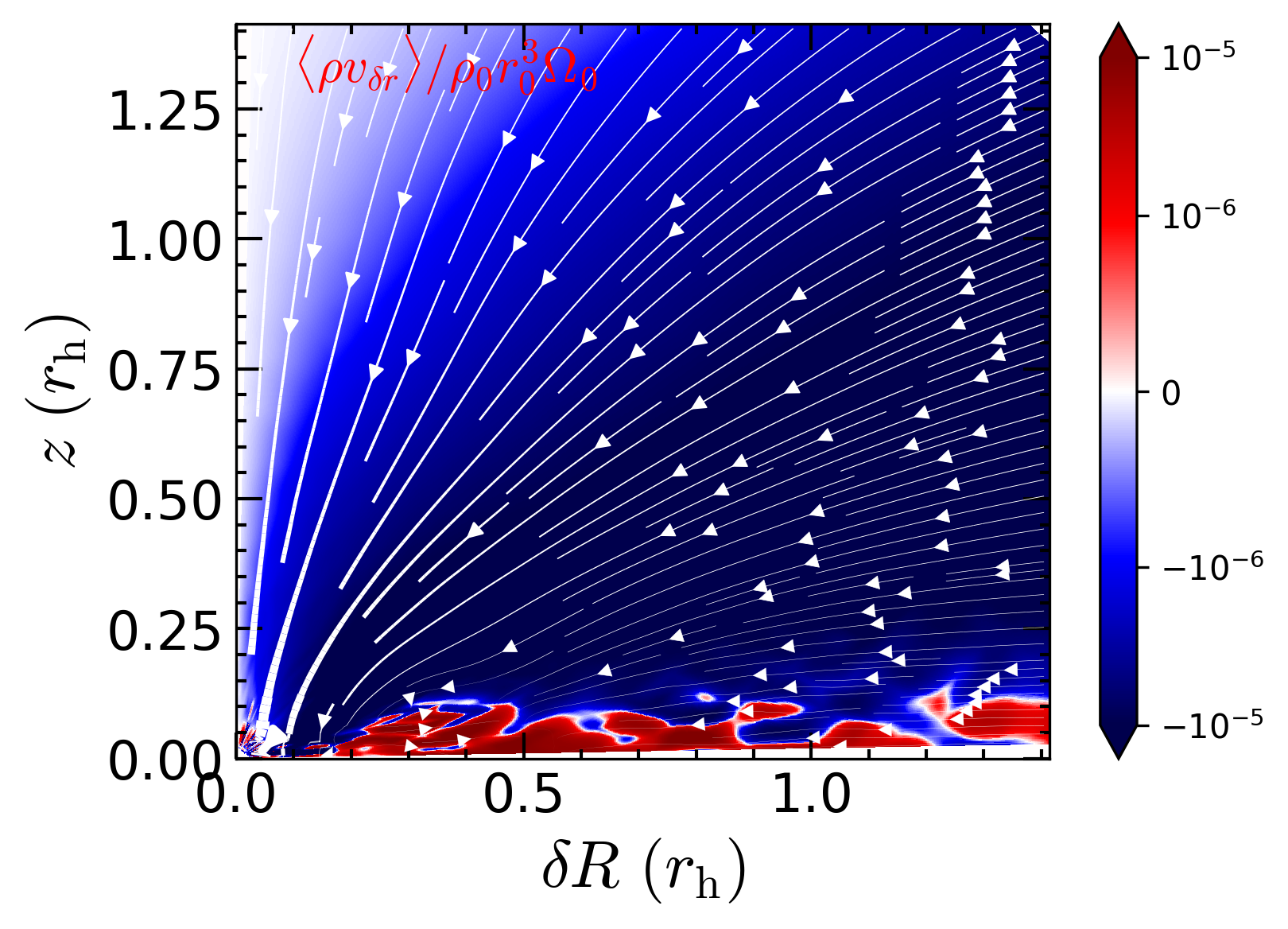}
\caption{Same as Figure~\ref{fig:fluxmass_fid} but for a planet with $e=0.1$. 
All the density and mass flux are time-averaged with sampling 100 equal-interval snapshots within one orbit.} \label{fig:fluxmass_e01}
\end{figure*}

Near the midplane, the accreted material mainly comes from the upper right and lower left horseshoe regions around the planet, 
indicated as blue regions in the left panels of Figure~\ref{fig:fluxmass_fid}. The spiral arms contribute mostly the outflowing mass flux, 
shown as red regions. 
The shock dissipation associated with the spiral arm in the CPD region drives the material gradually inspiralling inward to the planet, and finally getting accreted.
We can clearly see that there is circum-planetary rotation around the planet. The size of the CPD (indicated by strong rotation) is around $0.2\sim 0.3\ r_{\rm h}$ for the circular planet as shown in Figure~\ref{fig:vphi_cpd}, 
where we plot the midplane rotation curves of materials around the planet, averaged along the azimuth direction with respect to the planet. We have confirmed that the vertical integrated rotation velocity shows a very similar profile.
Note that we specifically label the radial component of this cylindrical coordinate system centered on the planet to $\delta R$ instead of $\delta r$. 
We utilize both these systems centered on the planet in the analysis of CPD properties, 
and while we plot azimuthally and polar-angle averaged quantities' radial profiles in the $\delta r$ coordinate like Figure \ref{fig:fluxmass1d_fid}, we will plot radial profiles of quantities in the CPD midplane or quantities that involve azimuthally vertically average in $\delta R$.
This CPD size is roughly consistent with the isothermal simulation results ($\sim0.18\ r_{\rm h}$) with $q_{\rm th}=4$ planet by \citet{Fung2019}, but smaller than that of the super-thermal simulations in \citet{Machida2008}, where they found a CPD size of $0.5\ r_{\rm h}$. Moreover, we note that our CPD size is generally larger than the measurement of centrifugal radius based on the specific angular momentum of CPD \citep[e.g.,][]{Tanigawa2012}.

The azimuthally-averaged density distributions in the $\delta R-z$ plane are shown in the upper right panels of Figure~\ref{fig:fluxmass_fid}. 
The averaged fluid streamlines are overlaided in the same plot. 
There exists a circulation pattern around the midplane of CPD, 
which drives the disc outflow. 
This outflow collides with the outer inflow material, 
which determines the outer boundary of the meridional circulation eddy, 
although such a boundary differs for different planet masses as we will show below. Note there is a mismatch between the outflowing mass flux and incoming streamline line near the midplane at $\delta R \gtrsim r_{\rm h}$. 
This is because that mass flux
is essentially density-weighted gas velocity, and it can be seen from the left panels of Figure~\ref{fig:fluxmass_fid} that in this region near the midplane, incoming streamlines (blue regions) has large inflow velocities that could dominate the azimuthally-averaged velocity, 
but they carry very low density and are unable to dominate the azimuthally-averaged mass flux.

In the presence of an embedded object in the disc, the thickness of CPD follows from the vertical hydrostatic equilibrium, and we have \citep{Dempsey2022}
\begin{equation}\label{eq:h_cpd}
h_{\rm CPD}(\delta R) = \frac{H_{0}}{\sqrt{3}}\left(\frac{\delta R}{r_{\rm h}}\right)^{3/2}\left[1+\frac{1}{3}\left(\frac{\delta R}{r_{\rm h}}\right)^{3}\right]^{-1/2}.
\end{equation}
At a distance of $\sim0.3\ r_{\rm h}$ from the planet, the thickness of the CPD is on the order of $\sim0.1H_{0}$ or equivalently $\sim0.07r_{\rm h}$ for $q=10^{-3}$. There does not appear to be any $h_{\rm cpd}$ dependence of the exact value of $\rho_{\rm p}$.

We present the CPD scale height in Figure~\ref{fig:hcpd} based on the vertical density distribution in Figure~\ref{fig:fluxmass_fid}, 
which shows that our simulations are well consistent with the vertical density distribution. 
After obtained the azimuthally averaged surface density $\Sigma(\delta R)$ and midplane density $\rho_{\rm mid}(\delta R)$, the disc scale height in the CPD region can be obtained through $h_{\rm CPD}=\Sigma/\sqrt{2\pi}\rho_{\rm mid}$.
The disc aspect ratio for CPD $h_{\rm cpd}/\delta R \sim 0.3$ within $r_{\rm h}$ from the planet, which is much larger than $h_0=0.05$ of the PPD. CPD's scale density height $h_{\rm cpd}$ transitions to $H_0$ at $\delta R \gtrsim 2 r_{\rm h}$.

In the lower right panel of Figure~\ref{fig:fluxmass_fid}, we show the mass flux in the $\delta R-z$ plane. 
Most of the inflow material is channeled through mid latitudes, rather than near the midplane or through the polar region, 
even though the inflow velocity from the polar region is a large fraction of the free-fall speed.
This dichotomy between infall speed and mass flux is mainly due to tenuous gas density from the polar region.
Most of the gas settles down to the midplane to form the CPD. 
Such a mid-latitude inflow coupled with midplane inflow near the planet (at $\delta R \lesssim 0.2 r_{\rm h}$) 
have also been found in previous 3D simulations \citep[e.g.,][]{Machida2008,Fung2015,Lambrechts2017,Schulik2019}.
Note that the streamlines in the outer part of midplane ($\delta R \gtrsim 1 r_{\rm h}$) is incoming, 
although the mass flux therein is still positive (i.e. outward). Such a mismatch is again due to that the mass flux 
is additionally weighted by the density before the azimuthal averaging procedure is applied, and the incoming gas streams 
in certain azimuthal directions have large velocity but are low in gas density.

In Figure~\ref{fig:fluxmass1d_fid}, we quantitatively evaluate the radial distribution of mass flux from different polar angle regions plotted as solid lines of different color. 
The $\dot{m}(r)$ profile gradually converges to the total accretion rate as the 
extent latitude range approaches to the midplane at $90^{\rm o}$.
{We observe that the accreted materials are mostly channeled through mid-latitude polar angles ($\theta=30^{\circ}\sim60^{\circ}$)
rather than via the polar ($0^{\rm o}$).  A general decrease in the magnitude of $\dot{m}(r)$ between $80^{\rm o}$ and $90^{\rm o}$ 
implies an outflow near the midplane regions. }

\subsection{Effect of Planetary Eccentricity}

To investigate how the accretion depends of the orbital eccentricity of the planet, we perform a run similar to the fiducial case but the planet has a fixed eccentricity of $e=0.1$. Completely analogous to the circular case, we plot the evolution of accretion rate in Figure~\ref{fig:mdot_ecc}, the rotation curve in Figure~\ref{fig:vphi_cpd}, and the CPD scale height in Figure~\ref{fig:hcpd}. In the upper panel of Figure~\ref{fig:mdot_ecc}, the accretion rate for the eccentric planet $e=0.1$ is smoothed over one orbital timescale to compare with the stable accretion rate profile of the circular case. It can be seen that the averaged accretion rate for the $e=0.1$ case is a factor of $2-3$ higher than that of circular case. Apart from this, the accretion structure and CPD profiles are similar.

\begin{figure}
\centering
\includegraphics[width=0.45\textwidth,clip=true]{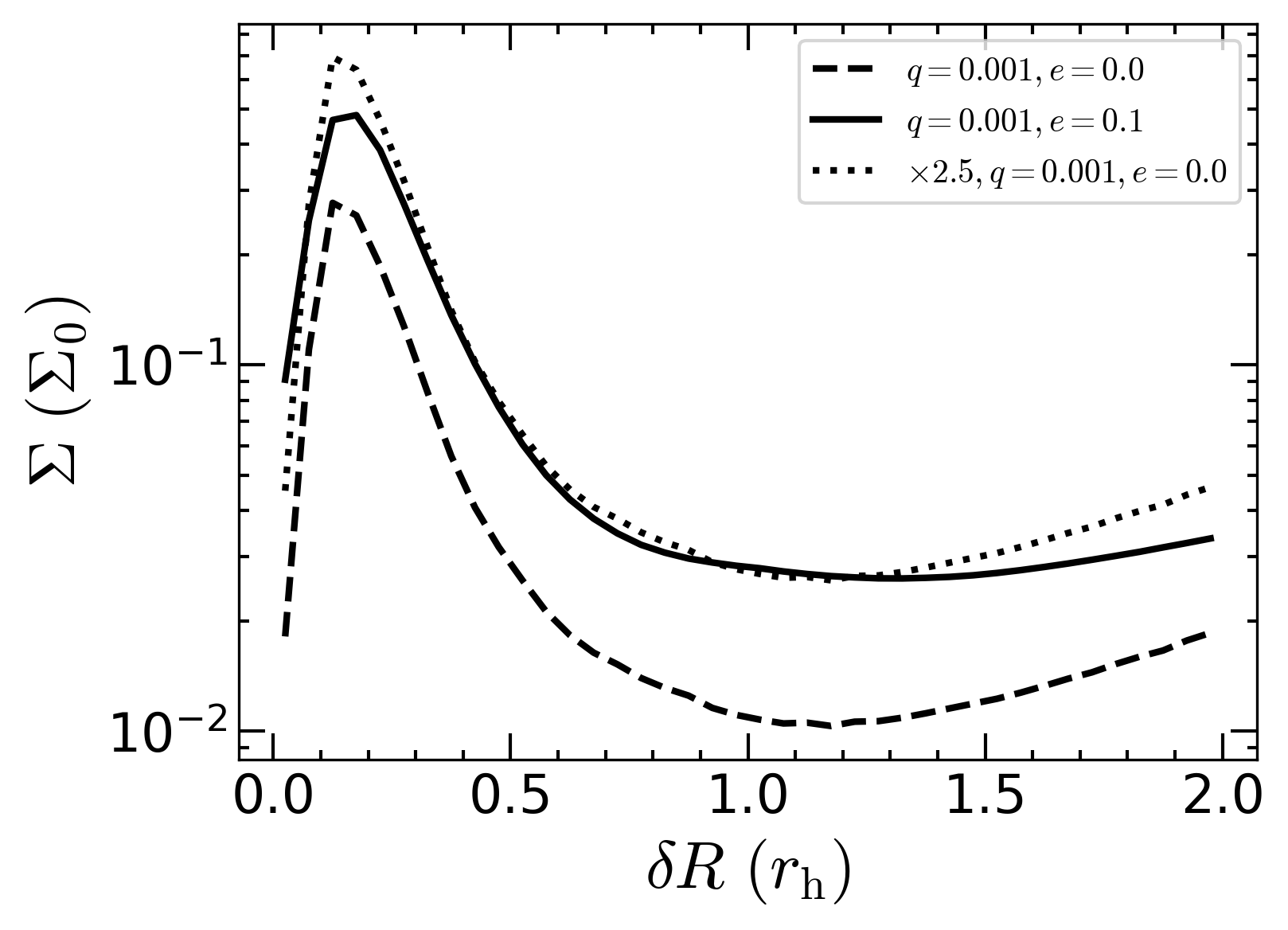}
\includegraphics[width=0.45\textwidth,clip=true]{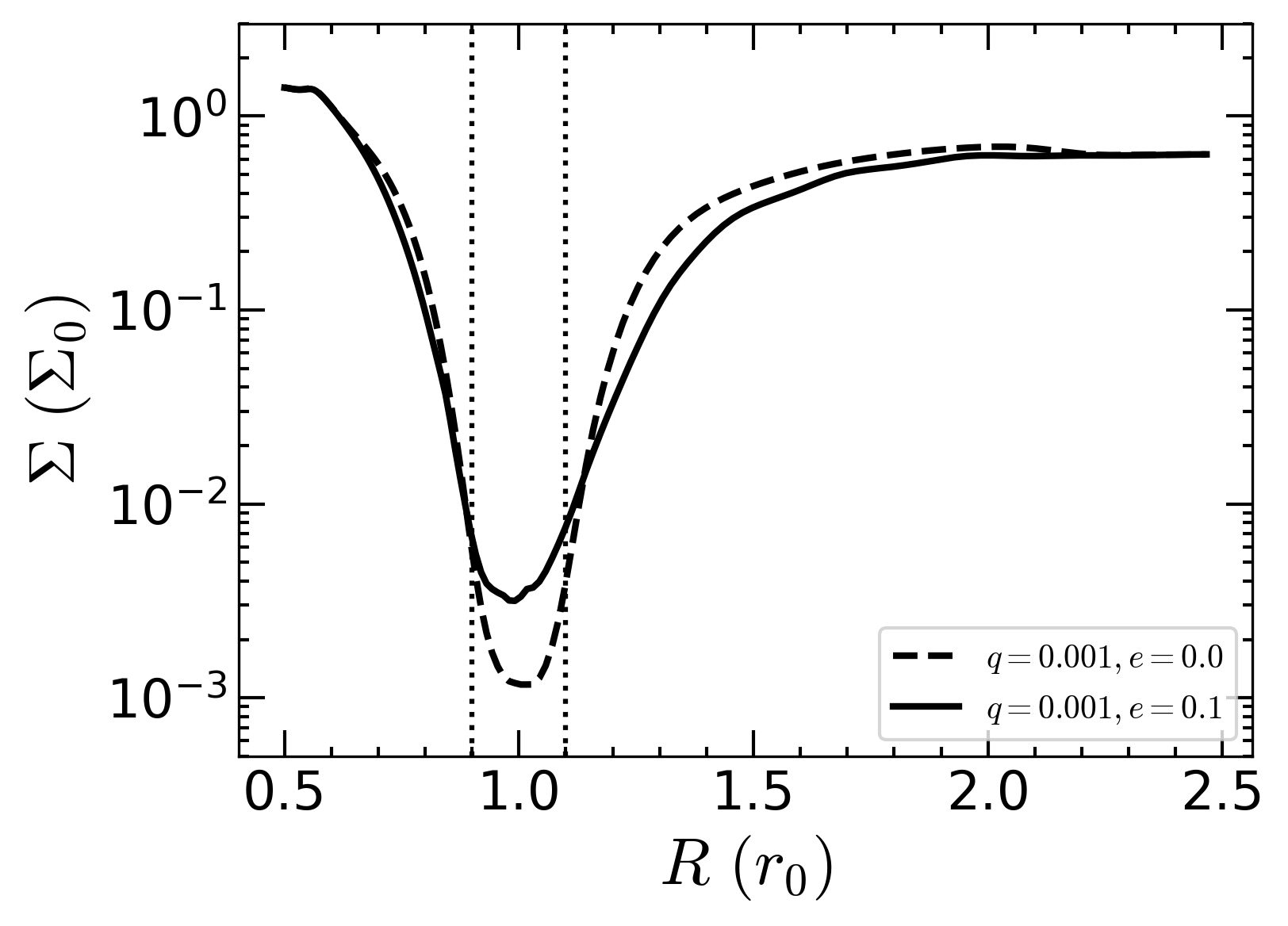}
\caption{Upper panel: time-averaged (over one global orbit at $r_{0}$), vertically-integrated 
surface density profile as a function of the cylindrical distance from the planet for $e=0.1$ (solid line)
and $e=0.0$ cases (dashed line). The surface density $\Sigma$ is averaged over azimuthal angle relative to the 
planet. Within $\delta R < 2 r_{\rm h}$, $\Sigma$ of the $e=0.1$ planet is uniformally enhanced from that
of the $e=0$ planet (dotted line). Lower panel: globally azimuthal-averaged (relative to the central star) surface density profile of the PPD, which shows that the eccentric planet carves a shallower gap. Note that the density enhancement in the CPD region is smoothed out here. The two vertical dotted lines show the extent of the eccentric planet's radial excursion (i.e., $r_{\rm p}\sim 0.9-1.1\ r_{0}$).}
 \label{fig:sigma_cpd}
\end{figure}

We show the midplane \& azimuthally averaged density and mass flux map in Figure~\ref{fig:fluxmass_e01}, similar to Figure~\ref{fig:fluxmass_fid}. 
It can also be seen that there exists a rotation-supported disc as in the circular planet, which is confirmed in Figure~\ref{fig:vphi_cpd}. 
But it should be expected that deviation from the Keplerian profile becomes more significantly as the planetary eccentricity increases further. 
Based on the density profiles, we calculate the azimuthally and time averaged disc scale height shown in Figure~\ref{fig:hcpd}. 
For the eccentric case, 
the disc scale height shows a similar profile, although the steady-state midplane density $\rho_{\rm p}$ is factor of $2\sim3$ higher.

The mass flux in the $x-y$ and $r-z$ plane are shown in the right panels of Figure~\ref{fig:fluxmass_e01}. Consistent with the circular case, the accretion is still dominated by mid-latitude inflow rather than the polar region. 
As shown in Figure~\ref{fig:fluxmass_e01},  the outflowing mass flux in the midplane region is weaker for the eccentric case when compared with that of the circular planet, although neither of them are significant compared to the total mass flux dominated by incoming streamlines at higher latitudes.

The different accretion rates for different planetary eccentricity is mainly due to different density in the CPD region. 
To quantify this effect, we plot the azimuthal-averaged surface density in the global disc and in CPD region in Figure~\ref{fig:sigma_cpd}. 
It can be clearly seen that the eccentric planet opens a shallower gap compared with the circular planet. 
The density enhancement in the CPD region is also a factor of $2\sim3$ higher for the $e=0.1$ planet compared to the circular case.

\begin{figure} 
\centering 
\includegraphics[width=0.45\textwidth,clip=true]{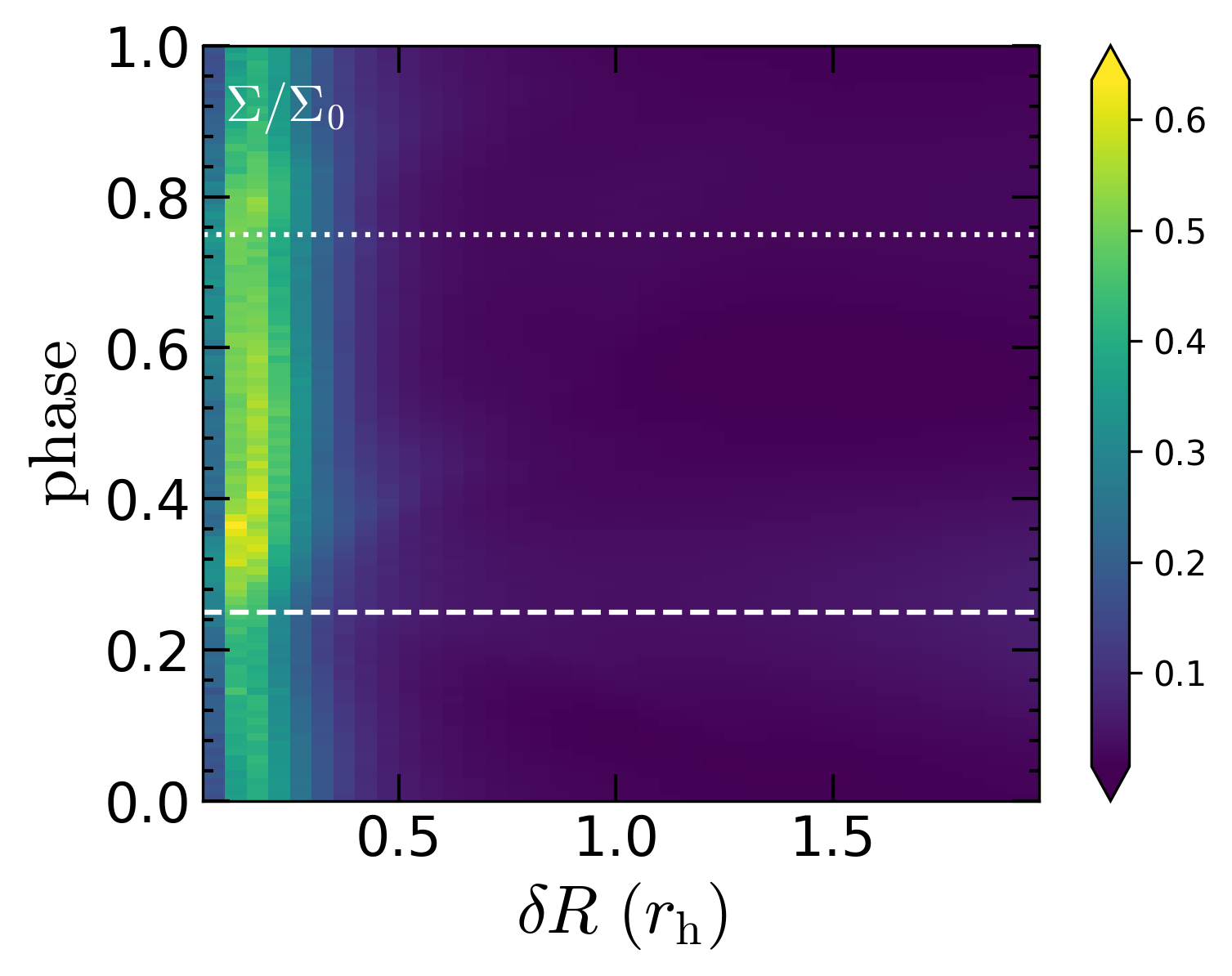}
\includegraphics[width=0.45\textwidth,clip=true]{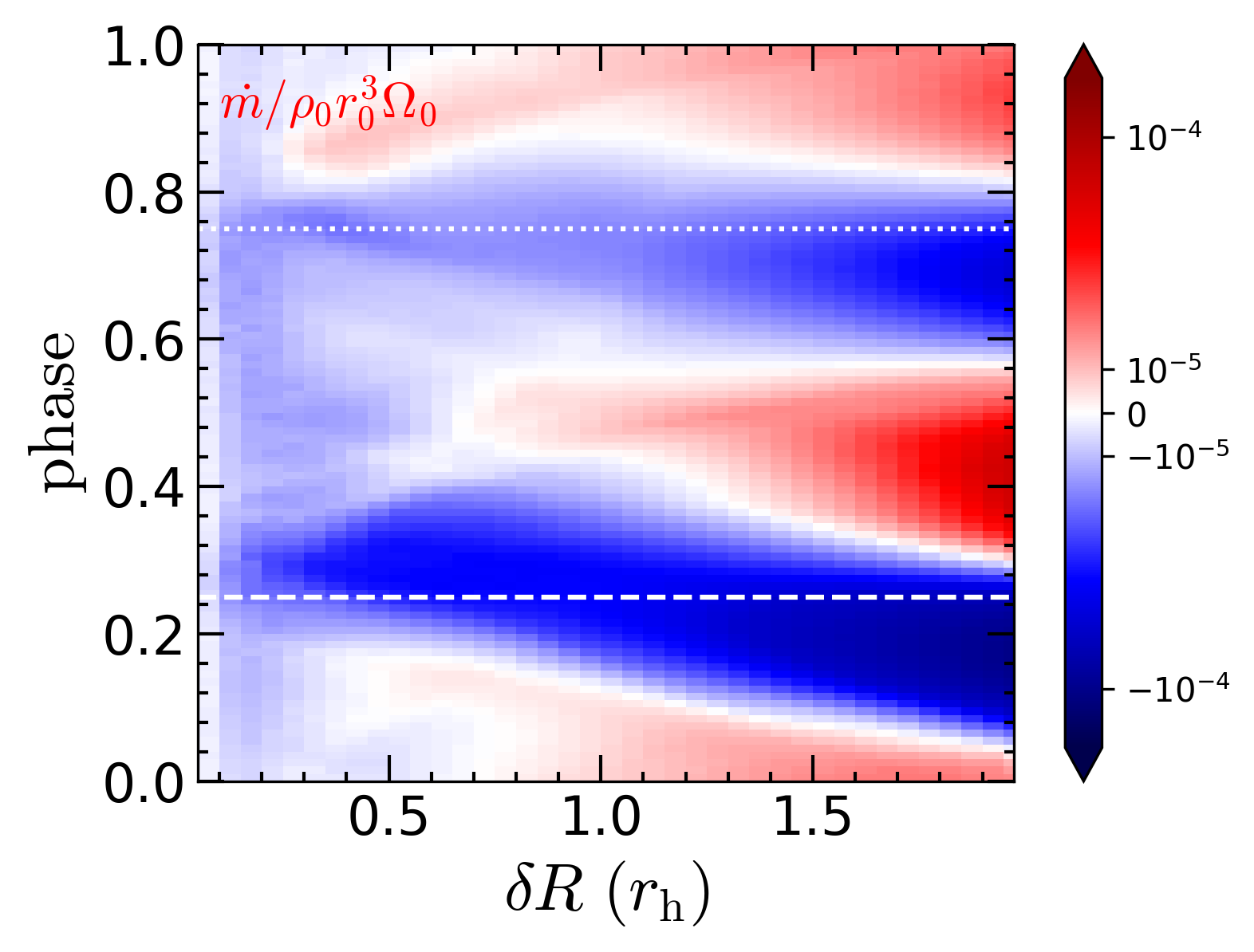}
\caption{Upper panel: phase evolution in one orbital period for the radial distribution of the surface density 
(upper panel) and mass flux (lower panel) distribution in the frame centered on the $e=0.1$ planet. Both 
the surface density and the mass flux are azimuthally vertically averaged.
The phases for pericenter and apocenter are indicated by dashed and dotted lines respectively.}
 \label{fig:mdot_phase}
\end{figure}

The eccentric planet further shows significant time variability with a period of about $\ P_{\rm b}$, where $P_{\rm b}\equiv2\pi/\Omega_{0}$ is the orbital period of the planet at its semi-major axis. 
The two peak of the accretion rate in one orbital phase correspond to the pericenter and apocenter of the eccentric orbit, while the accretion rate at the pericenter is slightly higher than that of the apocenter. This is due to the periodical density enhancement around the pericenter and apocenter where the eccentric planet is closer to the edge of the gap. Such a periodical density enhancement can be seen in the upper panel of Figure~\ref{fig:mdot_phase}, where we show the azimuthal-averaged surface density as a function of the distance to the planet at different phases within one orbit. The azimuthally and vertically-averaged mass flux at different orbital phases is shown in the lower panel of Figure~\ref{fig:mdot_phase}. 
The modulation of the mass accretion rates at different orbital phases can be clearly identified.

\subsection{Effect of $q_{\rm th}$}

\subsubsection{High-Mass Cases with Large $q_{\rm th}$} \label{sec:higq}

\begin{figure}
\centering
\includegraphics[width=0.45\textwidth,clip=true]{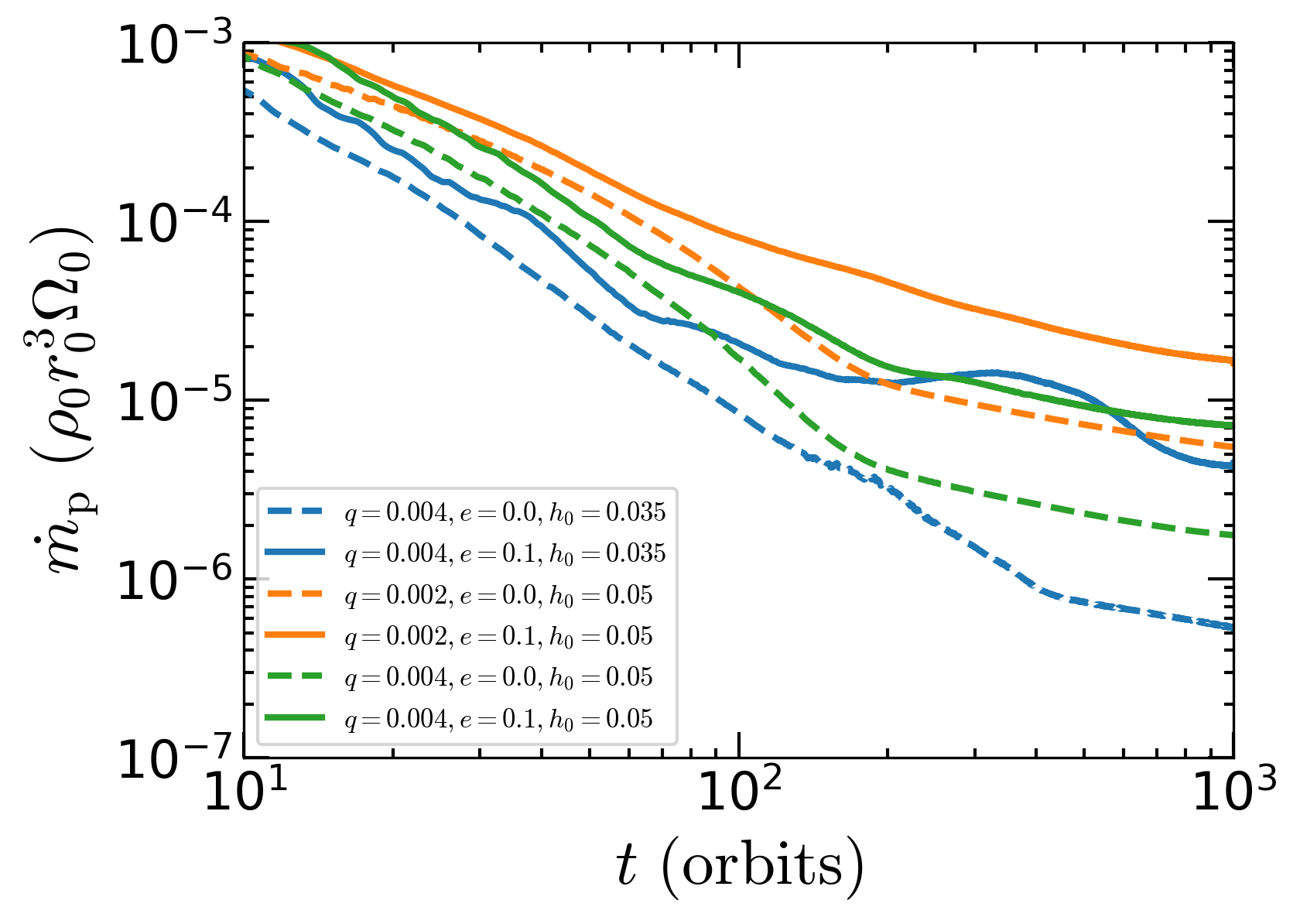}
\includegraphics[width=0.45\textwidth,clip=true]{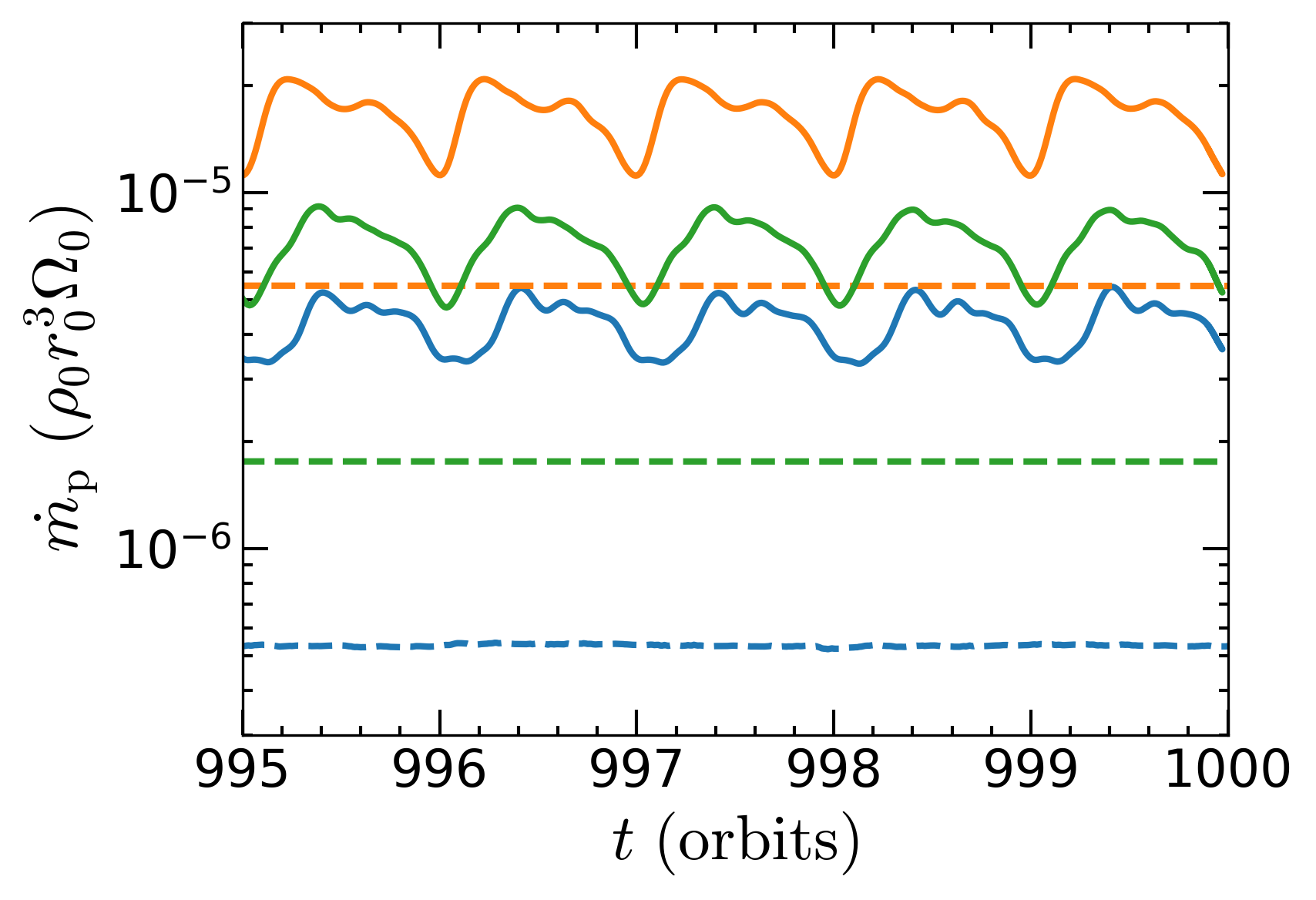}
\caption{The evolution of planetary accretion rate, measured in scale-free unit $\rho_{\rm 0}r_{\rm 0}^3\Omega_{\rm 0}$. 
The time-averaged over $\sim1$ orbits  accretion rates for the $e=0.1$ (solid lines) and $e=0$ (dashed line) planets is shown in the upper panel.  The instantaneous accretion rate for the $e=0.1$ planet 
in the lower panel shows a factor-of-two variability albeit that for the $e=0$ planet remains
constant.}
 \label{fig:mdot_highq}
\end{figure}

In 2D simulations, it has been found that a very massive planet can excite strong disc eccentricity, which then induces accretion burst onto the planet \citep{Papaloizou2001,Kley2006,Li2021,Tanaka2022}. It's natural to ask whether such kind of eccentric mode transition exist in 3D simulations. The excitement of disc eccentricity is not only closely related to the asymptotic mass of mature planetary system as protoplanet grows in mass, but also direction of type II migration \citep{Dempsey2021}. 
To this end, we explore the accretion rates for very massive planet with extremely high $q_{\rm th}$. 
We carry out simulations with higher planet mass up to $q=0.004$ and/or lower disc scale height down to $h_{0}=0.035$. With such an extreme parameter combination, $q_{\rm th}$ can be as large as 93. The long-term and short-term accretion rates for some example cases (with planet on both circular and eccentric orbits) are shown in Figure~\ref{fig:mdot_highq}. 
We can see that planetary accretion rate decreases with increasing $q_{\rm th}$ without apparent enhancement(outburst) of accretion rates compared to that of $q_{\rm th}=8$.  Both the long-term and short-term accretion rates are quite smooth for the circular planet. This stable accretion history has been verified by extending the simulations beyond 2000 orbits. This is significantly different from 2D simulations, which found a reversed dependence on planet mass once $q>0.003$ is high enough  where the disc eccentricity instability can be excited \citep{Kley2006,Li2021,Tanaka2022}. 

\begin{figure}
\centering
\includegraphics[width=0.45\textwidth,clip=true]{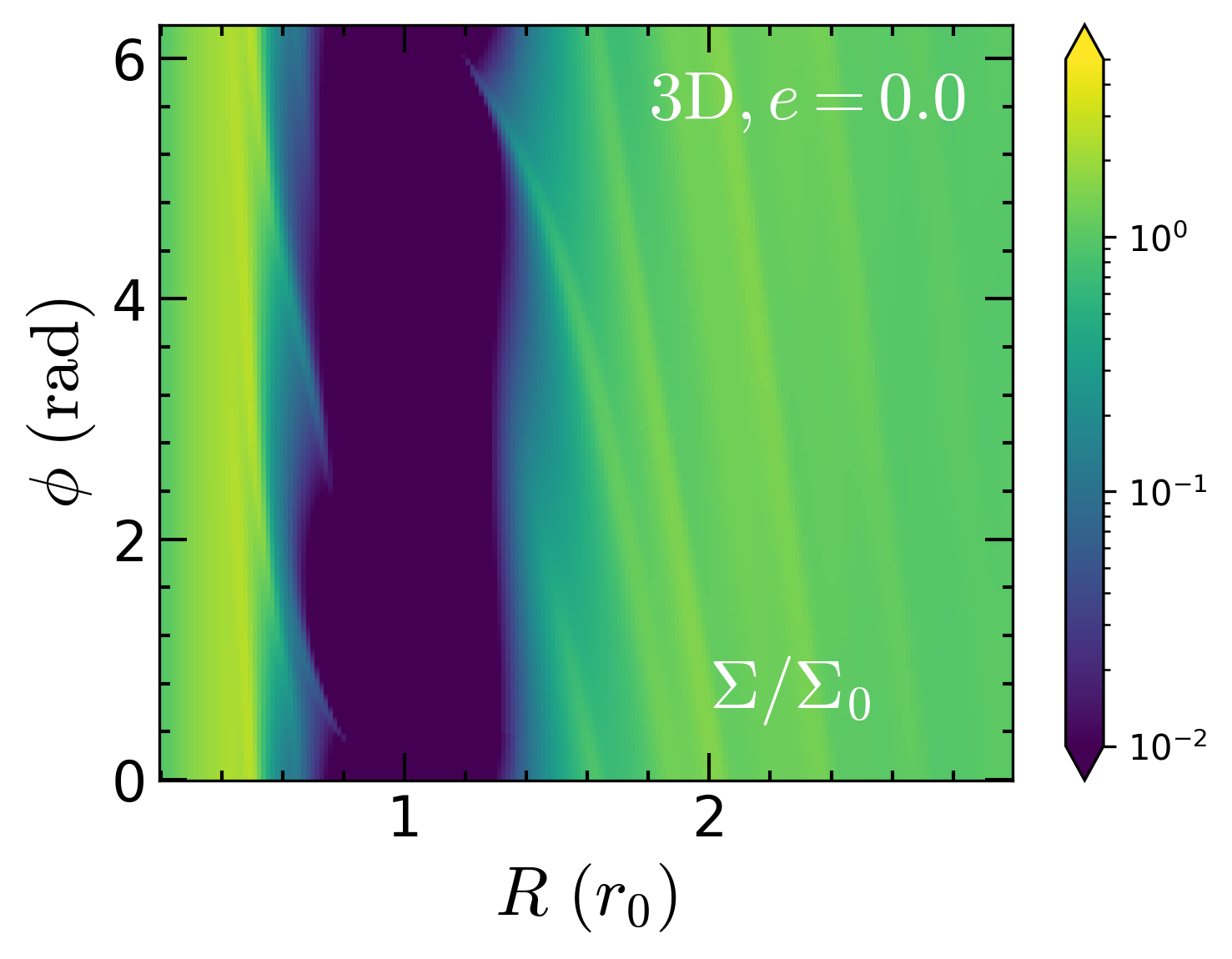}
\includegraphics[width=0.45\textwidth,clip=true]{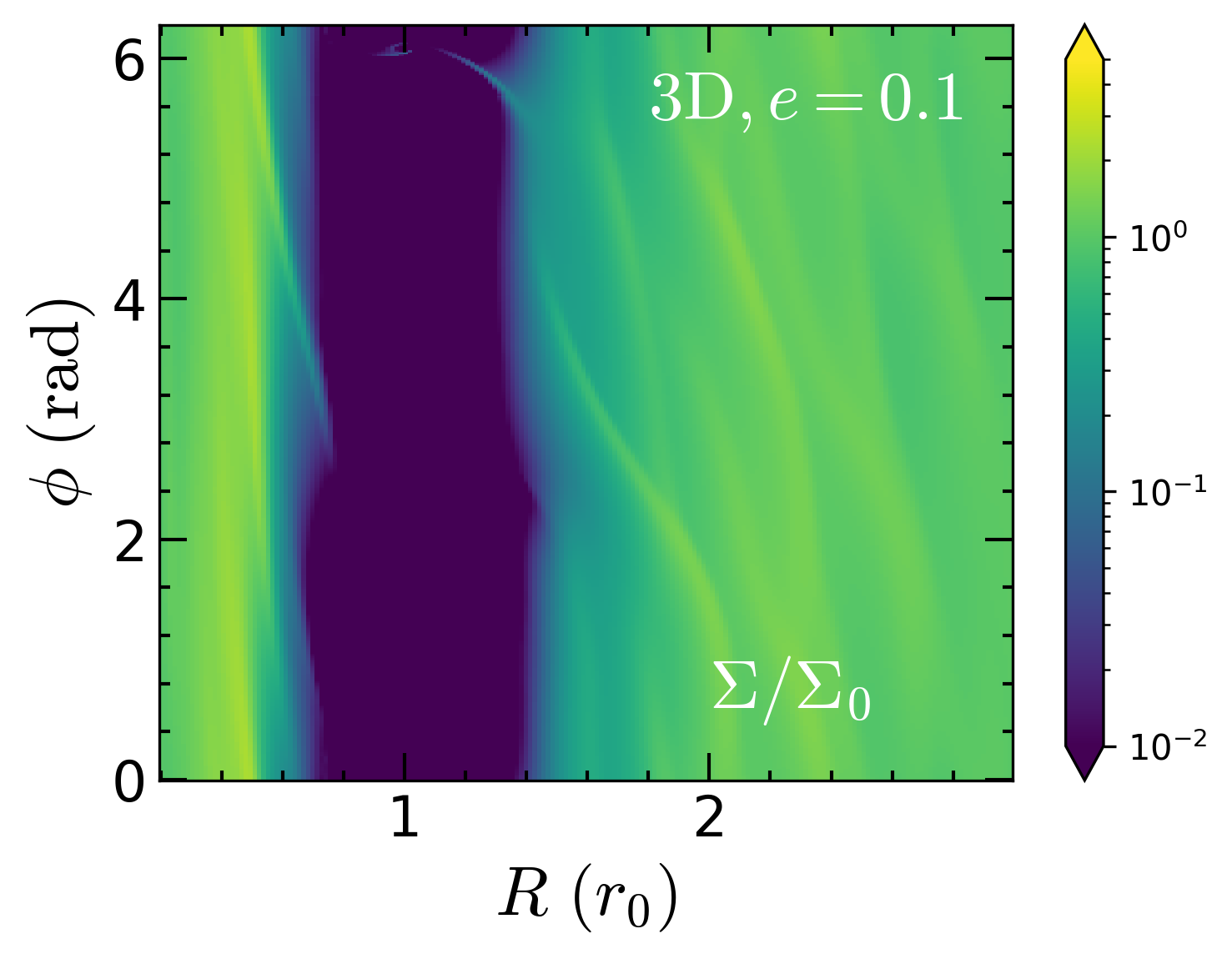}
\caption{The global distribution of surface density  for the $e=0.0$ (upper panel) and $e=0.1$ 
planets (lower panel) with $q=0.004$, $h_0=0.035$, {and $\alpha=0.001$}.
} \label{fig:sigma_rphi}
\end{figure}

Because both the eccentricity damping and excitation resonance depends sensitively on the gap profile, it is thus should expected that such a transition from circular to eccentric disc may correlate with gap width. 
The gap width has been found empirically to correlates with $K^\prime=\frac{q^2}{\alpha h_{0}^{3}}$ parameter \citep{Kanagawa2016}. 
\citet{Dempsey2021} found that the transition to eccentric disc occurs for $K^\prime=\frac{q^2}{\alpha h_{0}^{3}}\gtrsim 20$ based on a series of 2D simulations, 
although they did not explore its effect on the planetary accretion rates. 
Specifically, the disc eccentricity is expected to be excited by outer eccentric Lindblad 1:3 resonance at $r=2.08r_{\rm p}$ \citep{Lubow1991a}, but be damped predominately by the outer 1:2 resonance at $r=1.58r_{\rm p}$ and co-orbital resonance \citep{Goldreich2003}. 
As the planet mass increases, the gap becomes wider and deeper. 
The disc eccentricity can be maintained only if the two damping resonances cannot suppress the excitation due to the wider gap induced by the massive planet. Otherwise, a circular disc would be observed.
The disc surface densities for very high $q_{\rm th}$ ($q=0.004, h_{0}=0.035$) are shown in Figure~\ref{fig:sigma_rphi}. We can see that, both for the circular and eccentric planets, the outer gap edge in $r\lesssim1.5r_{0}$, which thus results in very circular disc and gap profiles for both for $e=0.0$ and $e=0.1$ planets.
Such a relatively narrow gap profile in 3D simulation compared with that in 2D (Fig. \ref{fig:sigma_comp2d})
thus causes the different disc eccentricity and thus accretion history.

\begin{figure}
\centering
\includegraphics[width=0.45\textwidth,clip=true]{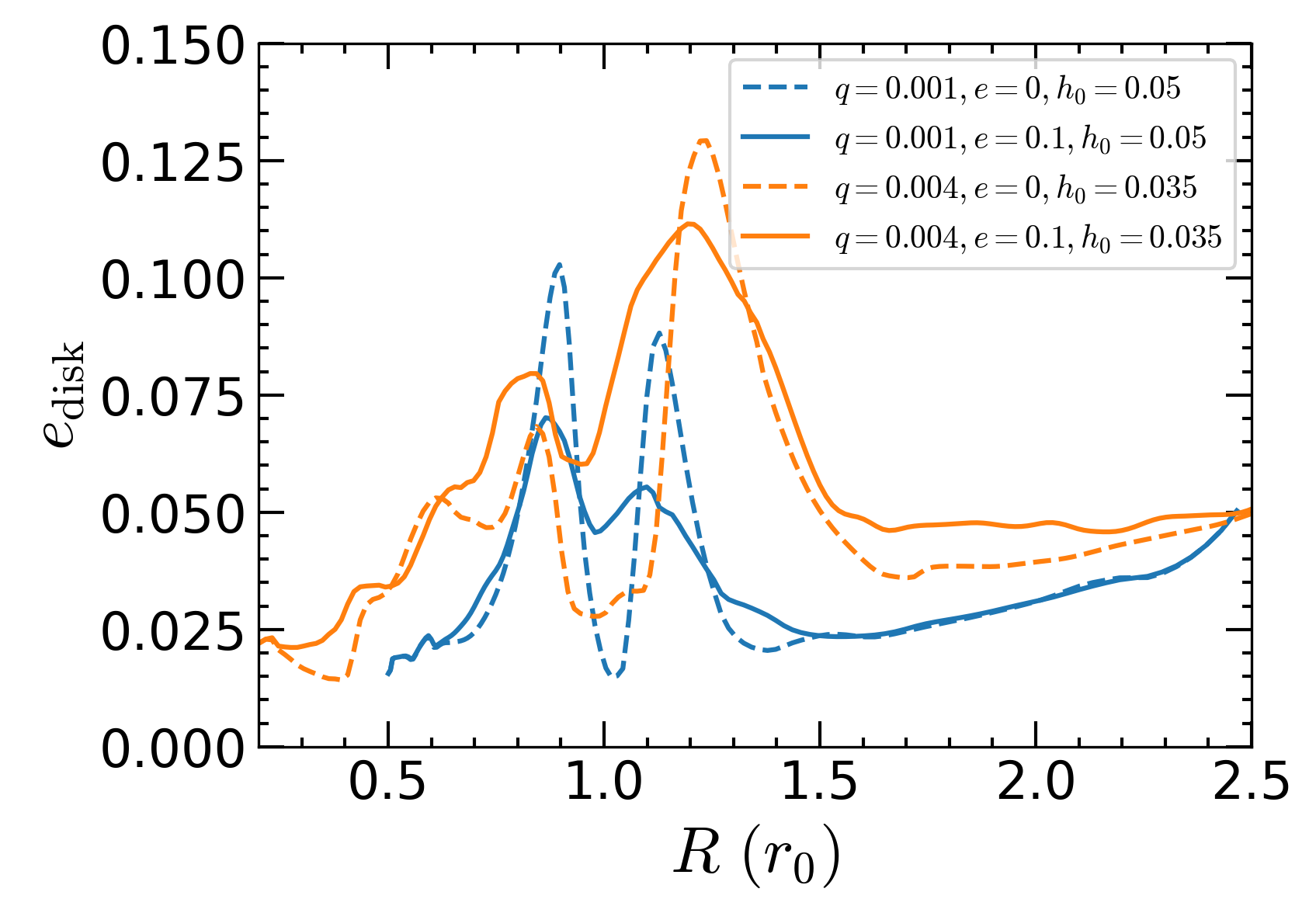}
\caption{The azimuthal- and vertical-averaged radial distribution of global-disc
streamline-eccentricity at 1000 orbits.  
Note that for the $q=0.004$ runs, we have extended the inner edge to $0.2\ r_{0}$, as discussed in Appendix~\ref{app:bd}.}
 \label{fig:edisc}
\end{figure}

Moreover, we show a comparison of the radial profile of disc streamline eccentricity in Figure~\ref{fig:edisc}. For circular planets, there is an eccentricity gap sandwiched by two peaks around the planet location, with the peak streamline line eccentricity being lower than $\lesssim0.1$ for both high and low $q_{\rm th}$ models, confirming the low disk eccentricity as discussed above. 


Regarding the cases of planetary eccentricity $e=0.1$,  accretion rates are also larger than the circular cases and highly variable on the orbital period timescale. 
However, as the $q_{\rm th}$ increases, the double peaks in one orbital period gradually merger into one broad peak as shown in the lower panel of Figure~\ref{fig:mdot_highq}. This is associated with the disappearance of the eccentricity gap around the eccentric planet as shown in solid lines in Figure~\ref{fig:edisc}. 
For small $q_{\rm th}$, the two well-separated eccentricity peaks enhance the accretion rates as the eccentric planet comes across the eccentric disc streamline. 
As $q_{\rm th}$ increases, the eccentricity gaps becomes wider and shallower, 
such an periodical enhancement between the pericenter and apocenter of accretion rates thus disappears as long as the enhancement of disc streamline eccentricity is still weak.

\begin{figure}
\centering 
\includegraphics[width=0.45\textwidth,clip=true]{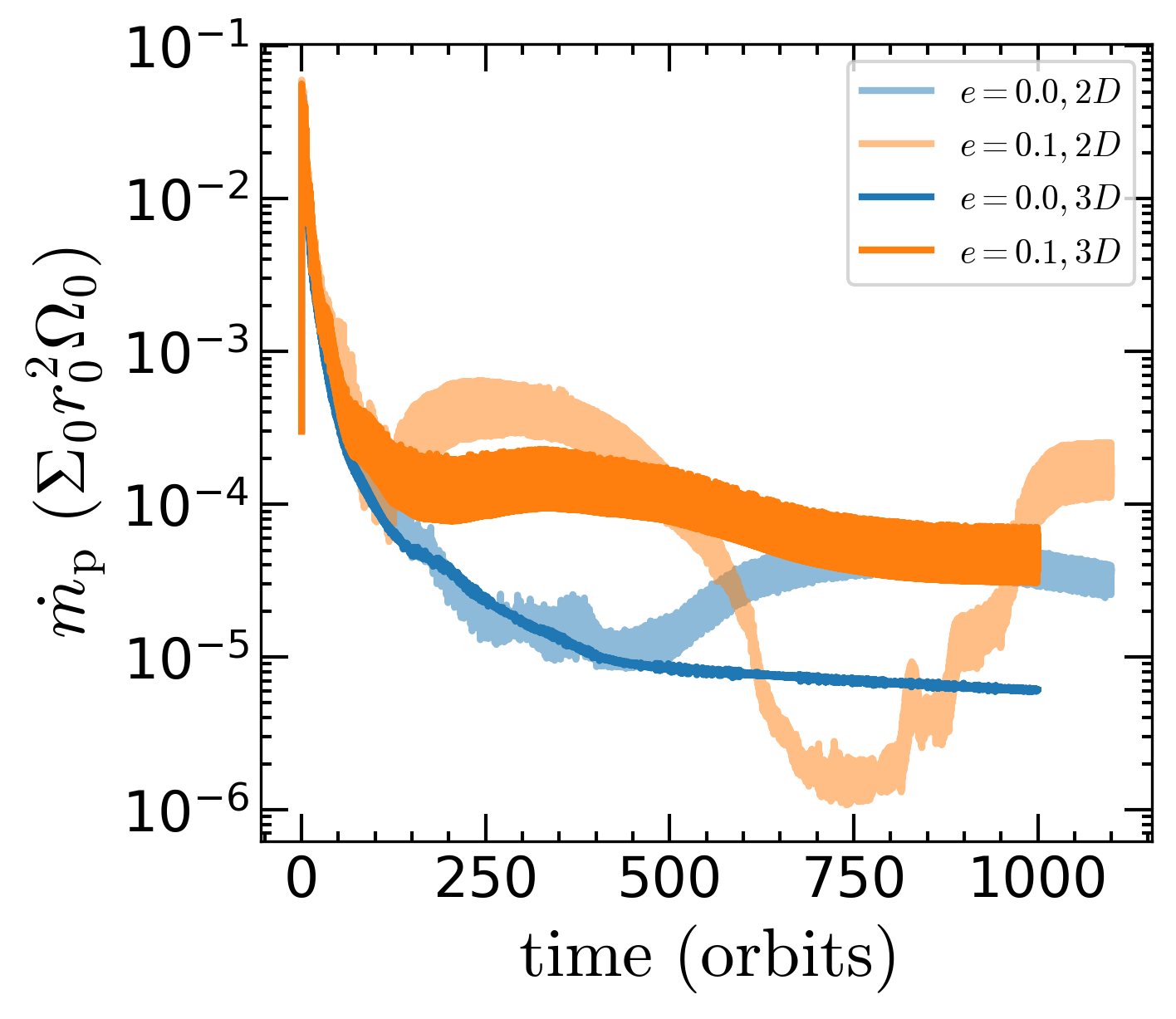}
\includegraphics[width=0.45\textwidth,clip=true]{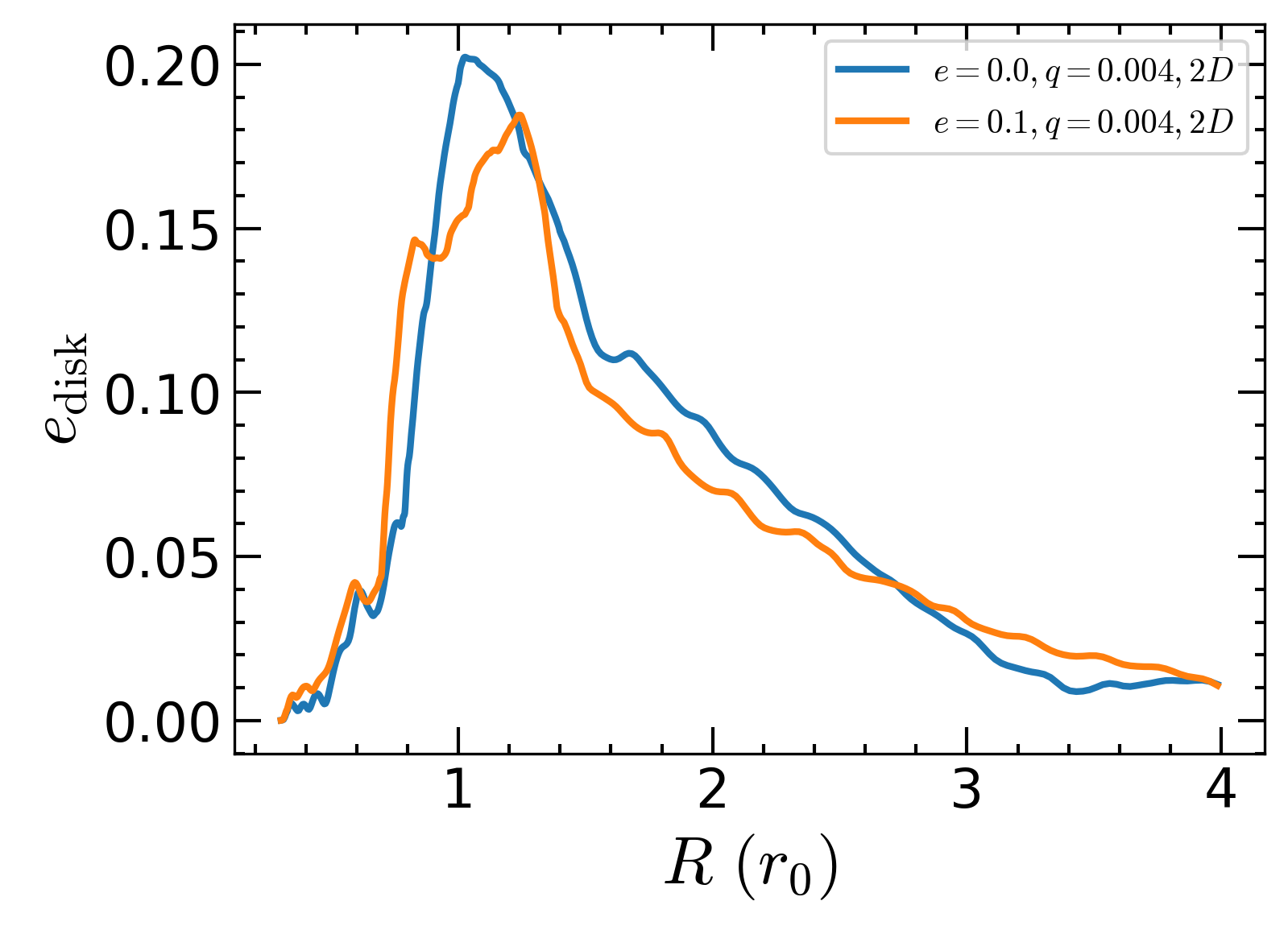}
\caption{Upper panel: Comparison of planetary accretion rates between 3D and 2D simulations. Lower panel: 
{ the azimuthal-averaged} disc-eccentricity distribution for 2D models. 
The disc streamline eccentricity is much higher than those of 3D counterpart shown in Figure~\ref{fig:edisc}.} \label{fig:mdot_comp2d}
\end{figure}

In our 3D simulations, we have found that the disc streamline eccentricity cannot be excited up to $q_{\rm th}\sim90$ and $K^{\prime}\sim370$. This result is contrary to what have been found in previous 2D simulations \citep[e.g.,][]{Papaloizou2001,Kley2006,Li2021,Dempsey2021,Tanaka2022}. To confirm this discrepancy more robustly, we carry out 2D simulations with $q=0.004$, $h_{0}=0.035$ and $\alpha=0.001$ to compare the 3D counterpart presented above. 
The accretion rates and disc eccentricity profiles are shown in Figure~\ref{fig:mdot_comp2d}. 
For such a large $K^{\prime}$, the accretion rates are highly variable even for the circular planet, 
accompanied with high eccentricity being excited. 
Before the long-term oscillation of accretion around 100 orbits, the 2D and 3D runs show similar accretion history. This demonstrates the necessity of long-term simulation to evaluate the accretion history, since simulations run for only a few orbits may not be adequate to confirm the absence of such effects \citep{Choksi2023}.

\begin{figure}
\centering 
\includegraphics[width=0.45\textwidth,clip=true]{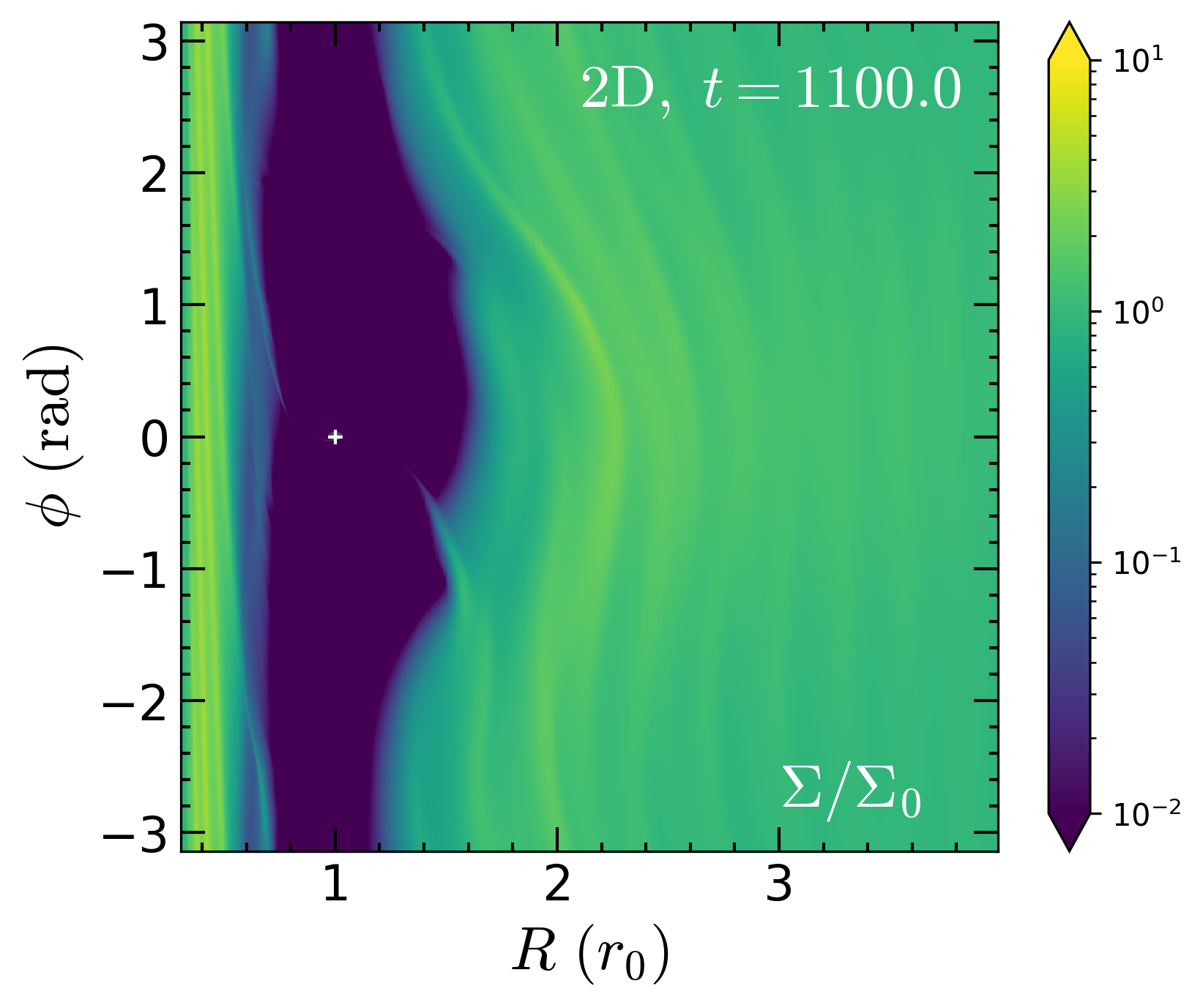}
\includegraphics[width=0.45\textwidth,clip=true]{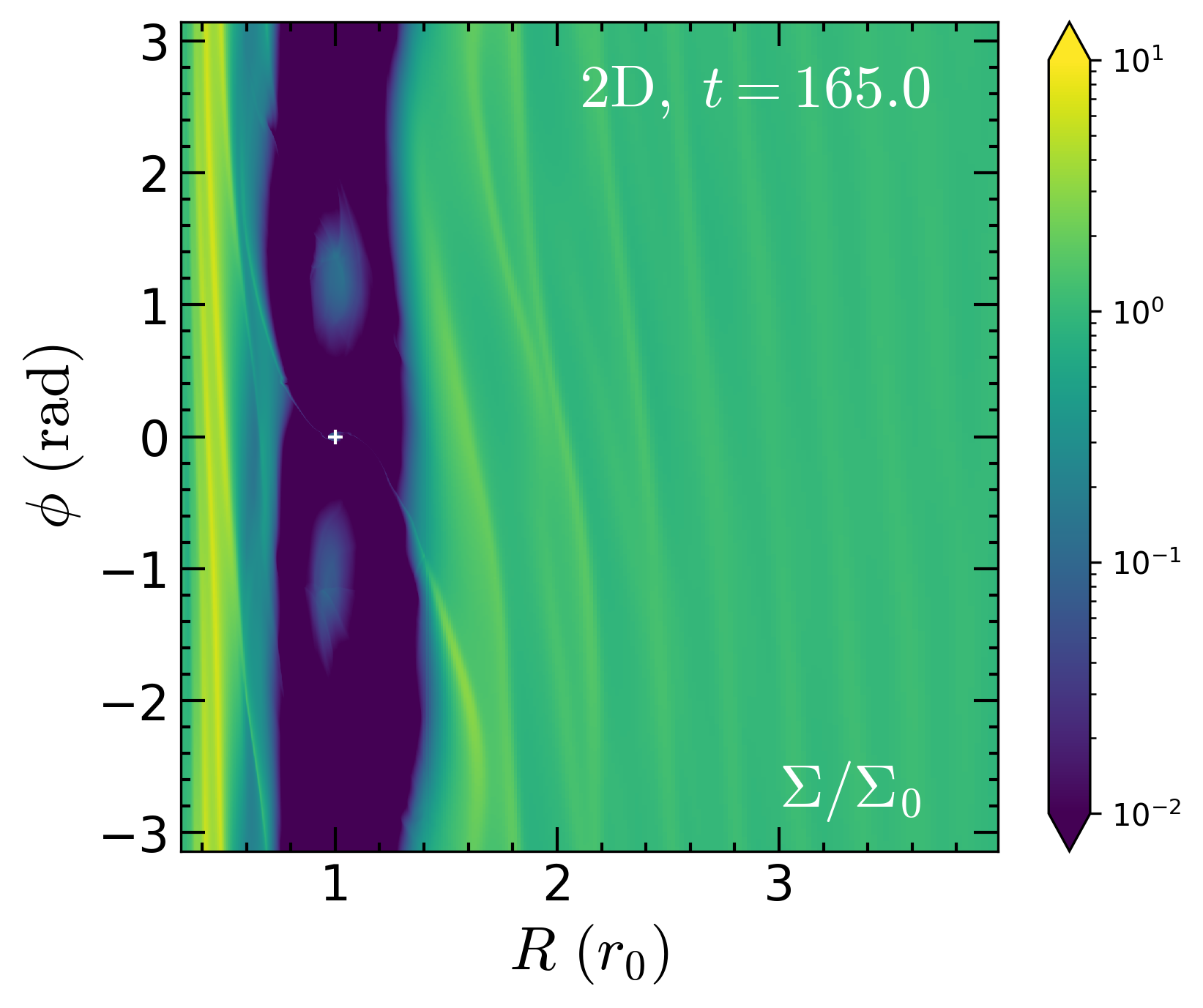}
\includegraphics[width=0.45\textwidth,clip=true]{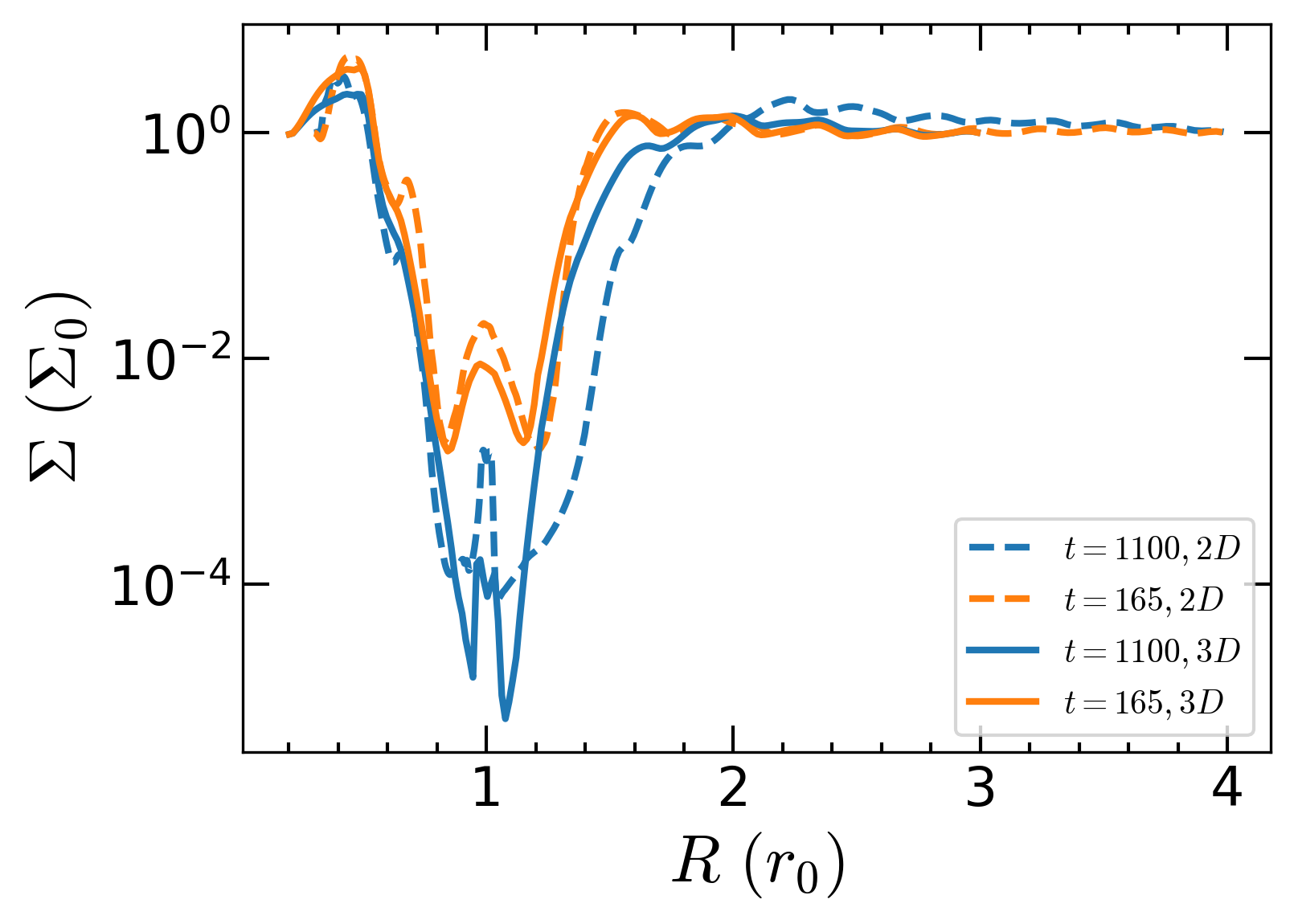}
\caption{Upper panel: surface density for a 2D run with the mass ratio $q=0.004$, $h_{0}=0.035$ and $\alpha=0.001$ at 1100 orbits. 
The planet is in a circular orbit. The white cross indicates the planet location. The eccentric gap can be clearly identified with a wider gap at certain azimuth $\phi$. Middle panel: similar to the upper panel but at t=165 orbits, when the accretion outburst has not been triggered yet. The lower panel shows the azimuthal-averaged surface-density distribution between 2D and 3D runs at different times. 
Note that for the eccentric gaps in the 2D simulation, the averaged value is evaluated around the planet azimuthal location [-1, 1] radian to reveal the gap width. The density gap after the eccentricity excitation for 2D runs are significantly shallower and wider than those of 3D, while they are similar before the eccentricity is excited. }
 \label{fig:sigma_comp2d}
\end{figure}

In Figure~\ref{fig:sigma_comp2d}, we plot the 2D global disc's surface density profile in the upper panel after the accretion outburst is triggered, and compare the azimuthal-averaged surface density profiles with the 3D results in the lower panel. Because in 2D the gap profile becomes strongly non-axisymmetric, we plot the radial surface density profile along $\phi=0$ which is more directly associated with resonance damping (although linear theories that assume axisymmetry may already start to break down in this context). In contrast to 3D, the 1:2 resonance loses its damping effect in 2D.
The 2D gap is also asymmetric in the radial direction, with the outer gap edge being much farther away from the planet than the inner one, and the outer gap edge is highly eccentric. 
This is consistent with the disc eccentricity profile we have shown in the lower panel of Figure~\ref{fig:mdot_comp2d}. The
precession of the coherent eccentric mode can contribute to the long-term oscillation of accretion rates.
As the disc eccentricity is excited, the gap center becomes much shallower than that of 3D as shown in the lower panel of Figure~\ref{fig:sigma_comp2d}, 
which then induce higher accretion rates on average. 
Right before the accretion outburst is triggered, e.g., around 165 orbits, we confirm the gap is nearly circular without significant disc eccentricity excitation. The 2D and 3D accretion rates at this stage are thus similar. Figure~\ref{fig:sigma_comp2d} also shows that at this time the 2D gap is only slightly wider than the 3D gap, but this difference is significantly amplified later 
by non-linear effects.

For our fiducial run with $q_{\rm th}=8$ and $K^{\prime}=8$, we have also explored the difference of accretion history between 2D and 3D \footnote{We have cross-check with our previous 2D simulations \citep{Li2021}, and the accretion rates there are a factor of three smaller than those reported here. This is mainly due to the a larger softening scale adopted in that work. The transition from circular to eccentric disc, however, is insensitive to the choice of softening scale.}. 
It turns out that the accretion rate are both stable without eccentricity excitation, although the accretion rates in 3D are slighter smaller than those in 2D simulations. 

\begin{figure*}
\centering
\includegraphics[width=0.32\textwidth,clip=true]{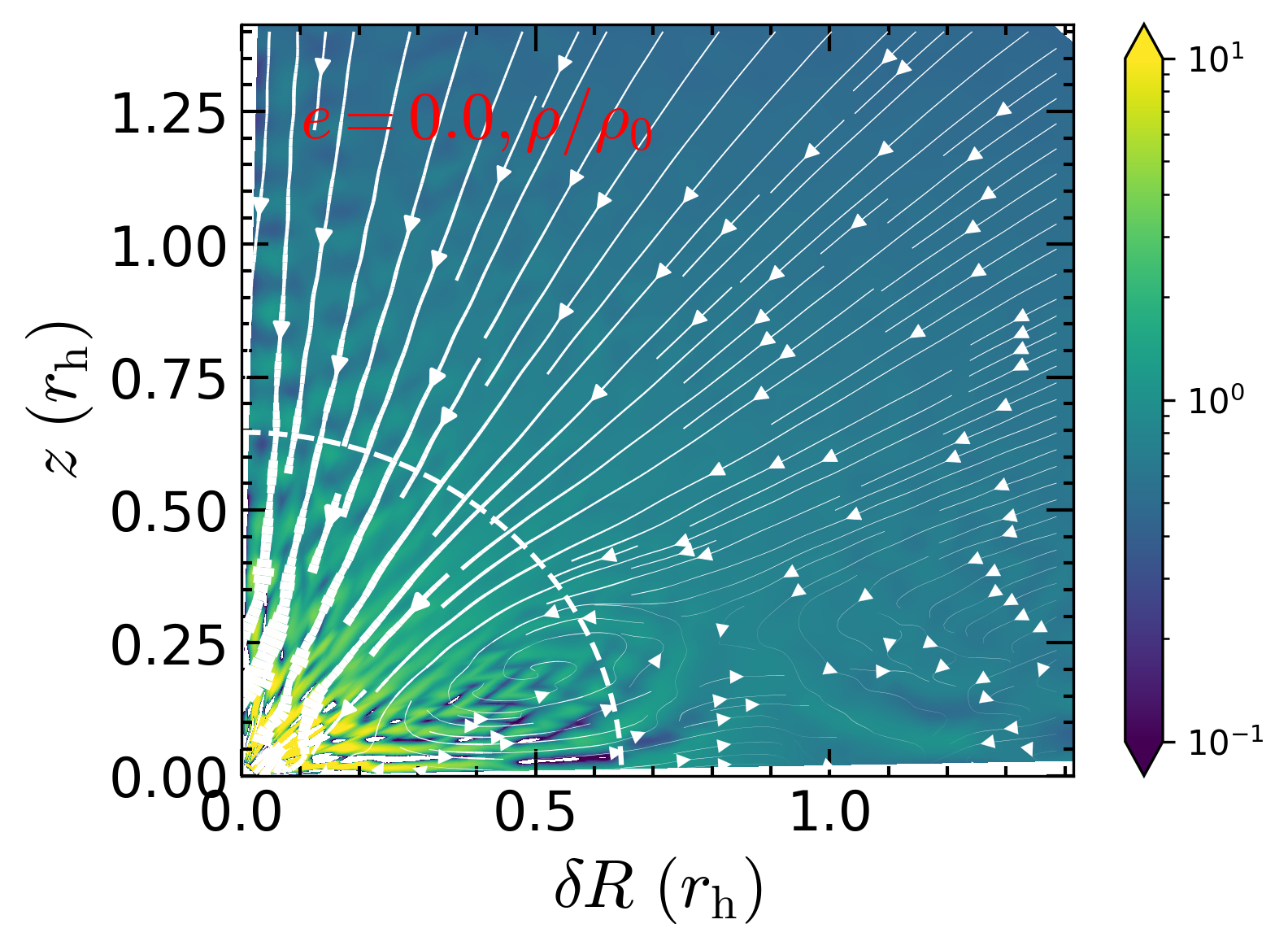}
\includegraphics[width=0.32\textwidth,clip=true]{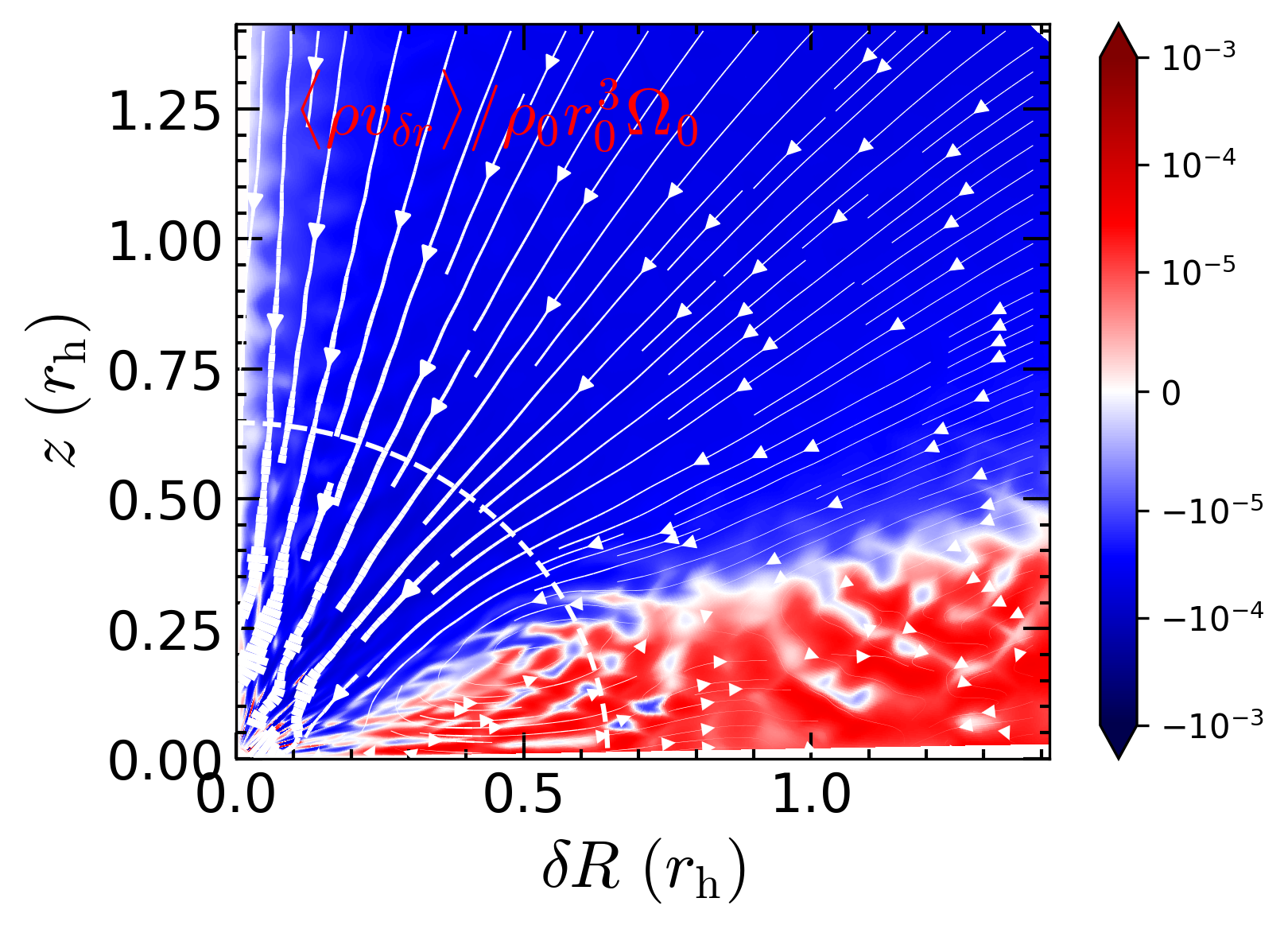}
\includegraphics[width=0.32\textwidth,clip=true]{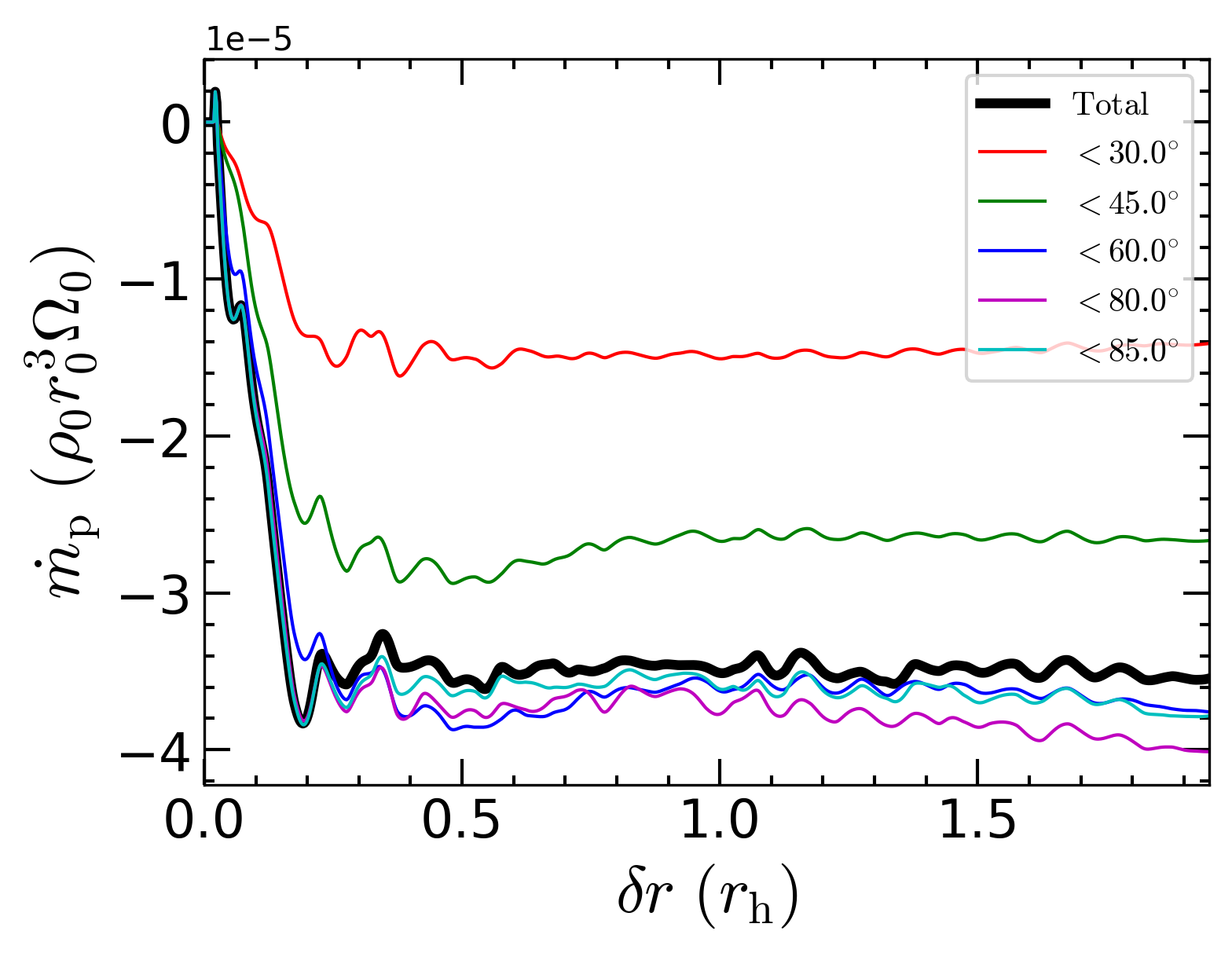}
\includegraphics[width=0.32\textwidth,clip=true]{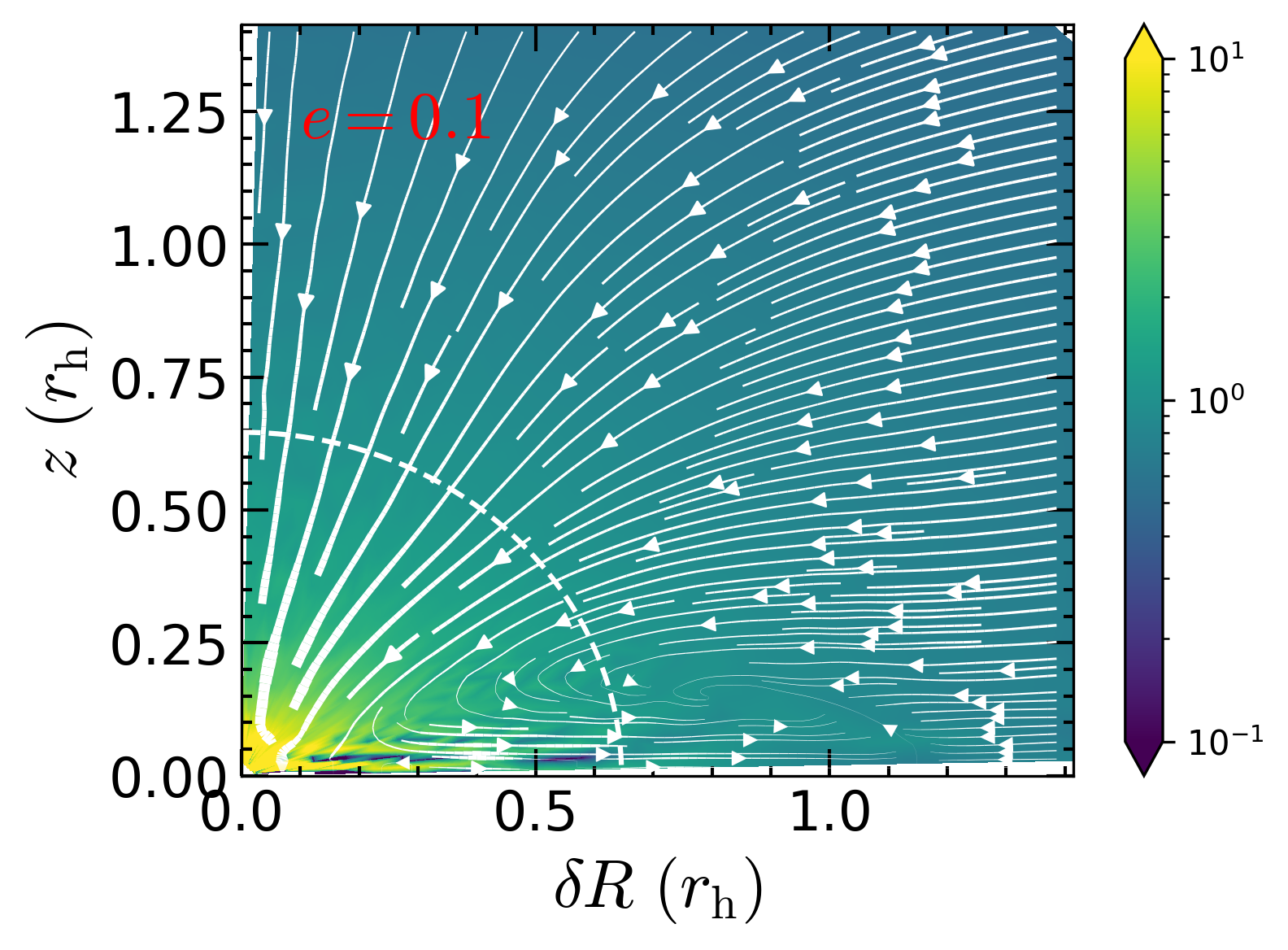}
\includegraphics[width=0.32\textwidth,clip=true]{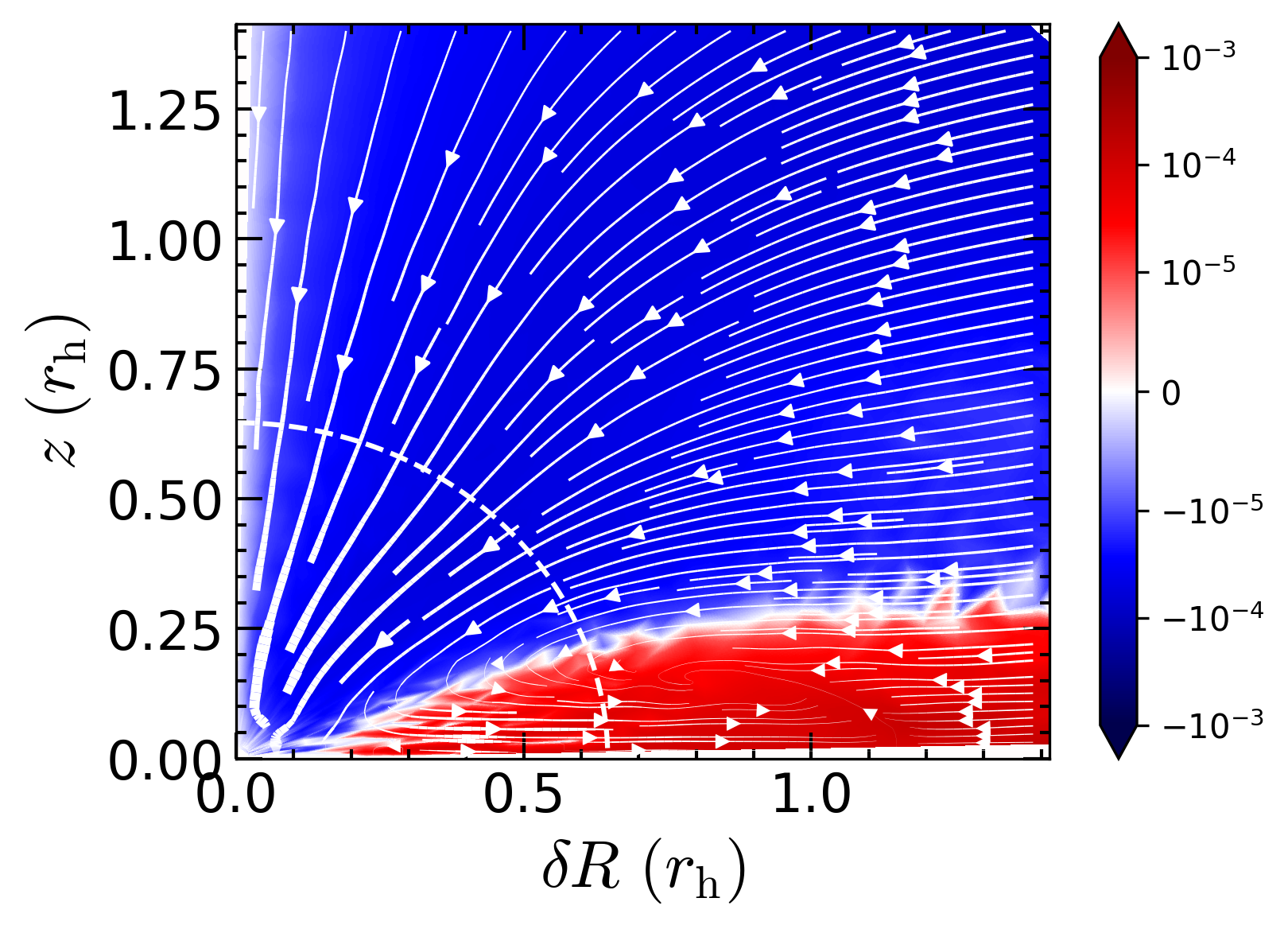}
\includegraphics[width=0.32\textwidth,clip=true]{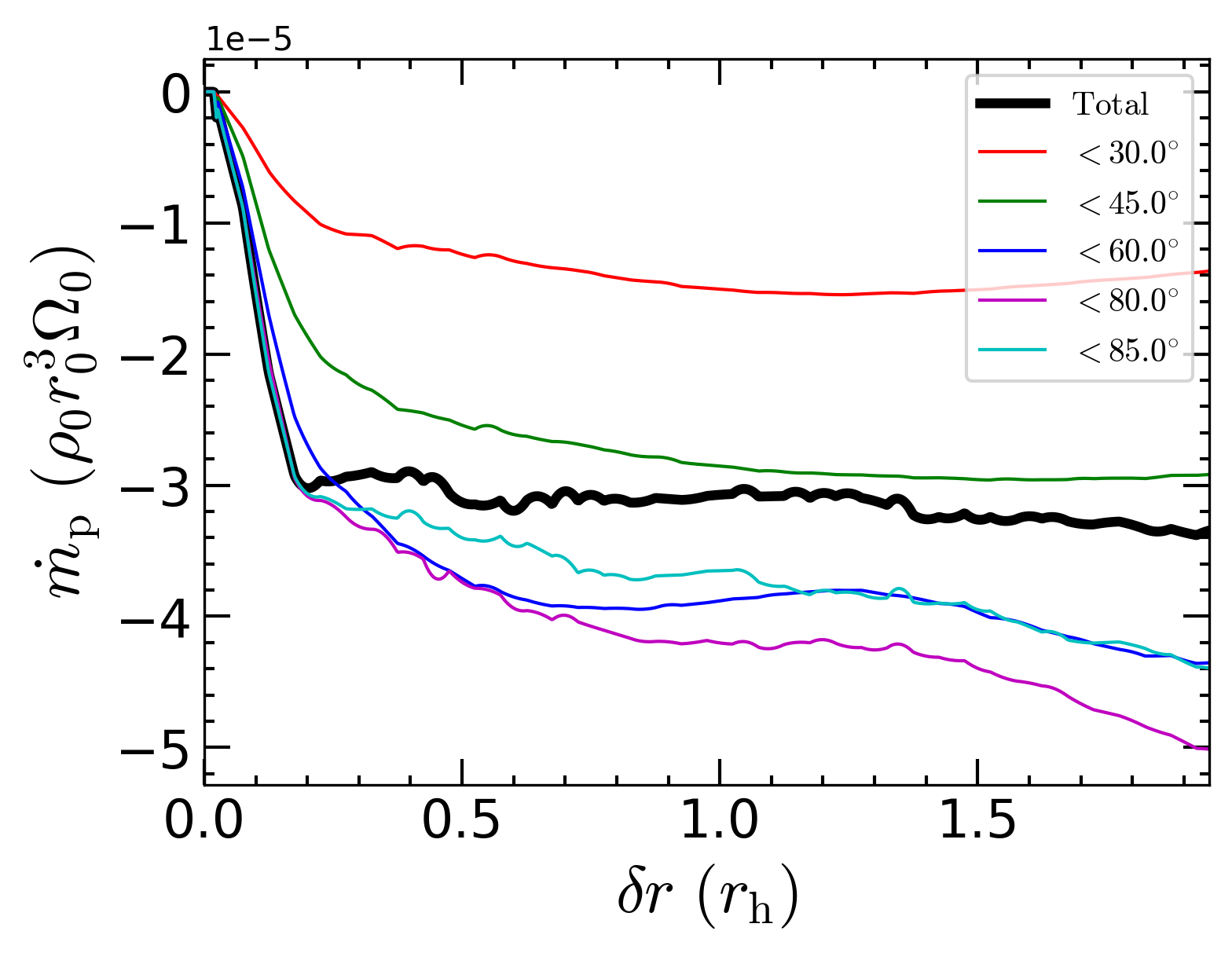}
\caption{
Left panels: time-, and azimuthally averaged density for $e=0.0$ (upper panel), and $e=0.1$ (lower panel) for 
{a sub-thermal planet with} $q=3\times10^{-4}$, $h_{0}=0.1$ and $\alpha=0.001$. Middel panels: time-, and azimuthally averaged mass flux onto the planet for $e=0.0$ (upper panel), and $e=0.1$ (lower panel). The arrows in these plots show the time-averaged and azimuthally averaged velocity in the frame co-moving with the planet. The dashed circle in the left and middle panels represent the Bondi radius $R_{\rm B} \simeq 0.65 r_{\rm h}$. Right panels: the $\delta r$ distribution of mass flux onto the planet for $e=0.0$ (upper panel), and $e=0.1$ (lower panel). The mass fluxes are vertical integrated from different polar angles based on the middle panels. Different lines in each plot show the contribution from different polar angles. 
}
 \label{fig:fluxmass2d_rtheta_lowqth}
\end{figure*}

\begin{figure}
\centering
\includegraphics[width=0.45\textwidth,clip=true]{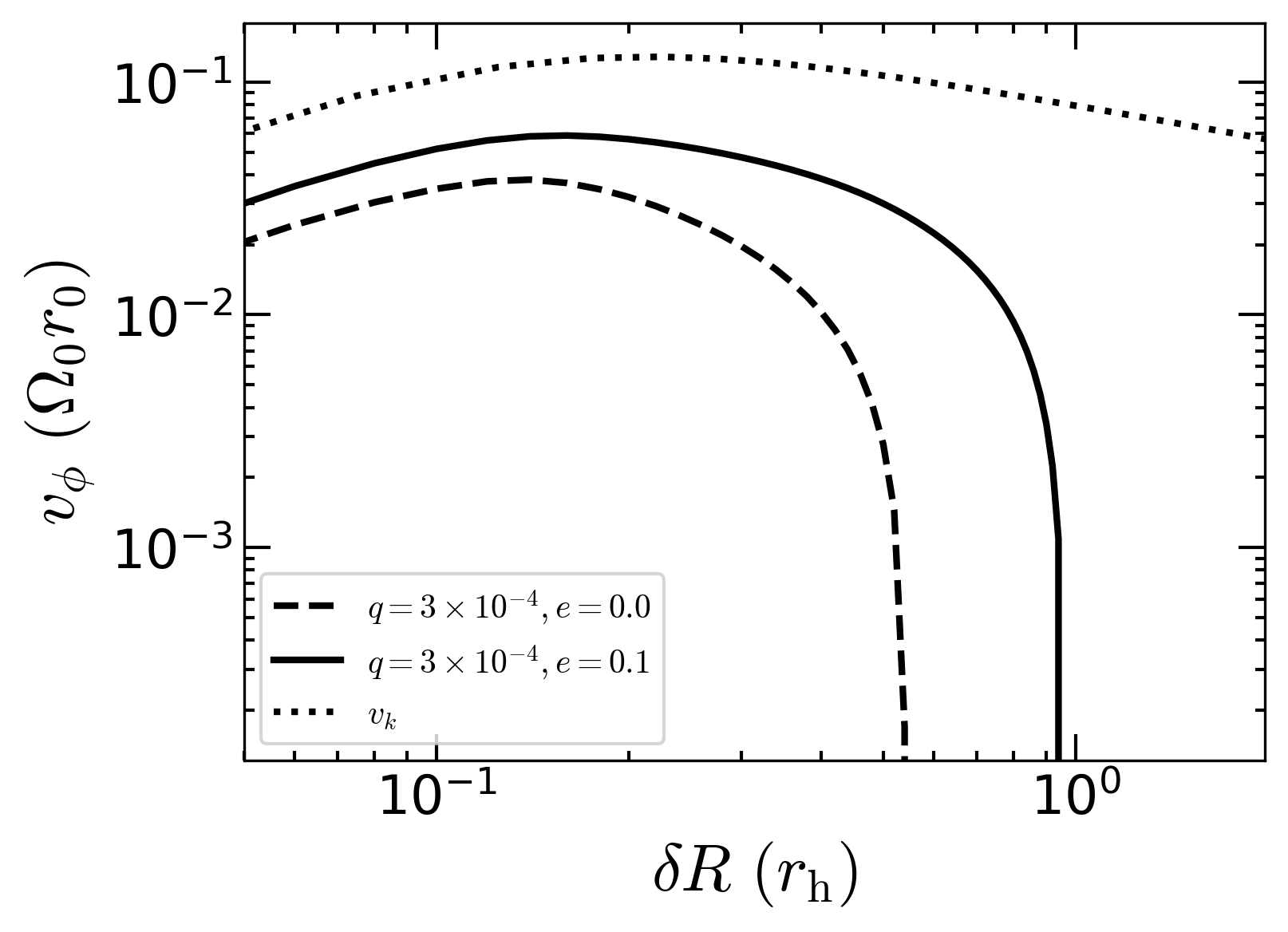}
\caption{In comparison with Figure~\ref{fig:vphi_cpd}, the 
CPD around a sub-thermal planet (with $q=3\times10^{-4}$ and $h_{0}=0.1$)
is not rotationally supported as its $v_{\phi} \lesssim 0.3 v_{\rm k}$. It 
is also truncated slightly inside the planet's Bondi radius $R_{\rm B} \simeq 0.65 r_{\rm h}$.
}
 \label{fig:vphi_cpd_lowq}
\end{figure}

\subsubsection{Low Mass Cases with Lower $q_{\rm th}$} \label{sec:lowq}

When $q_{\rm th}\lesssim3$ , or equivalently $r_{\rm h}<H$, the flow pattern around the planet essentially becomes 3D, and the vertical flow from the highest latitudes becomes more important.
We show two examples of { sub-thermal planets} with $q_{\rm th}=0.3$ with $e=0.0$ and $e=0.1$ in Figure~\ref{fig:fluxmass2d_rtheta_lowqth}. The corresponding planet mass ratio is $q=3\times10^{-4}$ and disc aspect ratio is $h_{0}=0.10$. The density is now not well settle down to middle plane and puffs up as a nearly spherical envelope. 
As expected, the rotation velocity is $20\%-30\%$ of the local Keplerian value {(relative to the planet)} as shown in Figure~\ref{fig:vphi_cpd_lowq}. The rotation-supported disc for sub-thermal cases, if exists, should be much smaller than our softening length $0.1\ r_{\rm h}$.
We have checked that the sub-Keplerian CPD disc is now $\sim50\%$ thicker than that predicated by Equation~\ref{eq:h_cpd}.

The region close to the polar axis can also contribute a sizeable fraction to the total accretion rate. Instead, the outflow from the midplane becomes more prominent compared to the high $q_{\rm th}$ cases.
In right panel of Figure~\ref{fig:fluxmass2d_rtheta_lowqth}, we plot the contribution of mass flux within different polar angles. We can see that the accretion flux are mainly attributed to the region $\theta<45^{\circ}$, while the contribution from the disc surface becomes insignificant. 
This is again confirms a different flow structure from the high $q_{\rm th}$ cases shown above.

\begin{figure}
\centering 
\includegraphics[width=0.45\textwidth,clip=true]{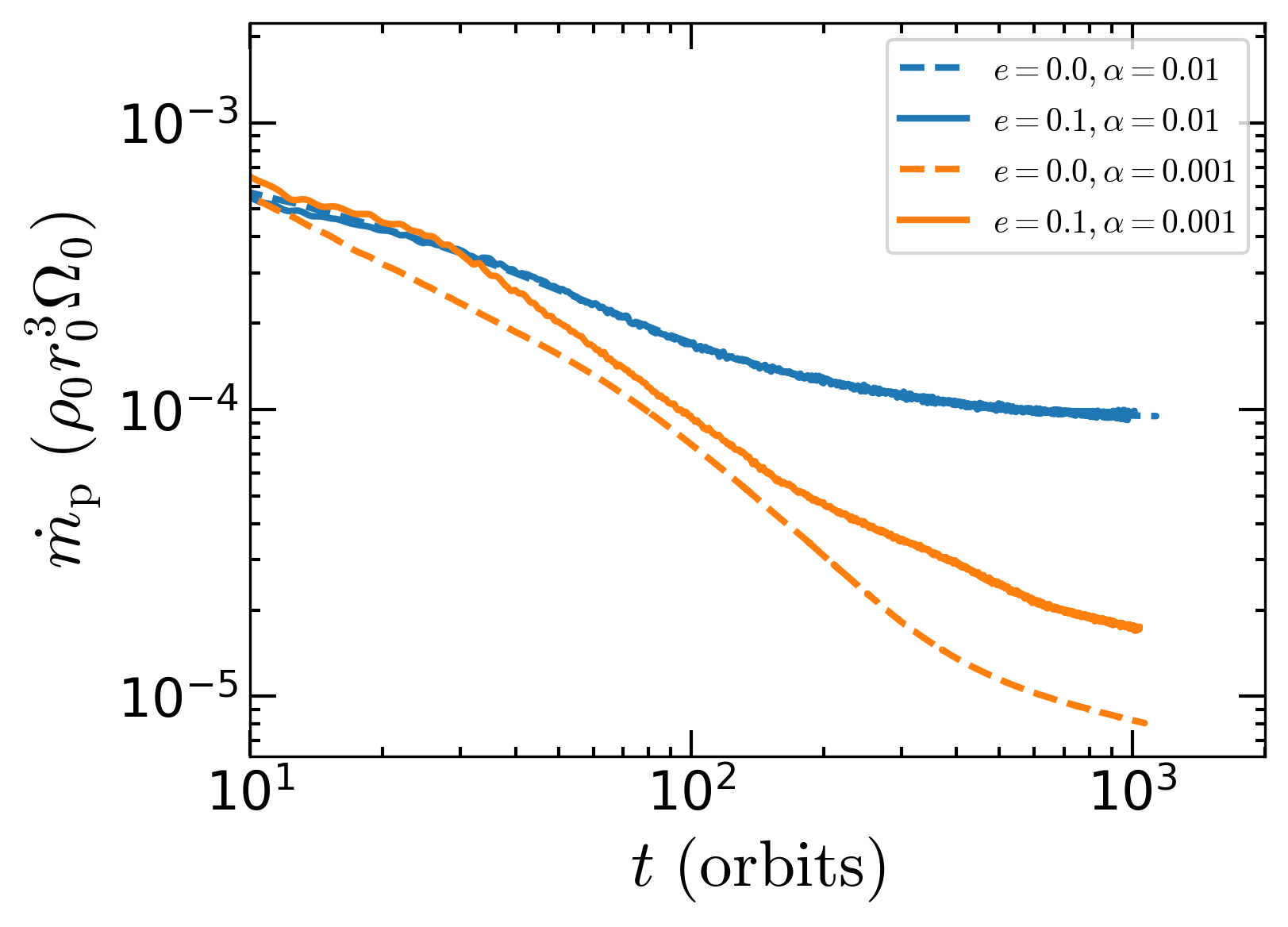}
\includegraphics[width=0.45\textwidth,clip=true]{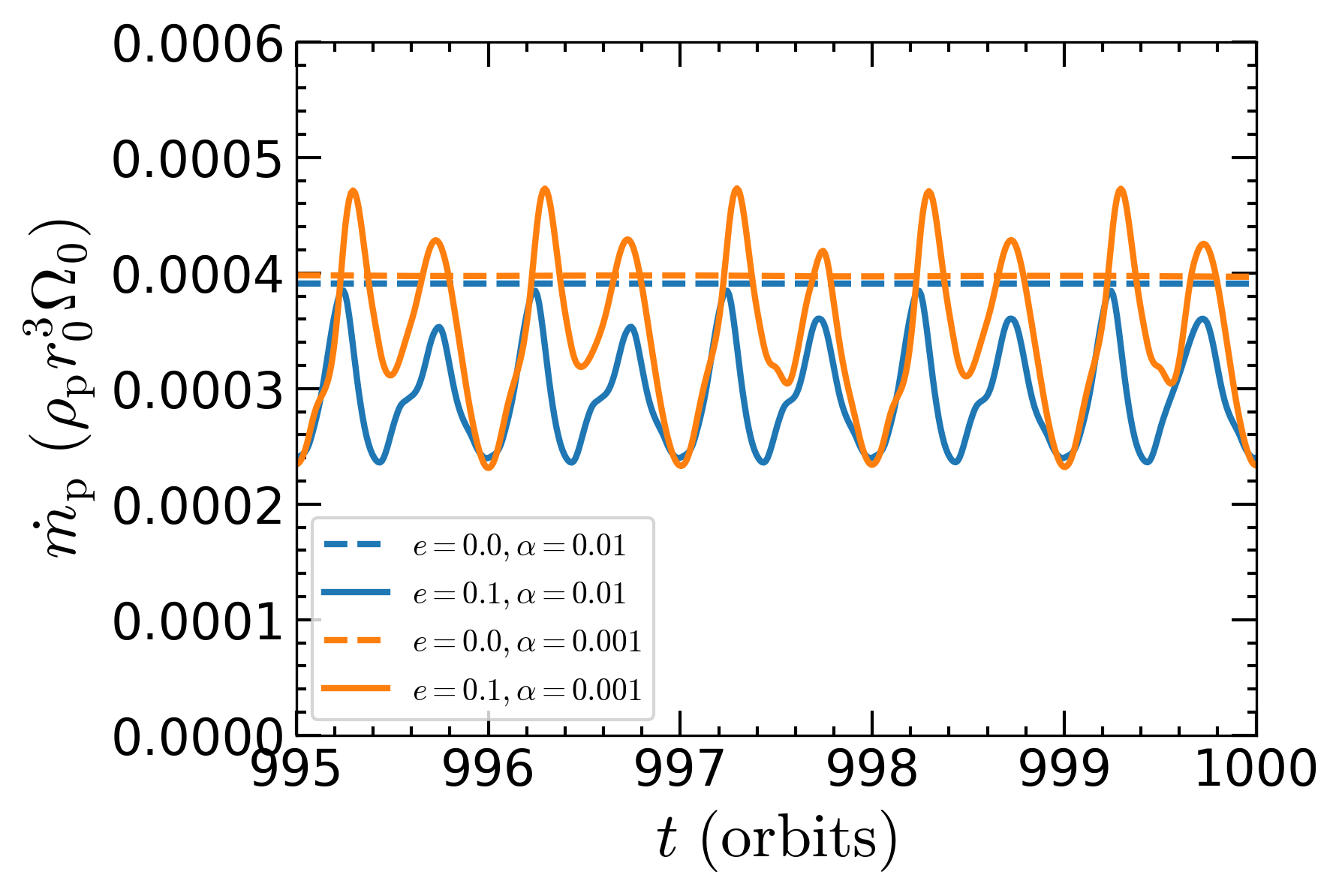}
\caption{
Upper panel: accretion rates for different disc viscosities and planetary eccentricities. The other model parameters are $q=0.001$, and $h_{0}=0.05$.
Note that the time-averaged accretion rates for high viscosity cases are almost the same for the two planetary eccentricities. Lower panel: accretion rates normalized by \textit{perturbed} density $\rho_{\rm p}$ at $1\ r_{\rm h}$ from the planet.  The normalized accretion rates are quite similar for different viscosities (identical for circular cases with two different viscosities) although the variability amplitudes are slightly different for eccentric planets.
}
 \label{fig:mdot_alpha}
\end{figure}

\subsection{Effect of Viscosity}\label{sec:vis}

Another factor which could affect the planetary accretion is the global disc viscosity. 
This is because viscosity plays an important role in shaping the gap profile. 
The accretion rates for different viscosities and planetary eccentricities are shown in Figure~\ref{fig:mdot_alpha}. As expected, the planetary accretion rates increase with increasing disc viscosity. In addition, we find that the time-averaged accretion rates are the insensitive to the planetary eccentricity for the high viscosity cases ($\alpha=0.01$) as the gap density for the high viscosity case does not increase significantly for the $e=0.1$ planet compared to the circular case. 

After normalization by the midplane density $\rho_{\rm p}$ at $r_{\rm h}$, the dependence of accretion rates on the disc viscosity 
essentially vanishes, for both the circular and eccentric planets, as shown in the lower panel of Figure~\ref{fig:mdot_alpha}. 
This result suggests that such a dependence is primarily due to the CPD density, which is related to by the gap profile induced by the planet. The dynamical flow structures of the CPD introduce no additional factors at all that are dependent on viscosity and low eccentricity.

\subsection{Scaling Relation}
\label{sec:scaling_relation}

\begin{figure}
\centering 
\includegraphics[width=0.45\textwidth,clip=true]{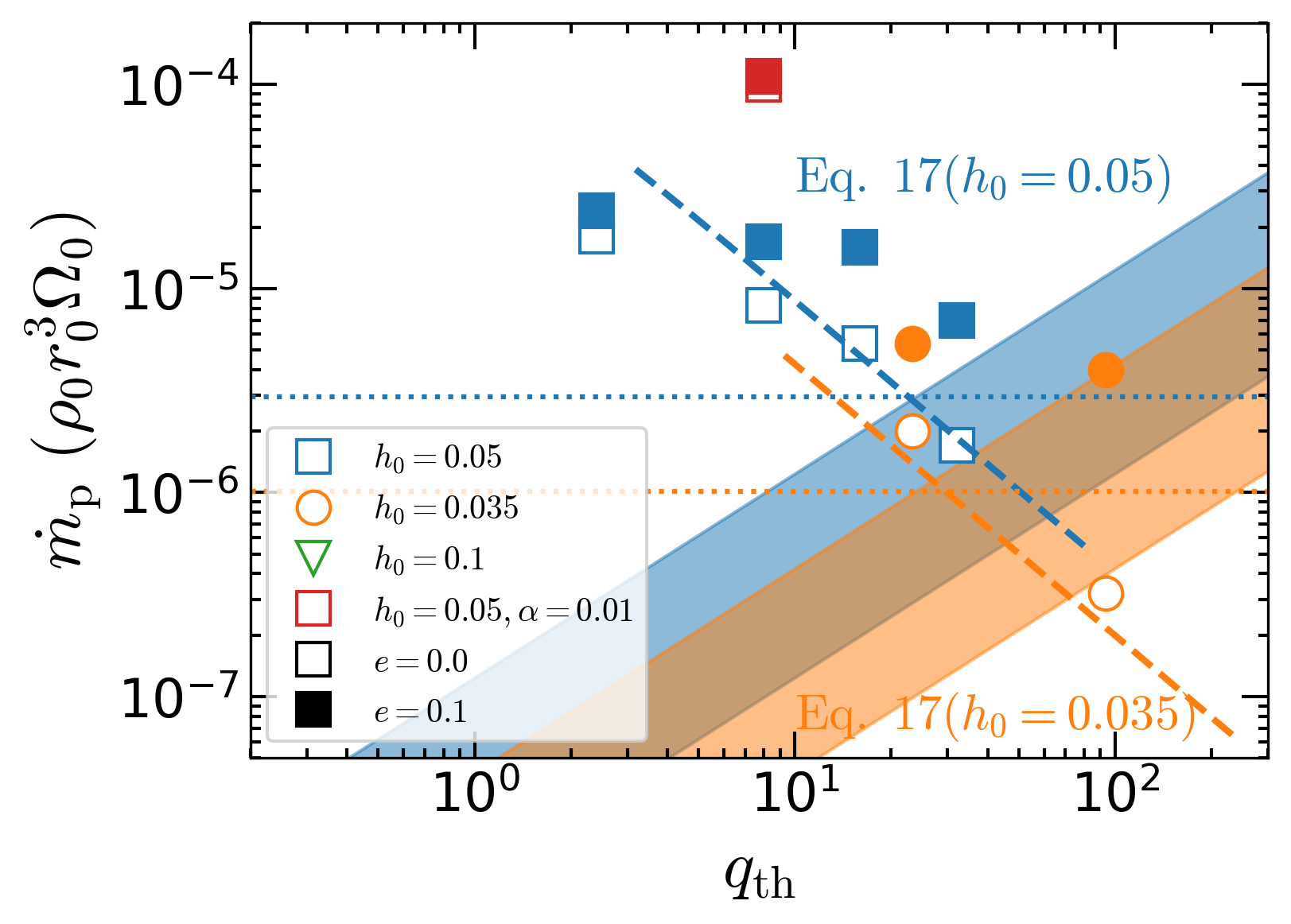}
\includegraphics[width=0.45\textwidth,clip=true]{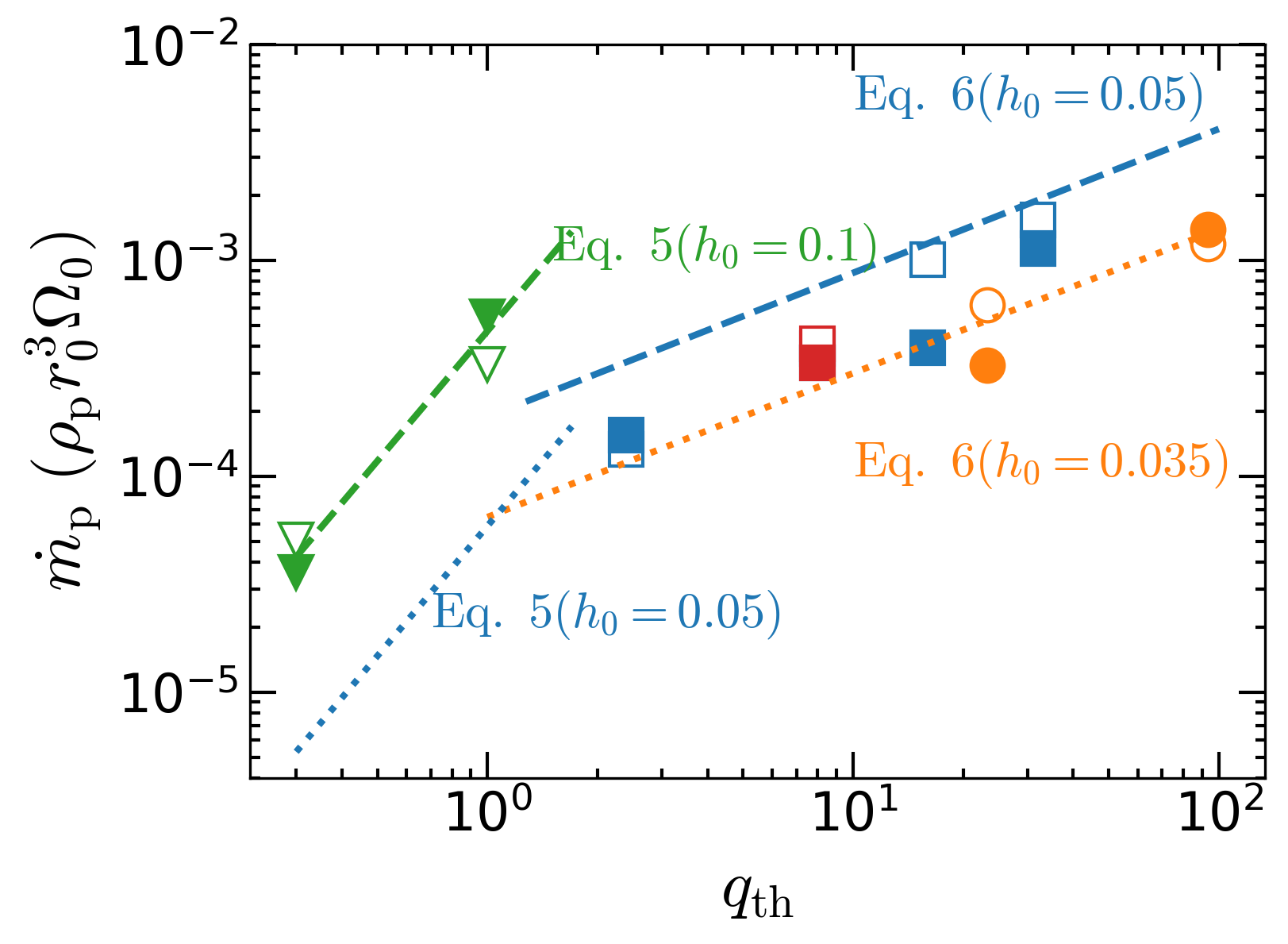}
\caption{
Scaling relation of the planetary accretion rate $\dot{m}_{\rm p}$ with the planet thermal mass $q_{\rm th}$. Upper panel: $\dot{m}_{\rm p}$ measured using initial density as a function of planet thermal mass. Different symbols correspond to different $h_{0}$ and/or $\alpha$. The disc viscosity for most models is $\alpha=0.001$ unless for two special cases represented by red squares. Open (filled) symbols indicate the $e=0.0$ ($e=0.1$) planets. The two dashed lines in the upper panel show the theoretical accretion rates based on the Hill accretion with $\alpha=0.001$ (Equation~\ref{eq:mdot_modified}), which can adequately explain the corresponding accretion rates for circular planets (open blue and orange symbols). 
The two dotted lines in the upper panel show the disc accretion rate with $\alpha=0.001$, $h_{0}=0.05$ and $h_{0}=0.035$.
The color bands indicate the regime where the mass-doubling timescale for the planet is $3$ Myr with the disc accretion rate between $10^{-9}-10^{-8}\ {M_{\odot}\ {\rm yr^{-1}}}$. Two different color bands correspond to different $h_{0}$ while $\alpha$ is fixed as 0.001.
Lower panel: $\dot{m}_{\rm p}$ measured with perturbed midplane density at ${\rm min}(r_{\rm h},R_{\rm B})$ from the planet $\rho(\delta r = {\rm min}(r_{\rm h},R_{\rm B}),\theta = \pi/2)$ (such that it's a ``background" density for Hill accretion) in the lower panel, averaged over the planet azimuth.
The two dashed lines show the power-law scaling of $q_{\rm th}^{2}$ and $q_{\rm th}^{2/3}$, which are based on Equations~\ref{eq:mdot_mb} and \ref{eq:mdot_mh}. The Bondi accretion rates based on Equation~\ref{eq:mdot_mb} (green lines) have been scaled down by a factor of $\sim7$ to match the simulation results. The two dotted lines are extrapolated from the the two dashed lines based on the $h_{0}^{3}$ scaling. We can see that the transition from $q_{\rm th}^{2}$ to $q_{\rm th}^{2/3}$ occurs around $q_{\rm th}\sim1-2$ for $h_{0}=0.05$.}
 \label{fig:mdot_mth}
\end{figure}

We compile the mass accretion rates onto the planet as a function of the thermal mass $q_{\rm th}$ (Eq. \ref{eq:qth}) 
in Figure~\ref{fig:mdot_mth}. In the upper panel of Figure~\ref{fig:mdot_mth}, the mass accretion rates are in unit of initial disc density $\rho_{0}r_{0}^{3}\Omega_{0}$, where all quantities are measured at $r_{0}$ at time $t=0$. We can see that the accretion rates decrease all the way up to $q_{\rm th}\simeq90$ for $q_{\rm th}\gtrsim1$ for a given viscosity (i.e., $\alpha=0.001$). This indicates that there is no significant disc eccentricity excitation and 2D-like accretion outburst even for very mass planet as we have mentioned above. 
In the sub-thermal cases, the planetary mass accretion rates show a reversed dependence on the thermal mass $q_{\rm th}$ as expected before gap opening \citep[e.g.,][]{DAngelo2003,Bodenheimer2013}.

Based on the modified Bondi accretion theory in Section~\ref{sec:theo}, we can first compare the theoretical accretion rates onto the planet with our simulation results in the ``unperturbed" unit $\rho_0r_0^3\Omega_0$. 
The key point is to predict the density at the gap location where the accretion material is supplied. 
The gap surface density is found in 2D simulations to be controlled by the $K\equiv q^2/\alpha h_{0}^{5}$ parameter \citep[e.g.,][]{Duffell2013,Kanagawa2018}, and

\begin{equation}\label{eq:sigma_gap}
\Sigma_{\rm p}=\frac{\Sigma_{0}}{1+0.04K},
\end{equation}

Based on the disc scale height of CPD, we can obtain the gap density $\rho_{\rm p}$ at $1r_{\rm h}$ from the super-thermal planet,

\begin{equation}
\rho_{\rm p}=\frac{\Sigma_{\rm p}}{\sqrt{2\pi}h_{\rm CPD}(1r_{\rm h})} \simeq \frac{2\rho_{0}}{1+0.04K},
\label{eq:rho_gap}
\end{equation}
where we have used $h_{\rm CPD}(1r_{\rm h})\simeq0.5H_{0}$ based on Equation~\ref{eq:h_cpd} \footnote{
However, the disc scale height for the sub-thermal planet may not be well described by Equation~\ref{eq:h_cpd} since it is now less rotation-supported as shown in Section~\ref{sec:lowq}.}.  
We show in Appendix~\ref{app:kgap} that such a scaling of the gap density with the $K$ parameter is well consistent with our 3D simulations ($K\gtrsim10$ and $q_{\rm th}\gtrsim0.3$).

The accretion rates onto the super-thermal planet then can be predicted based on  Equation~\ref{eq:mdot_mh} with $\rho_{\rm p}$ (Eq. \ref{eq:rho_gap}), i.e.

\begin{equation}
\dot{m}_{\rm H} = \pi\left(\frac{q_{\rm th}}{3}\right)^{2/3}h_{0}^{3}\rho_{\rm p}r_{0}^{3}\Omega_{0}
= {2\pi h_0^3 \rho_{0}r_{0}^{3}\Omega_{0} \over 1+0.04 q_{\rm th}^2 h_0/\alpha} \left(\frac{q_{\rm th}}{3}\right)^{2/3}.
\label{eq:mdot_modified}
\end{equation}

In comparison with the unperturbed steady-state accretion rate through the global disc, ${\dot M}_0 
= 3 \pi \Sigma \nu= \sqrt{18 \pi^3} \alpha h_0^3 \rho_0 \Omega_0 r_0^3$ (Eq. \ref{eq:steadymdot}), 

\begin{equation}\label{eq:mpdot_md}
    {\dot m}_{\rm H} = { \sqrt{2/9 \pi} \over (\alpha + 0.04 h_0 q_{\rm th}^2)} 
    \left( {q_{\rm th} \over 3} \right)^{2/3} {\dot M}_0,
\end{equation}
which reduces well below ${\dot M}_0$ in the highly super-thermal ($q_{\rm th} \gg 1$) limit.

The predicted accretion rates based on Equation~\ref{eq:mdot_modified} for different $h_{0}$ with fixed $\alpha=0.001$ are shown as dashed lines in the upper panel of Figure~\ref{fig:mdot_mth}. 
We can see that the theoretical rates can satisfactorily explain the $\propto q_{\rm th}^{-4/3}$ power law dependencies for the super-thermal cases when $K = q_{\rm th}^2 h_0/\alpha \gg 1$ (with Equation~\ref{eq:mdot_modified}). 
However, it is worth noting that in the mildly super-thermal cases, when $h_0^3 \lesssim q \lesssim 2.4 h_0^{9/4}$, we expect ${\dot m}_{\rm H} > {\dot M}_0$, the inward mass flow, from regions of the global disc outside the planet's orbit, 
is partially interrupted by the planetary accretion flow. 
We thus would expect the planetary accretion rate to be capped by the unperturbed disc accretion rate ${\dot M}_0$. 
This can be approximated as 

\begin{equation}
    \tilde{\dot m}_{\rm H}= \min({\dot{M}_{\rm 0},\dot{m}_{\rm H}}).
    \label{eq:mdot_cap}
\end{equation}

The capped accretion rates for two different $h_{0}$ are shown as dotted lines in the upper panel of Figure~\ref{fig:mdot_mth}. 
There are several mildly super-thermal models with ${\dot m}_{\rm p} > {\dot M}_0$. 
These value of 
excess ${\dot m}_{\rm p}$ is due to numerical artifacts of mass supply from the inner boundary to the the global disc.  
Since mass loss from the central star is not generally expected, 
appropriate boundary conditions need to be applied to 
prevent inflow into the computational domain. 
To explore the effect the inner disc depletion, we have shown in
Appendix~\ref{app:inner_bc} that the planetary accretion rates are limited by the disc accretion rates when 
adopting a more realistic inner boundary condition. 
Nevertheless, this disc depletion effect does not affect 
the scaling we show below when we normalized with the gap density $\rho_{\rm p}$ since this effect has already 
been absorbed into the depletion factor $\rho_{\rm p}/\rho_{0}$.

Beyond gap opening the mass doubling timescale $t_{\rm double} = M_{\rm p}/\dot{m}_{\rm p}$ continues to increase, until the planet reaches its asymptotic mass as $t_{\rm double}$ becomes comparable to the disc lifetime.
We plot in the upper panel of Figure~\ref{fig:mdot_mth} extra semi-transparent bands that indicate where the mass doubling of the planet reaches that of the disc lifetime. 
For a representative timescale of $t_{\rm double} = 3$Myr (upper limit for the lifetime of most discs), we calculate the required planetary accretion rates by converting from the physical unit to the code unit $\rho_{0}r_{0}^{3}\Omega_{0}$. 
To do this, we adopt disc accretion rates $10^{-9}{M_{\odot}\ {\rm yr^{-1}}} < \dot{M}_{\rm 0}< 10^{-8}{M_{\odot}\ {\rm yr^{-1}}}$, 
and further assume $\alpha = 0.001$, two different $h_0$ ($h_{0}=0.05$ and $h_{0}=0.035$ for two different color bands).
If accretion starts early in discs with $\dot{M}_{\rm 0}\sim  10^{-8}{M_{\odot}\ {\rm yr^{-1}}}$ (lower bounds of the color bands), the runaway process will be stalled at a few to ten Jupiter mass. 
Such a scenario would be difficult to reconcile with the apparent ceiling (a decline in numbers for planet mass larger than a few Jupiters) of the planetary mass distribution \citep{Cumming2008,Mayor2011,Petigura2018}. This might suggest that dynamical gas accretion occur 
in advanced stages of protostellar evolution \citep{Tanaka2020}, 
when photo-evaporation of the disc \citep{Owen2011} is effective or the accretion rates have typically declined to $10^{-9}{M_{\odot}\ {\rm yr^{-1}}}$. Notably, the mass for critical doubling timescale is also roughly where the planet accretion rate falls below $\dot{M}_{\rm 0}$ and can be approximated by Equation \ref{eq:mpdot_md}.

When we normalize the mass accretion with the perturbed density as shown in the lower panel of Figure~\ref{fig:mdot_mth} to isolate out the gap opening effect which is quite sensitive to $q_{\rm th}$ as well as $h_0$, 
the sub-linear scaling relation of $\dot{m}_{\rm p}-q_{\rm th}$ appears with a power law index of $2/3$ in the super-thermal regime due to the accretion structure. 
This is consistent with the theoretical expectation of Equation~\ref{eq:mdot_mh} based on modified Bondi accretion rate. This scaling relation is different from the $4/3$ power-law index based on the 3D local simulations for the marginally super-thermal planets $q_{\rm th}$ \citep{Maeda2022}. There are several reasons for this discrepancy. One reason is that the local shearing box simulations shown in \citet{Maeda2022} could have some boundary effects on the flow structures of the interface between the CPD and the global disk \citep{Dobbs-Dixon2007}, which is a major difference from our global simulations. The modification of the global disk structures due to the CPD boundary could significantly alter the accretion rates onto the planets.  
The second one is that they define the accretion streamlines as those reaching within $0.2r_{\rm h}$ from the planet even though their softening radius is much smaller. The choice of this accretion radius may affect the estimated accretion band width and its dependence on the planet thermal mass. 
For circular planets in the sub-thermal regime, the power law index of this scaling relation becomes $2$, which is also consistent with Equations~\ref{eq:mdot_mb}.

There should be another dependence of $\dot{m}_{\rm p}$ on the disc aspect ratio $\propto h_{0}^3$ in the super-thermal regime. We have two runs for $h_{0}=0.035$ with $q=0.001$ and $q=0.004$, shown as orange circles in Figure~\ref{fig:mdot_mth}. 
The accretion rates for those runs are expected to be a factor of $(0.05/0.035)^{3}\sim3$ lower than that of $h_{0}=0.05$ when fixed $q_{\rm th}$. 
However, as shown in lower panel of Figure~\ref{fig:mdot_mth}, 
our measured accretion rates are slightly higher than what we should expected from the $h_{0}$ scaling. 
We suspect that this is due to the slight accretion enhancement associated with the finite eccentricity excitation for those very high $q_{\rm th}$ planets (refer to Figure~\ref{fig:edisc}), 
even though the excited eccentricity is not significant enough to trigger the accretion outburst as shown in 2D counterparts.
When we extrapolate the $h_{0}=0.1$ models to $h_{0}=0.05$, shown as one dotted line in Figure~\ref{fig:mdot_mth}, the transition from the $q_{\rm th}^{2}$ 
to $q_{\rm th}^{2/3}$ scaling can be identified from the intersection point with the dashed blue line, which is roughly around $q_{\rm th}\sim1-2$. There can be an intermediate region with a scaling of $q_{\rm th}$ suggested by \citet{Choksi2023} which is not revealed in this work. 
This is due to the sparse sampling in our simulation data points, and complete parameter survey over $h_0$ is not the focus of the current work. 
On the other hand, quite apparent $h_0^3$ dependence of the $\rho_{\rm p}$-normalized accretion rate is shown in Figure 10 of \citet{Choksi2023}, 
which is measured in very short-term simulations before gap opening and excitation of global eccentric modes. 
We comment that although $\rho_{\rm p}$-normalization may isolate gap opening effects, global non-axisymmetric effects such as streamline eccentricity do exist to complicate the dependencies, 
and can only manifest when the simulations are run long enough for certain instability modes to develop.

\subsection{Case Comparison with Previous Simulations}

There are several global 3D simulations on the planetary accretion.
Most of them are focused on planet mass around and/or below Jupiter in circular orbits without planetary eccentricity. 
To compare them with the accretion rates in this work, we will mainly refer to the accretion rate for the circular planet with $q=0.001$, $\alpha=0.001$ and $h_{0}=0.05$. 
For this, we have the planetary accretion of $\dot{m}_{\rm p}\simeq 8\times10^{-6}\ \rho_{0}r_{0}^{3}\Omega_{0}$.

\citet{Kley2001} studied the accretion onto the Jupiter mass planet using 3D global simulations, and found $\dot{m}_{\rm p}\simeq 6.5\times10^{-5}\ \rho_{0}r_{0}^{3}\Omega_{0}$, which is slighter higher our accretion rate, after considering that they adopted a higher disc viscosity of $\alpha=0.004$ as discussed in Section~\ref{sec:vis}. We suspect that this is due to limited resolution around the planet in their simulations. \citet{DAngelo2003} further carried out several 3D simulations using nested grid for planet mass up to $\sim1$ Jupiter, and they found that the accretion rate for Jupiter mass planet is $\dot{m}_{\rm p}\simeq 2.5\times10^{-5}\ \rho_{0}r_{0}^{3}\Omega_{0}$. They also adopt a disc viscosity of $\alpha=0.004$. This accretion rate can be re-scaled to $\dot{m}_{\rm p}(\alpha=0.001)\simeq 0.6\times10^{-5}\ \rho_{0}r_{0}^{3}\Omega_{0}$, which thus corresponds well with the accretion rates in this work.  \citet{Bodenheimer2013} reported a similar planetary accretion rate of $2.4\times10^{-5}\ \rho_{0}r_{0}^{3}\Omega_{0}$ for the Jupiter mass planet embedded in a viscous disc with $\alpha=0.004$. We do note that when we raise our $\alpha$ to 0.01 the accretion rate becomes $\simeq 10^{-4} \rho_{0}r_{0}^{3}\Omega_{0}$ (see \S \ref{sec:vis}), so it's consistent that their numbers lie between our high and low viscosity measurements.
\citet{Bodenheimer2013} also showed the accretion accretion for very massive planets, e.g.,  $\dot{m}_{\rm p}\simeq 2.1\times10^{-6}\ \rho_{0}r_{0}^{3}\Omega_{0}$ for $q=0.004$ (with no sign of accretion burst). 
This value is close to what we obtain in this work, however, our viscosity parameter is a factor of 4 smaller than theirs. 

Another branch of simulation did not explicitly include viscosity for the disc evolution, and it is found that the accretion rate for a Jupiter mass planet is $\dot{m}_{\rm p}\simeq1.2\times10^{-5}\ \rho_{0}r_{0}^{3}\Omega_{0}$ \citep{Machida2010}, which is more consistent with our low viscosity results. 
Very recently, \citet{Choksi2023} performed inviscid simulations to explore the maximum accretion onto the super-thermal planets and measure an accretion rate of $\dot{m}_{\rm p}\sim 1.0\times 10^{-3}\ \rho_{\rm p}r_{0}^{3}\Omega_{0}$ at $q=0.001, h_0= 0.035$. To isolate the factor of gap opening, they run their simulations for a few tens of orbits and normalize their measurements with the evolving gap center density $\rho_{\rm p}$. In the same units, our measurements for this set of parameter after gap formation has reached a steady state is $\dot{m}_{\rm p}\sim 0.6 \times 10^{-3}\ \rho_{\rm p}r_{0}^{3}\Omega_{0}$. Since we discussed that viscosity only affects $\rho_{\rm p}/\rho_0$,
the reason for this order-unity discrepancy could be that they did not impose a sink hole around the planet to allow active accretion, or the disc has not evolved long enough to reach a global quasi-steady state.

Overall, the mass accretion rates measured in our models are in general consistent with recent 3D model using different numerical schemes.

\section{Observational Implications}\label{sec:imp}

\subsection{Accreting Protoplanets}

We can apply our simulation results to constrain properties of observed accreting protoplanets. 
Up to now, there is one observed accreting planetary system with two protoplanets PDS 70b, and PDS 70c embedded in a protostellar disc, 
which have been directly detected by near-infrared observations \citep{Keppler2018,Haffert2019}. 
The mass of the host star is $0.88\ M_{\odot}$ \citep{Keppler2019}, and the stellar age is $5.4\pm1.0\ \rm Myr$ \citep{Muller2018}. 
The two planets are located at $\sim22\ \rm au$ and $\sim34\ \rm au$ from the host star \citep{Keppler2018,Muller2018,Haffert2019,Wang2020,Benisty2021}, respectively, for PDS 70b and PDS 70c. The planet masses derived from near-infrared photometry and spectral energy distribution show large uncertainties, which are around $\sim2-17\ M_{\rm J}$ for PDS 70b, and $\sim0.5-12\ M_{\rm J}$ for PDS 70c, where $M_{\rm J}$ is the Jupiter mass \citep{Keppler2018,Muller2018,Christiaens2019,Mesa2019,Haffert2019,Wang2020,Wang2021}. 
The CPD surface density derived from molecular emission is constrained to be $8\times10^{-2}-8\times10^{-4}\ {\rm g\ cm^{-2}}$, although large uncertainty still exists \citep{Facchini2021,Choksi2022}. The stellar accretion rate is estimated to be $5.5\times10^{-8\pm0.4}\ {\rm M_{\rm J}\ yr^{-1}}$ \citep{Haffert2019} and the global disc's aspect ratios $h_0$ at 22 au and 34 au are estimated to be 0.07 and 0.08 \citep{Facchini2021,Choksi2022}.
The mass accretion rates for two planets are constrained to be $10^{-8}-7\times10^{-7}\ M_{\rm J}\ \rm yr^{-1}$, and $10^{-8}-5\times10^{-7}\ M_{\rm J}\ \rm yr^{-1}$ \citep{Wagner2018,Haffert2019,Wang2020}.

Based on the observational inferred values, we find $q_{\rm th} \simeq 6.6-56$ for PDS 70b and $\simeq 1-6$ for PDS 70c, qualifying both of them as super-thermal planets.  In this limit, the magnitude of ${\dot m}_{\rm p}$ is determined by Equation \ref{eq:mpdot_md} and Figure~\ref{fig:mdot_mth}.
Considering that the gas density is estimated at the planet locations, although we note that the upper limit are obtained around the gap edge, we need the accretion rates shown in the lower panel of Figure~\ref{fig:mdot_mth}. When we adopt $q=0.004$ for both planets for simplicity and $h_{0}=0.05$, $\dot{m}_{\rm p}\simeq 10^{-3}\ \rho_{\rm p}r_{0}^{3}\Omega_{0}$ \footnote{Note that the observation show a disc aspect ratio of $0.07$ and $0.08$ for two planets. According to the linear scaling of accretion rate on $h_{0}$ for super-thermal planets, this will slightly increase the accretion rates by $0.07/0.05$ and $0.08/0.05$. 
This slight increase of $\dot{m}_{\rm p}$ does not affect the conclusions drawn below.}. 
For PDS 70b and PDS 70c, this corresponds to $\dot{m}_{\rm p}({\rm PDS\ 70b})\simeq1.4\times10^{-8}-1.4\times10^{-6}\ M_{\rm J}\ \rm yr^{-1}$, and $\dot{m}_{\rm p}({\rm PDS\ 70c})\simeq1.5\times10^{-8}-1.5\times10^{-6}\ M_{\rm J}\ \rm yr^{-1}$, respectively. This large uncertainty arises from the surface density measurements. 
When the isothermal assumption is relaxed, the actual accretion rates could be smaller considering the suppression of the formation of CPD by heating around the protoplanets \citep{Szulagyi2017}, 
but even adiabatic CPDs can still become rotationally support at high $q_{\rm th}$ \citep{Fung2019}. 
Our simulations are, therefore, broadly consistent with the measured accretion rates. 
Furthermore, the mass doubling time for two protoplanets are $\sim3-300$ Myr, which is comparable or much larger than the planet age of $\sim5$ Myr. 
The upper limit corresponds to the lower bound of the gas surface density presented above, while the lower limit corresponds to the upper bound. 
Even for the lower limit of the mass doubling time, it suggests that the two protoplanets have largely finished their forming process. 
The upper limit of the mass doubling suggests that the two planets might have experienced outward migration in resonance to speed up the runaway growth.

Our estimated CPD size is $\lesssim 0.3\ r_{\rm h}$, which corresponds to $\lesssim1.0$ au and $\lesssim1.6$ au for the two protoplanets. These are roughly consistent with the observed outer CPD size and $<1.2$ au \citep{Wang2021,Benisty2021}. 
The inner CPD size is estimated to be $<0.3$ au, which may be associated with it having a large eccentricity $0.17\pm0.06$ \citep{Wang2021}. According to \citet{ChenYX2022}, for super-thermal planets with $e \gtrsim (q/3)^{1/3}$, the CPD size will no longer be determined by $r_{\rm h}$ as in the circular or mildly eccentric case, 
but by the much smaller Bond-Hoyle-Lyttleton radius 
that decreases steeply with eccentricity as $e^{-2}$ due to strong epicyclic motion.

\subsection{Accretion onto Stellar-mass Black Hole Embedded in AGN Discs}

AGN discs have emerged as rich factories for producing massive stars and their remnant stellar-mass black holes
(sBH), neutron stars (NSs), including some of the detected compact binary mergers by gravitational wave observations \citep[e.g.,][]{Artymowicz1993, McKernan2012,  Bartos2017,  Stone2017,  Leigh2018,  Grobner2020, Davies2020, Tagawa2020a,Li2021b,Li2022b,Dempsey2022,Kaaz2023,LiR2022a,LiR2023a,Lai2023}. Due to the ultra-dense environment around the embedded objects in AGN discs, their accretion and feedback could play an important role in shaping their dynamics and produce possible electromagnetic counterpart \citep{McKernan2019,Abbott2020b,Graham2020,Kimura2021,Perna2021,WangJM2021a,WangJM2021,ZhuJP2021,WangY2022,Fan2023,WangM2023}.

For the embedded compact object in AGN discs, the mass ratio between the central supermassive black hole (SMBH) and embedded sBHs is around $q\sim10^{-5}-10^{-7}$, the disc midplane density $\sim10^{-9}-10^{-10}\ {\rm g\ cm^{-3}}$, and the disc aspect ratio $h_{0}\sim0.001-0.01$ at $10^{3}\ R_{\rm g}$ from the SMBH \citep{Sirko2003,Thompson2005}, so the thermal mass $q_{\rm th}\sim10^{-1}-10^2$. Applying the accretion rates to embedded sBHs in AGN discs with $q_{\rm th}\sim10$, the accretion rates onto the sBHs are $\dot{m}_{\rm bh}\sim8\times10^{24}\ {\rm g\ s^{-1}}$, which is about 3-4 orders of magnitude higher than the Eddington mass accretion rates for a $\sim100\ M_{\odot}$ black hole. 
Such an accretion rate is about 2-3 orders of magnitude smaller than the Bondi accretion rates even after considering the gap depletion by the massive embedded sBH. 
This suggests that the Bondi accretion cannot be fully suppressed by the gap opening process of massive embedded object alone, as the gap is never totally depleted \citep[e.g.,][]{Chen2020b}. 
However, the accretion rates obtained here can be regarded as the supply rates around the Bondi radius of the embedded objects, as 1) our sink hole radius is much larger than the gravitational radii of sBHs and/or the size of massive stars, 
2) we have not included any possible radiative and/or mechanical feedback from sBH and stars. 
Strong outflow coupled with inflow could strongly suppress the accretion onto the sBHs, even though the inflow rates at Bondi radius is highly super-Eddington \citep{Pan2021,Chen2023}. 
Whether such a suppressed accretion for the sBH can be observable depends strongly on the strength of feedback \citep{Tagawa2022,Tagawa2023}, which will be studied numerically in subsequent works dedicated to sBH accretion in specific AGN-related environments.

\section{Conclusions and Discussions}\label{sec:con}

We perform a series of 3D simulations to study the dynamical accretion of massive embedded objects in discs. Our simulations are most relevant for the runaway accretion of gas giant in PPDs. 
The accretion is modelled as sink hole parametrised with the sink hole radius and removal rate. 
The convergence of accretion rate with respect to these parameters has been carefully verified and calibrated with previous 3D simulations.
Our findings are summarized as follows.

\begin{itemize}
\item The accretion mass flux mainly comes from the intermediate latitude above the disc midplane rather than the polar region for the super-thermal planet, while for sub-thermal planets the contribution of accretion flux is dominated by polar region. On average, the disc midplane hosts strong mass outflow, but the velocity and density distribution is highly non-axisymmetric with respect to the planet, 
and there exists inflow streams with high velocity but low density from directions perpendicular to the density waves (see Figure \ref{fig:fluxmass_fid}). 
The higher-latitude inflow coupled with midplane outflow drives meridional circulation as found in previous 3D simulations. 

\item For our super-thermal planets, a rotation supported CPD can form out to $0.2-0.3\ r_{\rm h}$ (see Figure~\ref{fig:vphi_cpd}). The CPD disc is much thicker than the global PPD and the disc aspect ratio is $\sim 0.2$ within the CPD region (see Figure~\ref{fig:hcpd}). Accretion of sub-thermal planets are spherical and Bondi-like (see Figure~\ref{fig:fluxmass2d_rtheta_lowqth}).

\item The moderate orbital eccentricity increases the planetary accretion rates by a factor of $2-3$ in most cases (see Figure~\ref{fig:mdot_ecc} and Figure~\ref{fig:mdot_highq}). 
However, such kind of enhancement depends on the planet mass, and which becomes weak for non-gap opening planets (see Figure~\ref{fig:mdot_alpha}). The accretion rate enhancement is due to the shallower gap and increased density in the CPD region induced by the eccentric planet. 
Raising viscosity also enhances the accretion rate only through mitigating the gap opening process and introduce no changes in the accretion structure (see Figure~\ref{fig:mdot_alpha}).

\item For very massive planet with $K^{\prime}\equiv q^2/\alpha h_{0}^3$ up to $\sim370$, the planet cannot excite strong disc eccentricity, and induce accretion outburst in our 3D simulations.
This is contrary to previous 2D results, and confirmed in our 2D simulations as well (see Figure~\ref{fig:mdot_comp2d}). 
This is found to be related to the narrower gap carved in 3D simulations which can damp the eccentricity more efficiently. 
We suspect that the disc eccentricity excitation could happen for more massive companions as the eccentric cavity has been found for 3D simulations of equal-mass circumbinary discs, but such mass ratios will be irrelavant for gas giants.

\item We explore the scaling of accretion rates with the thermal mass $q_{\rm th}$. After normalizing accretion rates with the local gap density we find that they follow a power-law of $q_{\rm th}^{2/3}$ for super-thermal planet but transform to a relation of $q_{\rm th}^{2}$ for very sub-thermal cases. (see Figure~\ref{fig:mdot_mth})

\end{itemize}

We have applied our simulation results to the observed accreting planetary system PDS 70b and PDS 70c. 
Our simulation results can satisfactorily reproduce the accretion rates and CPD size for these two protoplanets. Our simulation results can also be applicable to the accretion of sBHs embedded in AGN discs, 
although the feedback from these sBHs could suppress the accretion and thus in turn affect the electromagnetic signatures from these embedded accreting sBHs. 

\section*{Acknowledgements}
We thank the referee for helpful comments that improve the manuscript. We thank Nick Choksi and Zhaohuan Zhu for helpful exchanges.
This work is supported in part by the Natural Science Foundation of Shanghai (grant NO. 23ZR1473700), the Natural Science Foundation of China (grants 12133008, 12192220, 12192223, and 12373070), the science research grants from the China Manned Space
Project. The calculations have made use of the High Performance Computing
Resource in the Core Facility for Advanced Research Computing
at Shanghai Astronomical Observatory. 
Softwares: \texttt{Athena++} \citep{Stone2020}, \texttt{FARGO3D} \citep{Benitez-Llambay2016}, \texttt{Numpy} \citep{vanderWalt2011}, \texttt{Scipy} \citep{Virtanen2020}, \texttt{Matplotlib} \citep{Hunter2007}.

\section*{Data availability}
The data underlying this article will be shared on reasonable request to the corresponding author.

\bibliography{references.bib}{}
\bibliographystyle{mnras}

\begin{appendix}

\section{Convergence test: accretion parameters}\label{app:acc}
\begin{figure}
\centering
\includegraphics[width=0.45\textwidth,clip=true]{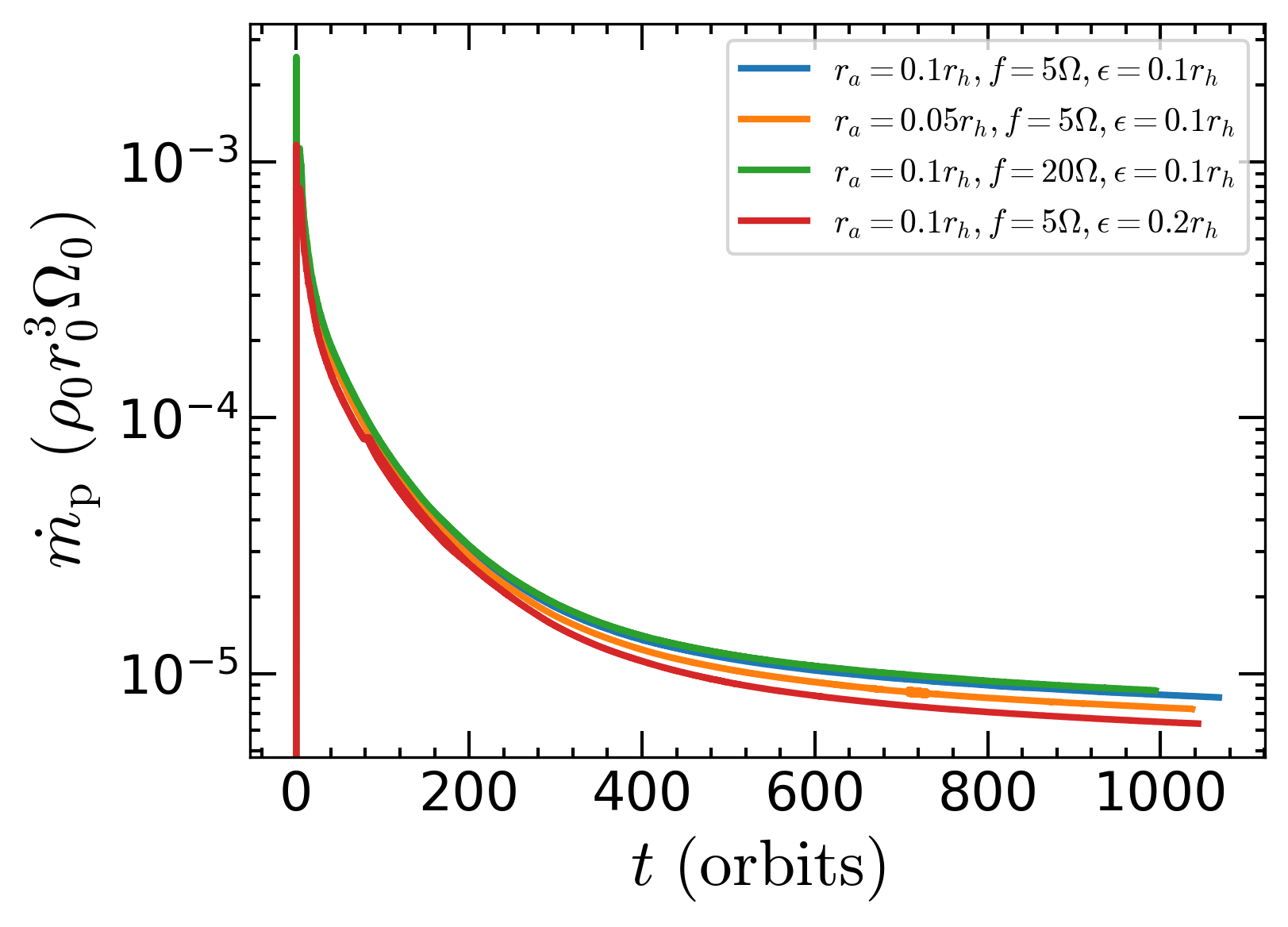}
\caption{The evolution of planetary accretion rate for different accretion parameters, measured in scale-free unit $\rho_{0}r_{0}^3\Omega_{0}$. The planet is on a circular orbit. The other model parameters are $q=0.001$, $h_{0}=0.05$, and $\alpha=0.001$.
}
 \label{fig:convergence}
\end{figure}

Here we test the dependence of planetary accretion rates on accretion parameters. There are a few parameters to control the accretion onto the planet, i.e., removal rate $f$, removal/sinkhole radius $r_{\rm a}$, and softening radius $\epsilon$. We show the effect of these parameters on the planetary accretion rates for the circular planet, which are shown in Figure~\ref{fig:convergence}. The other model parameters are $q=0.001$, $h_{0}=0.05$, and $\alpha=0.001$. There is only slightly decrease of $\dot{m}_{\rm p}$ with the increasing of $\epsilon=0.1\ r_h$ to $\epsilon=0.2\ r_h$, which is a good indication of convergence for our simulations. Similar tendencies are also seen for the dependence on $f$ and $r_{\rm a}$.  We also test the dependence for the eccentric planet of $e=0.1$, which shows similar convergence as the circular planet. Therefore, $\epsilon=0.1\ r_h$, $f=5\Omega_0$, and $r_{\rm a}=0.1\ r_h$ are chosen as our fiducial accretion parameters.

\section{Dependence on the global disc's vertical and radial extent}\label{app:bd}

\begin{figure}
\centering
\includegraphics[width=0.45\textwidth,clip=true]{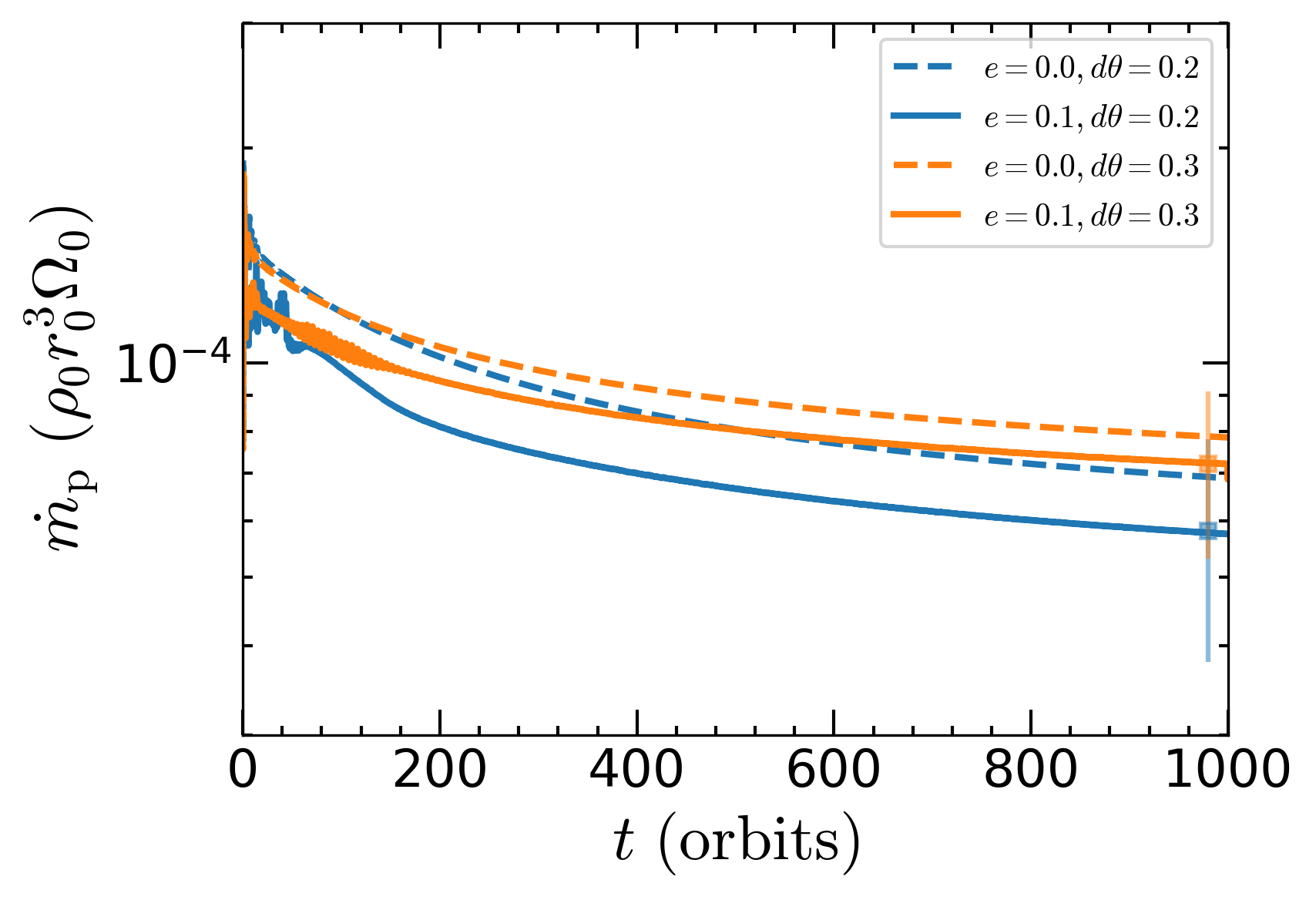}
\includegraphics[width=0.45\textwidth,clip=true]{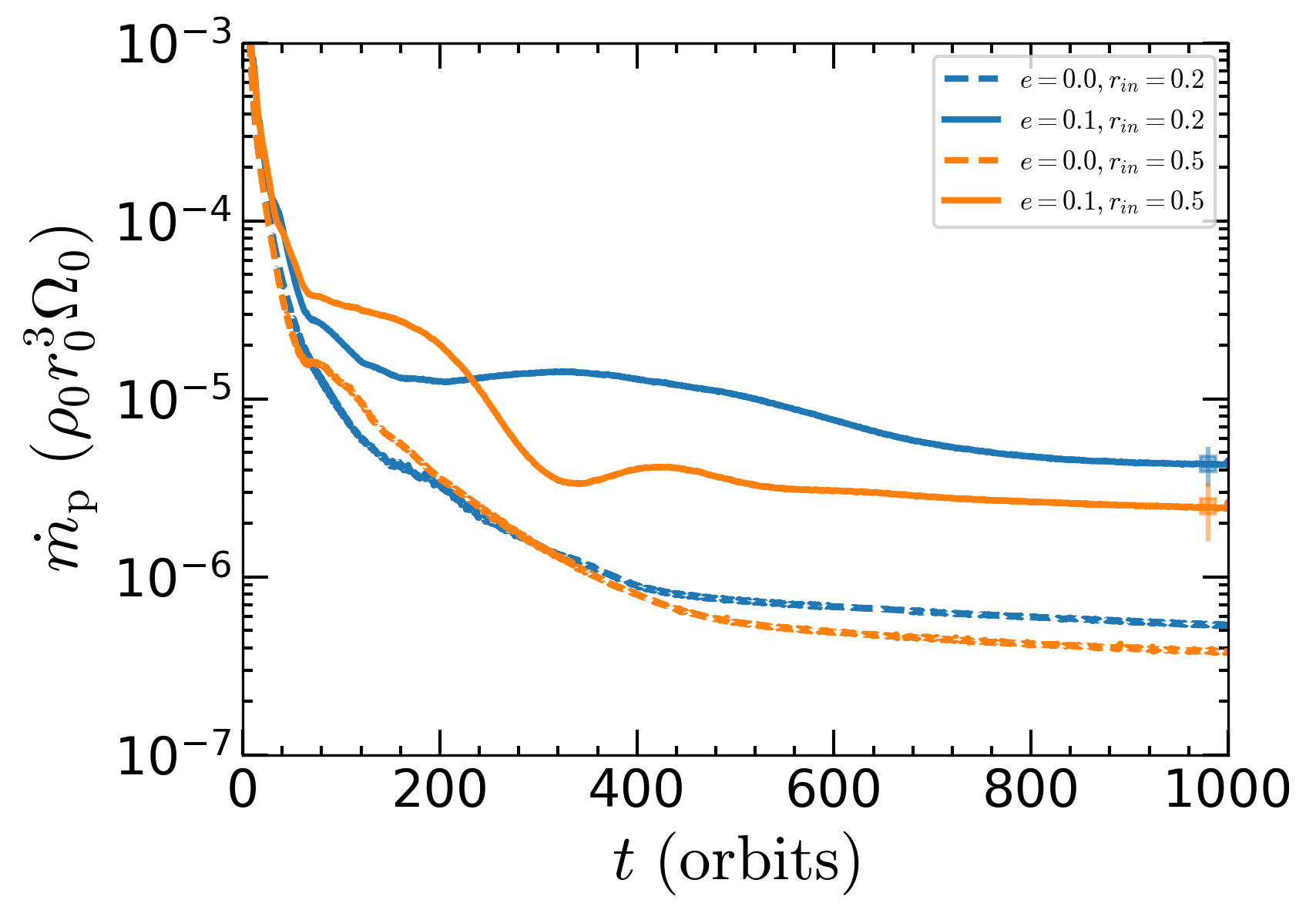}
\caption{The effect of different boundaries on the accretion rates. The upper panel tests the effect of vertical domain ($d\theta=0.2$ rad and $d\theta=0.3$ rad), while the lower panel is for the effect of inner boundary ($r_{\rm in}=0.2\ r_{0}$ and $r_{\rm in}=0.5\ r_{0}$). The error bars shows the typical variability amplitude for the eccentric cases. 
The other model parameters are $q=3\times10^{-4}$, $h_{0}=0.10$ for the upper panel,  while they are $q=0.004$, $h_{0}=0.035$ for the lower panel. The disc viscosity is fixed as $\alpha=0.001$. Note that $r_{\rm in}=0.5\ r_{0}$ and $d\theta=0.2$ rad for our fiducial run.  We can see that both the radial and vertical boundary imposes minor effect on the accretion rates. 
}
 \label{fig:convergence_bd}
\end{figure}

For our fidicial run, the vertical extension of the disc has $d\theta=0.2$ rad, i.e., $\theta_{\rm min}=\pi/2-0.2$. Such a vertical domain may impact the vertical flow pattern, and thus the accretion rates for some runs with $h_{0}=0.10$. To this end, we have several additional runs which extend the vertical domain to $\theta_{\rm min}=\pi/2-0.3$. We have fixed the other model parameters as $q=3\times10^{-4}$, $h_{0}=0.10$, and tests with both circular and eccentric cases. The planetary accretion rates are shown in upper panel of Figure~\ref{fig:convergence_bd}. We can see that the extension of the vertical domain of the disc impose negligible effect on the planetary accretion rates for both the circular and eccentric cases.

For the very high $q_{\rm th}$ cases, the inner edge of the global disc $r_{\rm in}=0.5\ r_{0}$ may be too close the  gap edge carved by the embedded planet. It is thus worthwhile justifying the effect of radial inner edge by extending the inner boundary further inward. We show the results of accretion rates with $r_{\rm in}=0.2\ r_{0}$ and $r_{\rm in}=0.5\ r_{0}$ in the lower panel of Figure~\ref{fig:convergence_bd}. To ensure a similar resolution as our fiducial runs, we increase the the radial grid number accordingly. After about 1000 orbits, the accretion rates for both the circular and eccentric planets do not show significant long-term accretion variability. We have also confirmed that disc eccentricity excitation is insignificant for both radial boundary in our 3D models. The averaged accretion rate onto both circular and eccentric planets are also comparable. All of these suggest a minor effect of the inner boundary.

\section{Testing the gap opening factor} \label{app:kgap}

\begin{figure}
\centering
\includegraphics[width=0.48\textwidth,clip=true]{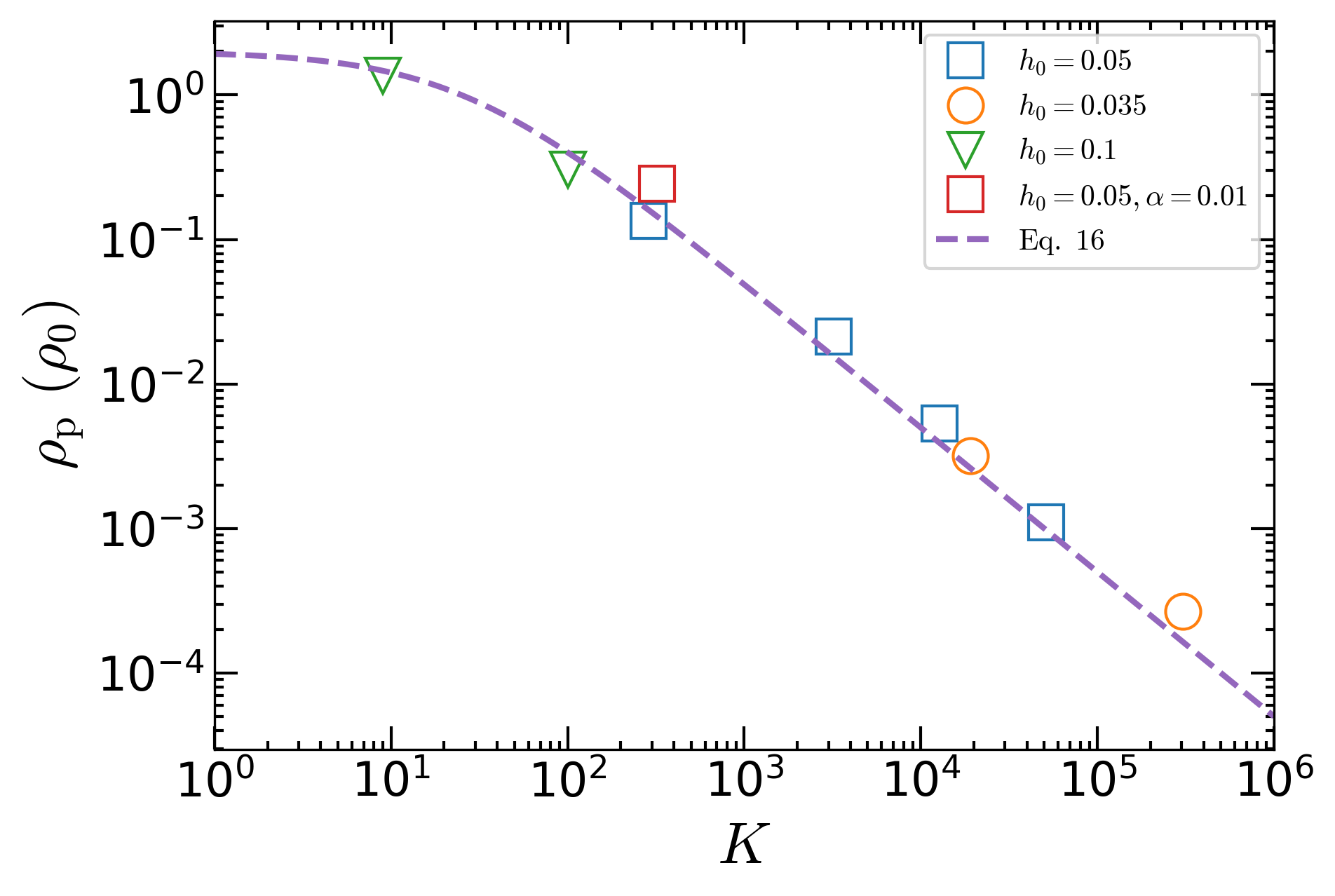}
\caption{The gap density $\rho_{\rm p}/\rho_{0}$ as a function of $K (\equiv q^2/\alpha h_{0}^{5}$) parameter. Here we only show the simulation results from circular planets. Except for the red circle, all other runs use $\alpha=0.001$. The dashed line shows the predicated gap density based on Equation~\ref{eq:rho_gap}.
}
 \label{fig:Kgap}
\end{figure}

As far as we know, this work is the first to apply 3D global simulation to study gas giants embedded in a PPD over a wide range of disc parameters including $q$, $h_0$ and $\alpha$ as well as over the gap-opening timescale. 
\citet{Fung2016} mainly discussed the dependence on $q$ fixing $h_0$ as well as $\alpha$.
Since the gap opening process have reached a quasi-steady state by the end of our runs, it's natural to compare the $\rho_{\rm p}/\rho_0$ results with the prediction by Equation \ref{eq:rho_gap} that has been established from 2D theory and simulations. 
We plot $\rho_{\rm p}/\rho_0$ from circular planet simulations (with fiducial boundary condition) in Figure~\ref{fig:Kgap}. 
We can see that the predicated gap densities based on Equation \ref{eq:rho_gap} are well consistent with our 3D simulation results both for super-thermal and sub-thermal cases. However, see Appendix \ref{app:inner_bc} for effect of more realistic 
global disc's inner boundary conditions.

\section{Effect of disc Depletion}\label{app:inner_bc}

As we have discussed in Section~\ref{sec:scaling_relation}, when $\dot{m}_{\rm p}$ approaches $\dot{M}_{\rm 0}$, 
the depletion of the inner disc should further suppress the planetary accretion rates, but this effect cannot be captured by our fiducial inner boundary condition. 
This effect should be important especially for the case when $ q \lesssim 2.4 h_0^{9/4}$ where ${\dot m}_{\rm H} > {\dot M}_{\rm 0}$. 
Under such circumstance, 
the inner disc could be servery depleted, which then modify the inner disc structure and the gap depth. 
To quantify this effect, we perform additional tests with a modified outflow inner boundary condition, 
where the inflow into the computation domain is not allowed so that the inner disc can freely evolve according to the supply rate from the outer disc, subtracted by the planetary accretion rate. 

\begin{figure}
\centering 
\includegraphics[width=0.45\textwidth,clip=true]{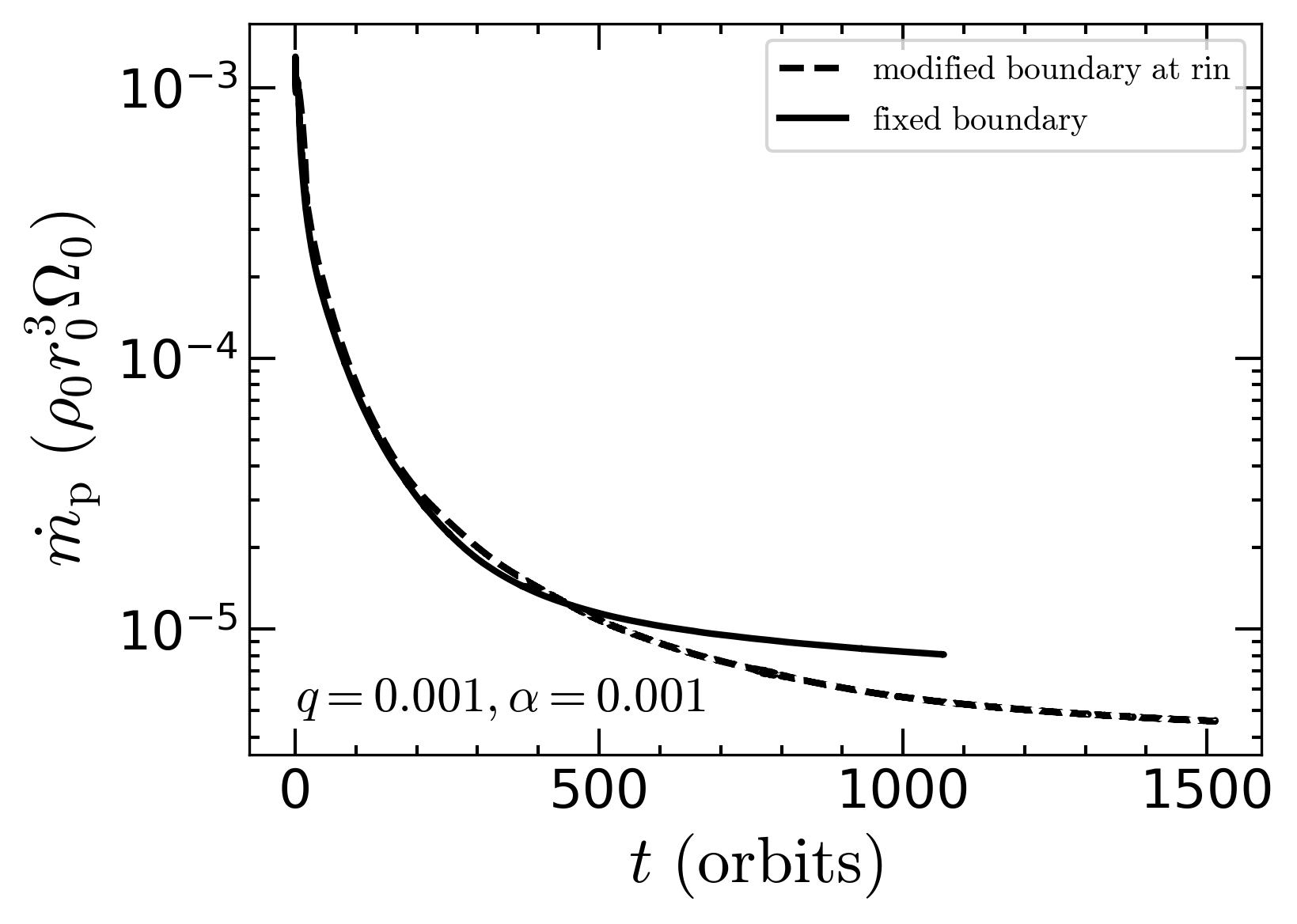}
\caption{
The effect of inner boundary on the planetary accretion rates. The model parameters are $q=0.001$, $\alpha=0.001$ and $h_{0}=0.05$. The planetary eccentricity is fixed as $e=0.0$. The dashed line uses the outflow boundary at the inner edge while inflow into the computation domain is prohibited, while the solid line shows the fixed boundary as in Figure~\ref{fig:mdot_ecc}.
}
 \label{fig:mdot_bd_lowa}
\end{figure}

Figure~\ref{fig:mdot_bd_lowa} shows the effect for our fiducial model parameter where $q=0.001$, $\alpha=0.001$ and $h_{0}=0.05$. 
After 1500 orbits, 
the planetary accretion rate is further suppressed by a factor of $\sim2$ compared to the fiducial inner boundary results and closer to the outer disc accretion rate $\dot{M}_{0} \approx 3\times 10^{-6}\rho_{0} r_{0}^3 \Omega_{0} $. 
However, the inner disc is still gradually depleting at this stage, indicating that the simulation requires more time to reach a steady state. 
This process could potentially take several viscous timescales (roughly $(2\pi \alpha h_0^2)^{-1}$ orbits) at $r_0$, presenting a significant computational challenge.
We thus resort to increase $\alpha$ which decreases the viscous timescale measured in orbits, 
while the accretion rate of the disc still conforms with the scaling relation with viscosity discussed in Section~\ref{sec:vis}.

\begin{figure}
\centering 
\includegraphics[width=0.45\textwidth,clip=true]{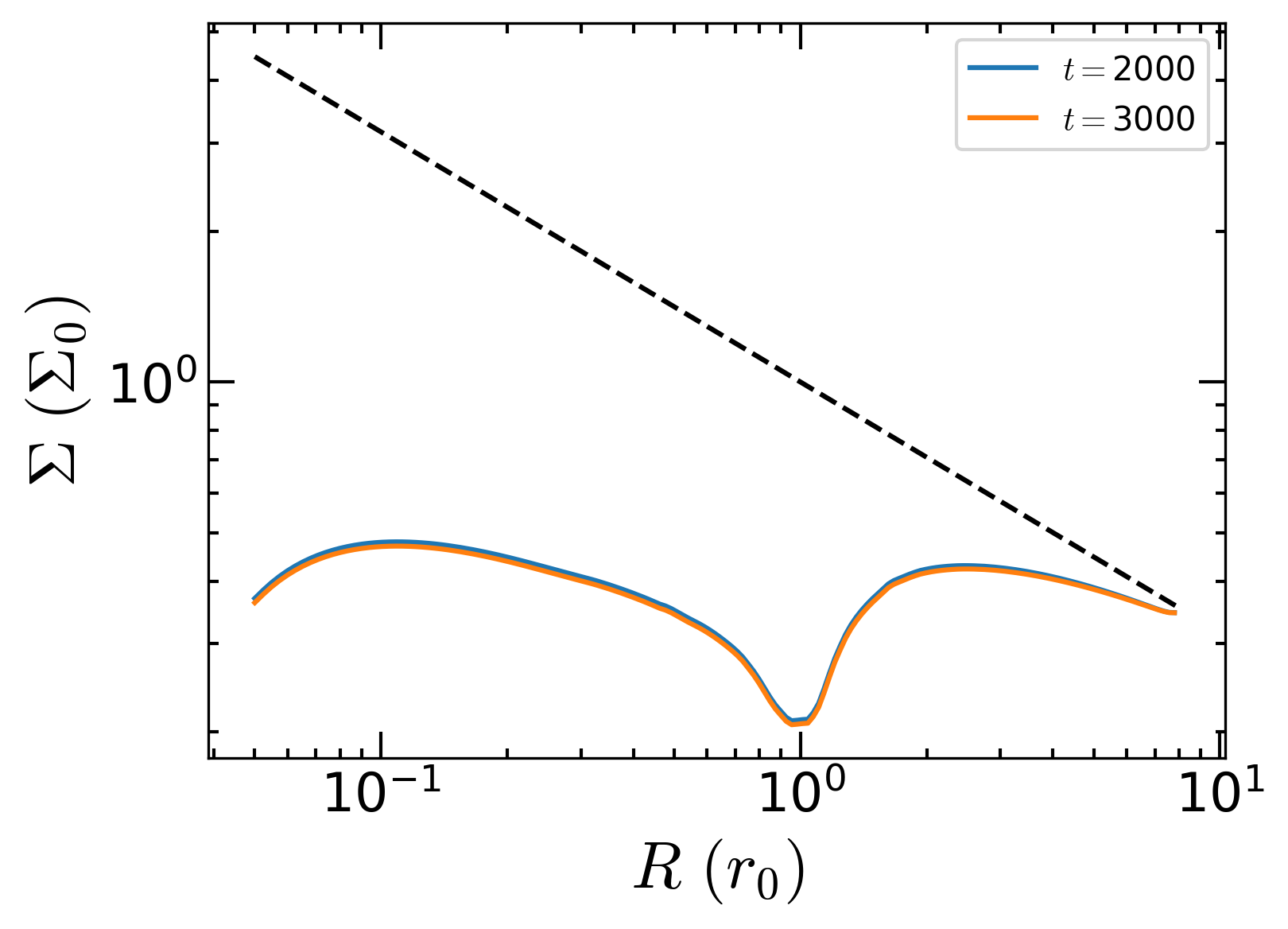}
\includegraphics[width=0.45\textwidth,clip=true]{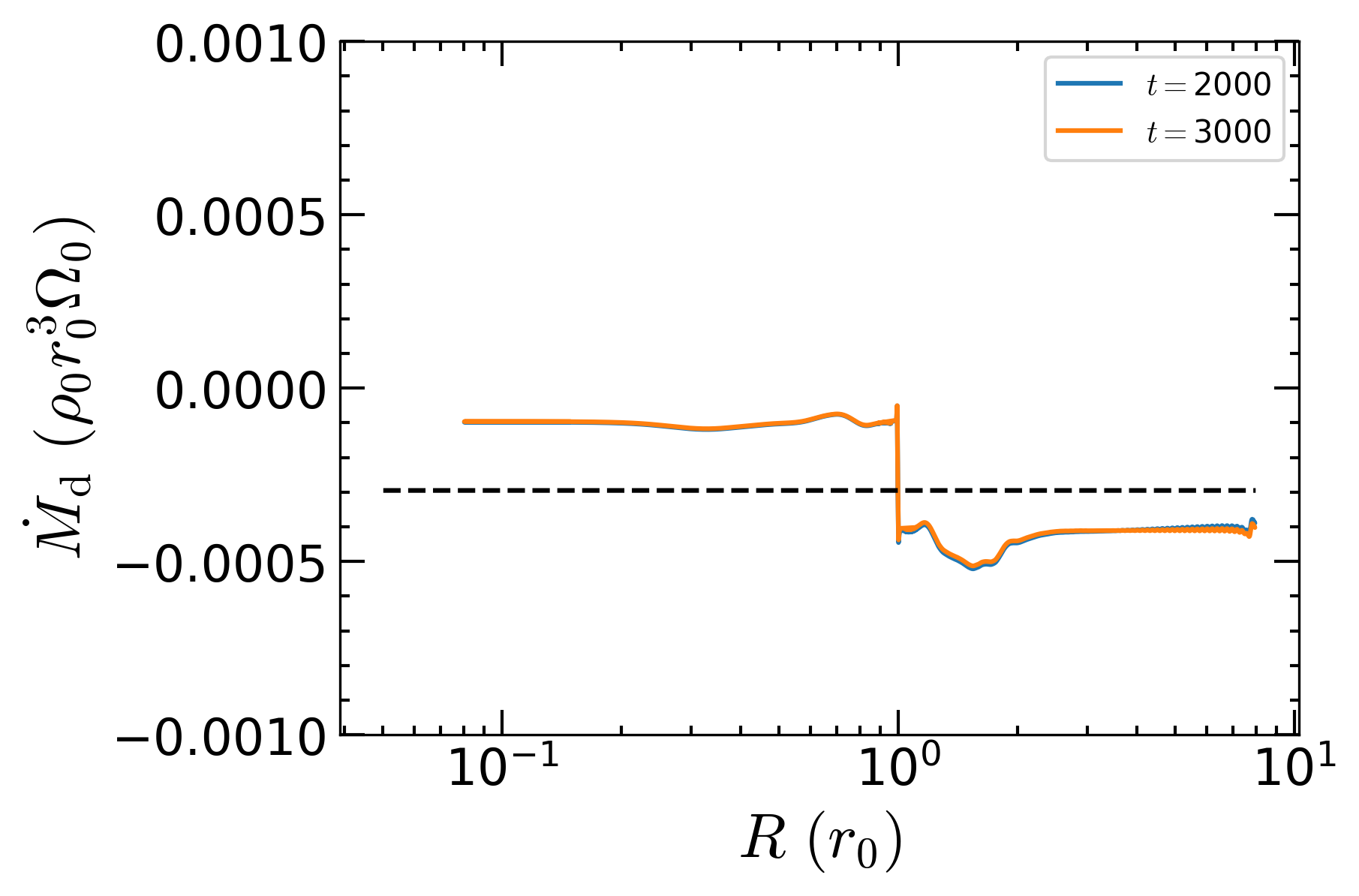}
\caption{
The disc surface densities (upper panel) and disc accretion rates (lower panel) at different times when adopting a modified inner boundary condition. The model parameters are $q=0.001$, $\alpha=0.1$ and $h_{0}=0.05$. The planetary eccentricity is fixed as $e=0.0$. The jump around $R=1.0$ suggests a planetary accretion rate of $3\times10^{-4}\rho_{0} r_{0}^{3}\Omega_{0}$, which is consistent with the measured accretion rates in the circumplanetary disc. The dashed lines show the initial surface density (upper panel) and mass flux (lower panel) across the disc.
}
 \label{fig:disc_bd_higa}
\end{figure}

Figure~\ref{fig:disc_bd_higa} shows the disc surface density profile and disc accretion rate with $\alpha=0.1$ and the modified inner boundary condition. 
A larger outer radial domain $r_{\rm out}=8$ and smaller $r_{\rm in}=0.05$ are adopted to minimize boundary effects. 
A logarithmic-uniform grid with mesh refinement is used to reach the similar resolution around the planet location $r=r_{0}$ as before. 
We have evolved the disc to $3000$ orbits, which is about a factor of 7 
longer than the viscous timescale at from $r_{0}$ to the inner boundary. 
As we can see in Figure~\ref{fig:disc_bd_higa}, 
the disc surface and accretion profile in the whole disc does not evolve after 2000 orbits. 
The disc accretion rate profile is almost radially constant except at $r_{0}$ where the jump of disc accretion rate matches the planetary accretion at the same orbital time. 
All of these suggest a viscous steady state of the global disc.

Compared with our fiducial cases, the inner disc in Figure~\ref{fig:disc_bd_higa} is significantly depleted. 
The planetary accretion rate (consistent with the magnitude of the jump in $\dot{M}_{\rm d}(r)$ at $r_0$) measured in this modified inner boundary is also capped by the disc accretion rate fed from the outer boundary, 
indicated by the dashed line in the lower panel to be ${\dot M}_0 = 3\times 10^{-4}\rho_0 r_0^3 \Omega_0$. 
Note that although $\dot{M}_{\rm d}$ in the outer boundary is supposed to be fixed at this initial unperturbed value, when the disc reaches a steady state the planet is drawing an additional mass flux from the outer boundary, such that the accretion rate in the outer disc increases to $\approx 4\times 10^{-4}\rho_0 r_0^3 \Omega_0$. Consequently, there is a residual inner disc accretion rate of $\approx 1\times 10^{-4}\rho_0 r_0^3 \Omega_0$, although the planet accretes nearly a hundred per cent of ${\dot M}_0$. This effect may be somewhat spurious because 
it's mitigated when we extend the outer boundary to even larger distance, 
but such simulations would require even longer time to reach viscous steady state.

If the accretion rate in the outer disc is accurately controlled, 
it would be possible use these modified inner boundary results to verify analytical estimations of planetary accretion rate constrained by the disc supply \citep[e.g.][see their Appendix A]{Rosenthal2020}, 
namely modifying $\dot{m}_{\rm H}$ in Equation \ref{eq:mpdot_md} to become

\begin{equation}
    \tilde{\dot{m}}_{\rm H} = \dfrac{\dot{m}_{\rm H} \dot{M}_{\rm d}}{\dot{m}_{\rm H} + \dot{M}_{\rm d}}
\end{equation}

While mild artificial mass generation prevents us from verifying this formula
to utmost precision, our results suggest that Equation \ref{eq:mdot_cap} is sufficient in effectively capturing the essence of the capping effect, in spite of numerical uncertainties.

Finally, it is important to recognize that the depletion of the inner disc can significantly affect the migration torque experienced by actively accreting planets. 
This particular aspect will be thoroughly examined in future studies.

\end{appendix}

\bsp	
\label{lastpage}

\end{document}